\documentclass[aps,10pt,notitlepage,twocolumn,prd,preprintnumbers,longbibliography,nofootinbib,showpacs]{revtex4-2}
\usepackage[colorlinks=true,urlcolor=blue,linkcolor=blue,citecolor=blue]{hyperref}
\usepackage{slashed,color,amsmath,amssymb,mathrsfs}
\usepackage{amsfonts,epsfig,amssymb}
\usepackage{graphicx}
\usepackage{longtable}
\usepackage{array}
\usepackage{accents}
\usepackage{tensor}
\usepackage{footmisc}
\usepackage{longtable}
\usepackage{booktabs}
\usepackage{multirow}
\usepackage{bm}
\usepackage{subfigure}
\usepackage{gensymb}
\usepackage{textcomp}
\usepackage{dcolumn}
\usepackage{tikz}
\usepackage{bm}

\allowdisplaybreaks

\begin{document}
\title[]{Study of transition form factors of the lightest pseudoscalars}
	\author{Yi-Hao Zhang$^{1,2,3}$}
	\author{Shao-Zhou Jiang$^{1}$}
	\email{jsz@gxu.edu.cn}    
	\author{Ling-Yun Dai$^{2,3}$}
	\email{dailingyun@hnu.edu.cn}
	\affiliation{$^{1}$ Key Laboratory for Relativistic Astrophysics, School of Physical Science and Technology, Guangxi University, Nanning 530004, People’s Republic of China}
	\affiliation{$^{2}$ School for Theoretical Physics, School of Physics and Electronics, Hunan University, Changsha 410082, China}
	\affiliation{$^{3}$ Hunan Provincial Key Laboratory of High-Energy Scale Physics and Applications, Hunan University, Changsha 410082, People's Republic of China}
	
\date{\today}

\begin{abstract}		
In this paper, we study the transition form factors of the lightest pseudoscalar mesons, $\pi^0$, $\eta$, and $\eta'$, within the framework of resonance chiral theory. Our analysis is performed based on the data of time-like and space-like singly-virtual and space-like doubly-virtual form factors, as well as the relevant cross sections and latest invariant mass spectra of $e^+e^-$ pair for the process of $P\to\gamma e^+ e^-$. The transition form factors of these pseudoscalars are obtained. 
Also, we evaluate their contributions to the light-by-light part of the anomalous magnetic moment of the muon.
Our two Fits give similar results, where Fit-A gives $a_\mu^{\pi^0 }=(61.6\pm 1.8)\times10^{-11}$, $a_\mu^{\eta }=(15.2\pm1.7)\times10^{-11}$, $a_\mu^{\eta'}=(16.0\pm 1.2)\times10^{-11}$, and the total contribution of neutral pseudo-scalar meson poles is $a_\mu^{\pi^0+\eta+\eta'}=(92.8\pm2.9)\times10^{-11}$.
	\end{abstract}
	
	\maketitle
	
 Quantum Chromodynamics (QCD) \cite{Greiner:2007zz} is considered as the fundamental theory of the strong interaction, which describes the interactions between quarks and gluons. However, the coupling $\alpha_s$ has a nature of asymptotic free, causing the challenge to apply perturbative QCD (pQCD) in the low energy region. There are several alternative approaches that are proposed to study the property of hadrons and their interactions, e.g., chiral perturbation theory (ChPT) \cite{Weinberg:1978kz,Gasser:1983yg},  Dyson-Schwinger equations (DSEs.)~\cite{Eichmann:2019tjk,Raya:2019dnh}, Nambu-Jona-Lasinio (NJL) model \cite{Nambu:1961tp,Klevansky:1992qe}, Vector-Meson-Dominance (VMD) model \cite{Sakurai:1969ss,Schildknecht:2005xr}, AdS/QCD \cite{Witten:1998qj,Cappiello:2010uy}, Lattice QCD (LQCD) \cite{Gupta:1997nd,Gattringer:2010zz}, and the dispersion relations (DRs)~\cite{Colangelo:2001df,Xiao:2000kx,Descotes-Genon:2006sdr,Dai:2014zta}. 
 Among them, ChPT is quite successful in studying the lightest pseudoscalars and their interactions. However, due to the momentum expansion, ChPT is only applicable within the low energy region. To describe hadron interactions involving resonances, one needs Resonance Chiral Theory (RChT) \cite{Ecker:1988te,Ecker:1989yg,Cirigliano:2006hb,Kampf:2006yf,Portoles:2010yt,Kampf:2011ty}. The guiding principles for constructing the RChT Lagrangian encompass chiral and discrete symmetries and Lorentz invariance. 

Recently, the transition form factors (TFFs) of the lightest pseudoscalars have attracted plenty of attention from the physics community. It  characters the property of non-perturbative QCD and is crucial for precisely calculating the hadronic contribution to the anomalous magnetic moment of the muon, $a_{\mu}=(g-2)_{\mu}/2$,
The experimental results on $(g-2)_{\mu}$ from the Fermi National Accelerator Laboratory (FNAL)~\cite{Muong-2:2021ojo,Muong-2:2025xyk} and the Brookhaven National Laboratory (BNL)~\cite{Muong-2:2006rrc} have achieved a remarkable precision of 124 ppb. On the theoretical side, the standard model (SM) predictions for $(g-2)_{\mu}$ can be categorized into three components: the electromagnetic part, the electroweak part, and the strong interaction part. The last one has the largest uncertainty and it can be divided into two parts: the hadronic vacuum polarization (HVP) \cite{Hoid:2020xjs,Benayoun:2021ody,Yi:2021ccc,Hoferichter:2021wyj,Colangelo:2022jxc,Dai:2013joa, Wang:2023njt,Qin:2024ulb,Aliberti:2025beg} and the hadronic light-by-light-scattering (HLbL) \cite{Roig:2014uja,Guevara:2018rhj,Raya:2019dnh,Estrada:2024cfy,Estrada:2025bty,Colangelo:2014dfa,Hoferichter:2018kwz,Holz:2024lom,Kadavy:2022scu,Gerardin:2023naa,ExtendedTwistedMass:2022ofm,ExtendedTwistedMass:2023hin,Hayakawa:1997rq,Cappiello:2010uy,Danilkin:2019mhd}, the HVP contribution is of order $\alpha^2$,\footnote{The largest uncertainty is from HVP.  The analyses from LQCD \cite{Gerardin:2023naa,ExtendedTwistedMass:2022ofm,ExtendedTwistedMass:2023hin} (and $\tau$ decays \cite{Miranda:2018cpf,Hoferichter:2023sli,GomezDumm:2013sib}) imply that there is no discrepancy between the theoretical and experimental $(g-2)_\mu$ values, while that of the data-driven method \cite{Keshavarzi:2018mgv,Colangelo:2019uex,Davier:2019can,Keshavarzi:2019abf,Qin:2020udp,Wang:2023njt,Fedotovich:2024dpk}, based on $e^+e^-$ annihilations \cite{KLOE:2008fmq,KLOE:2010qei,KLOE:2012anl,KLOE-2:2017fda,SND:2020nwa,BESIII:2015equ,Xiao:2017dqv,CMD-2:2005mvb,Aulchenko:2006dxz,CMD-2:2006gxt}, is different. The latest CMD-3 \cite{CMD-3:2023alj,CMD-3:2023rfe} and BaBar \cite{BaBar:2013jqz} measuements are compatible with the LQCD results. } while the HLbL contribution is of order $\alpha^3$.

In this analysis, we focus on the HLbL. 
The most significant contributions to HLbL arise from the energy region around the muon mass \cite{Jegerlehner:2009ry}, associated with the pion-pole and the pion-box, while the contributions from 500~MeV to 1000~MeV \cite{Bijnens:1995cc} are crucial, too. In this energy region, the $\eta$, $\eta'$-poles, kaon-box, and constituent quark loop contributions dominate. 
We calculate the TFFs of neutral pseudoscalar mesons ($\pi^0$, $\eta$, and $\eta'$) by RChT, where, for example, the interactions between vector and pseudoscalars are crucial. With these TFFs, one can evaluate the pseudoscalar pole contributions to $a_{\mu}^{\mathrm{HLbL}}$.
Studies on these TFFs can also be found in  LQCD~\cite{Gerardin:2023naa,ExtendedTwistedMass:2022ofm,ExtendedTwistedMass:2023hin}, DRs~\cite{Colangelo:2014dfa,Hoferichter:2018kwz,Holz:2024lom,Holz:2024diw,Aliberti:2025beg}, ChEFT~\cite{Guevara:2018rhj,Raya:2019dnh,Kadavy:2022scu,Estrada:2024cfy,Wang:2023njt}, Canterbury approximants~\cite{Masjuan:2017tvw}, NJL model~\cite{Bijnens:1995cc,Bijnens:2001cq,Bartos:2001pg}, AdS/QCD~\cite{Cappiello:2010uy}, DSEs~\cite{Eichmann:2019tjk,Raya:2019dnh}, VMD model~\cite{Hayakawa:1997rq}, and so on.
These approches, though successfully describe the TFFs of the lightest pseudoscalars in a certain energy region, do not give complete analyses in both the low and high energy region for the timelike and spacelike cases. For instance, the heavier vector resonance, appearing in the intermediate 1-2~GeV in the timelike region, are usually not considered. In contrast, in the present analysis the contribution of the heavier vector resonances can be estimated by extension of the Breit-Wigner propagators \cite{Dai:2013joa,Qin:2020udp,Wang:2023njt,Qin:2024ulb}, respecting the fact that the heavier multiplets have the same topologies/dynamics in Feynman diagrams as the lighest vector nonet \cite{Dai:2013joa}. 
To give a better description of the TFFs of $\eta$, and $\eta'$, we apply the $U(3)$ RChT instead of $SU(3)$ one. The $\delta$-expansion \cite{Kaiser:2000gs,Kaiser:2005eu}, a reasonable combination of the large-$N_C$ expansion and chiral counting rules,  is used to compensate for the power-counting. 
One can fix lots of the unknown couplings of RChT by matching with QCD at high energies \cite{Ecker:1988te,Ecker:1989yg,Ruiz-Femenia:2003jdx,Cirigliano:2006hb} and ChPT at the low energies \cite{Ecker:1988te,Ecker:1989yg,Kampf:2011ty}.


There are some processes that can be analyzed together to constrain the TFFs of the pseudoscalars, i.e., the Single-Dalitz decays of pseudoscalar ($P$) \cite{Landsberg:1985gaz} and the $e^+e^-$ annihilation processes \cite{Wang:2023njt} related to the time-like singly-virtual TFFs,  the two-photon scattering processes \cite{Danilkin:2019mhd} related to the space-like singly- and doubly-virtual TFFs. Besides, the LQCD data of the doubly-virtual TFFs are included in our anlaysis.
 The experimental datasets of $P \to \gamma l^+l^-$ are given as follows:  A2 \cite{A2:2016sjm} and NA62 \cite{NA62:2016zfg} for $\pi^0 \to  \gamma l^+l^-$; A2 \cite{Adlarson:2016hpp}, NA60 \cite{NA60:2016nad} and BESIII \cite{BESIII:2024pxo} for $\eta \to  \gamma l^+l^-$; and BESIII \cite{BESIII:2024pxo,BESIII:2015zpz} for $\eta' \to  \gamma l^+l^-$. 
For the very recent $\eta$ and $\eta'$ data from BESIII, the next-to-leading-order (NLO) radiative corrections have to be considered. In addition, we also consider the data of $\eta'\to\omega e^{+}e^{-}$ from BESIII \cite{BESIII:2015jiz}, as a supplement to give more constraints on the TFFs.  
 The $e^+e^-$ annihilation processes are taken from the following experiments: $e^+e^-\to\pi\gamma/\eta\gamma$ from Refs.~\cite{SND:2016drm,Achasov:2000zd,Achasov:2018ujw,Achasov:2003ed,Achasov:2006dv,Achasov:2013eli,CMD-2:2004ahv,CMD-2:2001dnv}. Note that these two processes have been studied in our previous work \cite{Wang:2023njt}, but with $SU(3)$ RChT. Moreover, the SND group has published their new measurement of the cross-section of $e^+e^-\to\eta'\gamma$ \cite{SND:2024qaq}, and we will include it in our analysis. The $U(3)$ RChT can describe the $\eta-\eta'$ mixing well. 
 The datasets of space-like singly-virtual TFFs are taken from CELLO \cite{CELLO:1990klc}, CLEO \cite{CLEO:1997fho}, BaBar \cite{BaBar:2009rrj}, Belle \cite{Belle:2012wwz} and BESIII  \cite{Redmer:2019zzr} for $\pi^0\gamma\gamma^*$, CELLO \cite{CELLO:1990klc}, CLEO \cite{CLEO:1997fho} and BaBar \cite{BaBar:2011nrp} for $\eta\gamma\gamma^*$, and CELLO \cite{CELLO:1990klc}, CLEO \cite{CLEO:1997fho}, LEP \cite{L3:1997ocz} and BaBar \cite{BaBar:2011nrp} for $\eta'\gamma\gamma^*$. The only available data of doubly-virtual TFFs is from BaBar \cite{BaBar:2018zpn} for $\eta'$. To complete our analysis, we also include the LQCD results, e.g., BMW \cite{Gerardin:2023naa}, ETM \cite{ExtendedTwistedMass:2023hin}, and Ref.~\cite{Lin:2024khg} for $\pi^0\gamma^*\gamma^*$,  BMW \cite{Gerardin:2023naa} and ETM \cite{ExtendedTwistedMass:2022ofm} for $\eta\gamma^*\gamma^*$, and BMW \cite{Gerardin:2023naa} for $\eta'\gamma^*\gamma^*$. 
A combined analysis of all the above processes can fix the unkown couplings reliably, resulting in strong constraints on the TFFs .

 The paper is organized as follows: In Sec.~\ref{sec2}, we give a brief introduction to the theoretical framework of RChT. Based on it, we construct the doubly-virtual TFFs. In Sec.~\ref{sec3}, we give a comprehensive analysis of all the data discussed above and get the time-like singly-virtual, space-like singly-virtual, and space-like doubly-virtual TFFs. In Sec.~\ref{sec4}, we evaluate their contributions to $a_{\mu}^{\mathrm{HLbL}}$. Finally, we give our conclusion in Sec.~\ref{sec5}.

\section{Theoretical framework}\label{sec2}
	\subsection{Construction of the \textbf{U(3)} resonance chiral effective Lagrangian}
As discussed above, ChPT is founded on the principle that the lowest pseudoscalar octet are Goldstone bosons generated by the spontaneous breaking of chiral symmetry of the light quarks, $u$, $d$, and $s$. It is very successful in describing the low energy interactions of these pseudoscalars. In the higher energy regions, another light meson, $\eta'$, and resonances ($\rho$, $\omega$, $\phi$, etc.) emerge, and their roles can not be ignored. 
The $U(3)$ RChT \cite{Gasser:1984gg, Gasser:2020mzy,Chen:2012vw} is a theory that can include both the ninth Goldstone boson and the other lightest resonances. It is grounded on the Large-$N_C$ QCD. In the chiral and Large-$N_C$ limits, the $\eta'$ becomes massless and can be regarded as the ninth Goldstone particle \cite{Kaiser:2005eu}.     
 The effective interaction Lagrangian between the Goldstone nonet and the resonances of our interest is
\begin{equation}
\mathcal{L}^{\mathrm{int} } =   \mathcal{L} _{(2)}^{\mathrm{GB} }  +\mathcal{L} _{\mathrm{WZW} }+\mathcal{L} _{\mathrm{kin} }^{\mathrm{R} }+\mathcal{L} _{(2)}^{\mathrm{R} }  +\mathcal{L}_{(4)}^{R}+\mathcal{L}_{(2)}^{RR}\,, \label{Eq:L;int}
\end{equation}
where the subscripts ``2,4" in the bracket indicate the chiral counting about the lightest pseudoscalars, $O(p^{2})$ and $O(p^4)$, respectively.   $\mathcal{L} _{(2)}^{\mathrm{GB} }$ is the lowest order Lagrangians of $\mathrm{U}(3)$ ChPT, 
\begin{equation}
\mathcal{L} _{(2)}^{\mathrm{GB} } = \frac{F^{2} }{4} \left \langle \tilde{u} _{\mu } \tilde{u} ^{\mu }   \right \rangle +\frac{F^{2} }{4}\left \langle \tilde{\chi} _{+}    \right \rangle +\frac{F^{2} }{3}M_{0}^2 \mathrm{ln}^{2}\mathrm{det}  \tilde{u},  
\end{equation}
where the last term corresponds to the $\mathrm{U_{A}}(1)$ anomaly \cite{Kaiser:2005eu}, and one has $M_0^2\propto 1/N_C $~\cite{Witten:1979vv}. In this work, we use $M_0=900$~MeV as suggested in Ref.~\cite{Chen:2012vw}. 
The Goldstone nonet is represented by
\begin{equation}
\tilde{u} =\exp \left(\frac{i \tilde{\Phi}}{\sqrt{2} F}\right),
\end{equation}
where $F$ is the pion decay constant, given as $F\approx 92.2$~MeV \cite{ParticleDataGroup:2024cfk}. 
The $\mathrm{det}  \tilde{u}$ is given as $\mathrm{det}  \tilde{u}=\exp\left(\frac{i\sqrt{3}}{\sqrt{2}F}\eta_1\right)$, where $\eta_1$ is the pseudoscalar singlet field.
The physical fields of $\eta$ and $\eta'$ in the two-angle mixing scheme \cite{Feldmann:1998vh,Feldmann:1999uf,Guo:2015xva} is
   \begin{equation}
       \binom{\eta}{\eta^{\prime}}=\frac{1}{F}\left(\begin{array}{cc}
F_8 \cos \theta_8 & -F_0 \sin \theta_0 \\
F_8 \sin \theta_8 & F_0 \cos \theta_0
\end{array}\right)\binom{\eta_8}{\eta_1}.
   \end{equation}   
The matrix of the nonet, $\tilde{\Phi}$, is given as 
   \begin{equation}
        \tilde{\Phi}=   \left(\begin{array}{ccc}
            \frac{\eta^{\prime} C_q{ }^{\prime}+\eta C_q+\pi_0 }{\sqrt{2}} & \pi^{+} & K^{+} \\[2mm]
            \pi^{-} & \frac{\eta^{\prime} C_q{ }^{\prime}+\eta C_q-\pi_0 }{\sqrt{2}} & K_0 \\[2mm]
 K^{-} & \bar{K}_0 & \eta^{\prime} C_s{ }^{\prime}-\eta C_s
        \end{array}\right),
    \end{equation}
 where 
\begin{eqnarray}
    C_q     & = & \frac{1}{\sqrt{3} \cos \left(\theta_8-\theta_0\right)}\left(\frac{\cos \theta_0}{f_8}-\frac{\sqrt{2} \sin \theta_8}{f_0}\right),\nonumber \\
    C_q^{\prime} & = & \frac{1}{\sqrt{3} \cos \left(\theta_8-\theta_0\right)}\left(\frac{\sqrt{2} \cos \theta_8}{f_0}+\frac{\sin \theta_0}{f_8}\right), \nonumber\\
    C_s     & = & \frac{1}{\sqrt{3} \cos \left(\theta_8-\theta_0\right)}\left(\frac{\sqrt{2} \cos \theta_0}{f_8}+\frac{\sin \theta_8}{f_0}\right),\nonumber \\
    C_s^{\prime} & = & \frac{1}{\sqrt{3} \cos \left(\theta_8-\theta_0\right)}\left(\frac{\cos \theta_8}{f_0}-\frac{\sqrt{2} \sin \theta_0}{f_8}\right),
\end{eqnarray}
with $F_8=f_8\cdot F$ and $F_0=f_0\cdot F$ are decay constants of octet-singlet bases, and $\theta_0 $ and $\theta_8$ are corresponding mixing angles.
In this study, we do not focus on the theoretical details regarding the masses of Goldstone bosons; instead, we utilize their experimental values from PDG \cite{ParticleDataGroup:2024cfk}. The definitions of chiral building blocks can be found in Ref.~\cite{Chen:2012vw}, 
\begin{eqnarray}
    \tilde{u}_\mu & = & i\left[\tilde{u}^{\dagger}\left(\partial_\mu-i r_\mu\right) \tilde{u} -\tilde{u} \left(\partial_\mu-i \ell_\mu\right) \tilde{u}^{\dagger}\right], \nonumber \\
    \tilde{\chi}_{\pm} & = & \tilde{u}^{\dagger} \chi \tilde{u}^{\dagger} \pm \tilde{u}  \chi^{\dagger} \tilde{u}, \nonumber \\
    \chi & = & 2 B_0(s+i p)=\mathrm{diag} (m_\pi,~ m_\pi,~ 2m_K-m_\pi).
\end{eqnarray}
$\mathcal{L} _{\mathrm{WZW} }$ is the Wess-Zumino-Witten (WZW) term, of which the complete terms are given in Ref.~\cite{Witten:1983tw,Wess:1971yu}. The lowest order contribution relevant to this work is
\begin{equation}
\mathcal{L} _{\mathrm{WZW} }=   -\frac{\sqrt{2}N_C }{8\pi^2F} \varepsilon _{\mu \nu \rho \sigma } \left \langle \tilde{\Phi } \partial^{\mu }  v ^{\nu }\partial^{\rho }  v ^{\sigma }   \right \rangle ,
\end{equation}
the external vector current $v ^{\mu }$ is given as $v ^{\mu }=-eQA^{\mu}$, and $Q=\mathrm{diag}\left \{ \frac{2}{3}, -\frac{1}{3} , -\frac{1}{3}\right \} $ is the electric charge matrix of the three light flavor quarks.
    
The third term of Eq.~(\ref{Eq:L;int}), $\mathcal{L} _{\mathrm{kin} }^{\mathrm{R} }$ is the  kinetic term of the vector mesons,
    \begin{equation}
            \mathcal{L} _{\mathrm{kin} }^{\mathrm{R} }= - \frac{1}{2}\left\langle\nabla^\lambda V_{\lambda \mu} \nabla_\nu V^{\nu \mu}-\frac{M_V^2}{2} V_{\mu \nu}V^{\mu \nu}\right\rangle.
    \end{equation}
 In the Large-$N_C$ limit, the resonance octet and singlet also become degenerate, and could be collected as a nonet, whose physical fields are defined as 
\begin{eqnarray}
 V\!_{\!\mu\nu}\!=\!
    \left(
    \begin{array}{c c c}
    \frac{\rho^0}{\sqrt{2}}\!+\!\frac{\omega_8}{\sqrt{6}}\!+\!\frac{\omega_0}{\sqrt{3}}     &      \rho^+    & K^{\ast +} \\
    \rho^-     & -\!\frac{\rho^0}{\sqrt{2}}\!+\!\frac{\omega_8}{\sqrt{6}}\!+\!\frac{\omega_0}{\sqrt{3}}    & K^{\ast 0} \\
 K^{\ast -} &      \bar{K}^{\ast 0}     & -\!\frac{2\omega_8}{\sqrt{6}}\!+\!\frac{\omega_0}{\sqrt{3}} \\
    \end{array}
    \!\right)_{\mu\nu}\!, \nonumber
\end{eqnarray}  
with $\omega-\phi$ mixing given as 
\begin{eqnarray}
\label{eq:V;thetaV}
\left ( \begin{array}{c} 
\omega_{8} \\
\omega_{0}
\end{array}
\right ) &= \left ( \begin{array}{cc}
\cos\theta_{V} & \sin\theta_{V} \\
-\sin\theta_{V} & \cos\theta_{V}
\end{array}    \right )
\left (  \begin{array}{c} \phi \\ 
     \omega
\end{array}
\right ) .
\end{eqnarray}  
    
 The fourth term of Eq.~(\ref{Eq:L;int}) is the lowest order interaction Lagrangian with one resonance involved, 
    \begin{equation}
        \mathcal{L}_{(2)}^{\mathrm{R} }=\frac{F_V}{2 \sqrt{2}}\left\langle V_{\mu \nu} \tilde{f}_{+}^{\mu \nu}\right\rangle, 
    \end{equation}
Here, $\tilde{f}_{+}^{\mu \nu}$ is defined as $ \tilde{f}_{+}^{\mu \nu}=\tilde{u}  F_L^{\mu \nu} \tilde{u} ^{\dagger} + \tilde{u} ^{\dagger} F_R^{\mu \nu} \tilde{u}$, with $F_L^{\mu \nu}=F_R^{\mu \nu}=-eQ(\partial^\mu A^\nu-\partial^\nu A^\mu)$. 
The higher order term with one resonance, $\mathcal{L}_{(4)}^{\mathrm{R} }$, and the lowest order term with two resonances, $\mathcal{L}_{(2)}^{\mathrm{RR} }$, of our interest are
\begin{eqnarray}
\mathcal{L}_{(4)}^{R }&=& \mathcal{\tilde{O}} _{\mathrm{VJ}}+   \sum_{j=1}^{7} \frac{\tilde{c} _{j} }{M_{V} } \mathcal{\tilde{O}} _{\mathrm{VJP}} ^{j}+\tilde{c} _{8}M_{V}\mathcal{\tilde{O}} _{\mathrm{VJP}} ^{8} ,   \nonumber \\
\mathcal{L}_{(2)}^{RR }&=&\sum_{i=1}^{4} \tilde{d} _{i} \mathcal{\tilde{O}} _{\mathrm{VVP}} ^{i}+\tilde{d} _{5}M_{V}^2 \mathcal{\tilde{O}} _{\mathrm{VVP}} ^{5}.
\end{eqnarray}
Here, the first term is a higher order term for $\gamma V$ vertex, which is essential to study the phenomenology of vector decays \cite{Cirigliano:2006hb,Dai:2013joa}, 
    \begin{equation}
        \mathcal{\tilde{O}} _{\mathrm{VJ}}=\frac{\alpha _{V}F_{V}  }{M_{V}^2 }\left \langle V_{\mu \nu } \left \{ \tilde{f}_{+}^{\mu \nu } ,\tilde{\chi }_{+}    \right \}   \right \rangle\,.
    \end{equation}
The VJP and VVP operators are given in Ref.~\cite{Chen:2012vw},
\begin{eqnarray}
    \mathcal{\tilde{O}}_{V J P}^1 & = & \varepsilon_{\mu \nu \rho \sigma}\left\langle\left\{V^{\mu \nu}, \tilde{f}_{+}^{\rho \alpha}\right\} \nabla_\alpha \tilde{u}^\sigma\right\rangle, \nonumber \\
    \mathcal{\tilde{O}}_{V J P}^2 & = & \varepsilon_{\mu \nu \rho \sigma}\left\langle\left\{V^{\mu \alpha}, \tilde{f}_{+}^{\rho \sigma}\right\} \nabla_\alpha \tilde{u}^\nu\right\rangle, \nonumber \\
    \mathcal{\tilde{O}}_{V J P}^3 & = & i \varepsilon_{\mu \nu \rho \sigma}\left\langle\left\{V^{\mu \nu}, \tilde{f}_{+}^{\rho \sigma}\right\} \tilde{\chi}_{-}\right\rangle, \nonumber \\
    \mathcal{\tilde{O}}_{V J P}^4 & = & i \varepsilon_{\mu \nu \rho \sigma}\left\langle V^{\mu \nu}\left[\tilde{f}_{-}^{\rho \sigma},\tilde{ \chi}_{+}\right]\right\rangle, \nonumber \\
    \mathcal{\tilde{O}}_{V J P}^5 & = & \varepsilon_{\mu \nu \rho \sigma}\left\langle\left\{\nabla_\alpha V^{\mu \nu}, \tilde{f}_{+}^{\rho \alpha}\right\} \tilde{u}^\sigma\right\rangle, \nonumber \\
    \mathcal{\tilde{O}}_{V J P}^6 & = & \varepsilon_{\mu \nu \rho \sigma}\left\langle\left\{\nabla_\alpha V^{\mu \alpha}, \tilde{f}_{+}^{\rho \sigma}\right\} \tilde{u}^\nu\right\rangle, \nonumber \\
    \mathcal{\tilde{O}}_{V J P}^7 & = & \varepsilon_{\mu \nu \rho \sigma}\left\langle\left\{\nabla^\sigma V^{\mu \nu}, \tilde{f}_{+}^{\rho \alpha}\right\} \tilde{u}_\alpha\right\rangle, \nonumber \\
    \mathcal{\tilde{O}} _{\mathrm{VJP}} ^{8} & = & -i\sqrt{\frac{2}{3} } \varepsilon _{\mu \nu \rho \sigma } \left \langle V^{\mu \nu }\tilde{f}^{\rho \sigma } _{+}    \right \rangle \ln (\mathrm{det\tilde {u}} ).
\end{eqnarray}
\begin{eqnarray}
    \mathcal{\tilde{O}}_{V V P}^1 & = & \varepsilon_{\mu \nu \rho \sigma}\left\langle\left\{V^{\mu \nu}, V^{\rho \alpha}\right\} \nabla_\alpha \tilde{u}^\sigma\right\rangle, \nonumber \\
    \mathcal{\tilde{O}}_{V V P}^2 & = & i \varepsilon_{\mu \nu \rho \sigma}\left\langle\left\{V^{\mu \nu}, V^{\rho \sigma}\right\}\tilde{ \chi}_{-}\right\rangle, \nonumber \\
    \mathcal{\tilde{O}}_{V V P}^3 & = & \varepsilon_{\mu \nu \rho \sigma}\left\langle\left\{\nabla_\alpha V^{\mu \nu}, V^{\rho \alpha}\right\}\tilde{ u}^\sigma\right\rangle, \nonumber \\
    \mathcal{\tilde{O}}_{V V P}^4 & = & \varepsilon_{\mu \nu \rho \sigma}\left\langle\left\{\nabla^\sigma V^{\mu \nu}, V^{\rho \alpha}\right\}\tilde{ u}_\alpha\right\rangle, \nonumber \\
    \mathcal{\tilde{O}} _{\mathrm{VVP}} ^{5} & = & -i\sqrt{\frac{2}{3} } \varepsilon _{\mu \nu \rho \sigma } \left \langle V^{\mu \nu }V^{\rho \sigma }     \right \rangle \ln (\mathrm{det\tilde {u}} ).
\end{eqnarray}

In order to include $\rho-\omega$ mixing, we apply the momentum-dependent mixing mechanism given in Refs.~\cite{Gasser:1982ap,Wang:2023njt},
    \begin{equation}
        \binom{\left|\bar{\rho}^{0}\right\rangle}{|\bar{\omega}\rangle} =\left(\begin{array}{cc}
            \cos \delta & -\sin \delta_\omega(q^2) \\
            \sin  \delta_\rho(q^2) & \cos \delta
        \end{array}\right)\binom{\left|\rho^0\right\rangle}{|\omega\rangle} .
    \end{equation} 
where $\bar{\rho}^0$ and $\bar{\omega}$ are the physical states and  $\delta$ is the $\rho-\omega$ mixing angle, and the non-diagnonal parts are given as 
\begin{eqnarray}
    \sin \delta_\omega (q^2) & = & -\sin\delta \frac{M_V\Gamma _V(q^2)}{\Delta^\ast_V(q^2) }, \nonumber \\
    \sin \delta_\rho (q^2) & = & \sin\delta \frac{M_V\Gamma_V(q^2)}{\Delta_V(q^2)  }.
\end{eqnarray}
Here, $\Delta_V(x)=M_V^2-x-i M_V \Gamma_V(x)$ is the denominator of the Breit-Wigner propagator, and one can set $V=\rho$ for simplicity. Notice that the $\Gamma_\rho(x)$ will be multiplied by a step function to make sure that it vanishes when the lowest threshold of its decay channels is not open. 

Besides, to extend the above analysis to a higher energy region, one has to include the heavier vector resonance multiplets, $V'$ and $V''$. Following Ref.~\cite{Wang:2023njt}, we apply the extension to the Breit-Wigner (BW) propagators 
\begin{equation}
\mathrm{BW}(V,x)= \frac{1}{\Delta _{V}(x) } \longrightarrow \frac{1}{\Delta _{V}(x) } +\frac{\beta _{P\gamma\gamma} '}{\Delta _{V'}(x) } +\frac{\beta _{P\gamma\gamma} ''}{\Delta _{V''}(x) } , \label{Eq:BW}
\end{equation}
where $P=\pi$, $\eta$, and $\eta'$. The details of the BW propagators of the vector mesons are shown in the Appendix~\ref{app:BW}.

\subsection{TFFs of the lightest neutral pseudoscalars in the time-like region}
\begin{figure}[!htb]
\centering
\includegraphics[width=1\linewidth,,height=0.07\textheight]{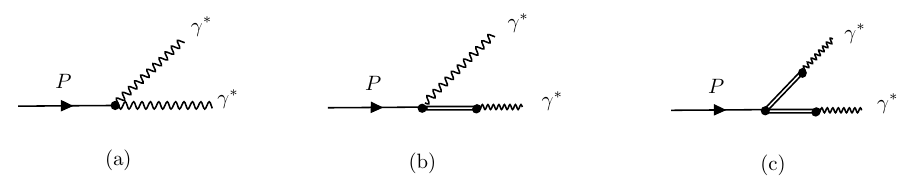}
\caption{Feynman diagrams of $P\to \gamma^*\gamma^*$. The double line represents vector resonance.}
\label{Fig:Feynman}
\end{figure}

The decay amplitudes of $P\to\gamma^*\gamma^*$ are defined as 
\begin{equation}
\mathcal{M} _{P\gamma ^{*} \gamma ^{*}  } =ie^2\varepsilon ^{\mu \nu \rho \sigma } q_{1\mu } q_{2\nu }\epsilon _{1\rho } \epsilon _{2\sigma } \cdot \mathcal{F} _{P \gamma^* \gamma^{*}  }(q_1^2,q_2^2).
\end{equation}
where $q_{1,2}$ are the momenta of the two photons.
$\mathcal{F} _{P \gamma^* \gamma^{*}  }(q_1^2,q_2^2)$ is the doubly-virtual TFF. 
It is obtained through the hadronization of two electromagnetic currents, in terms of the vector current $\mathcal{V}_{\nu}^i=\bar{q} (\lambda ^{i}/2)q$, 
\begin{eqnarray}
    &&\left \langle P|(\mathcal{V}_{\mu}^3+ \mathcal{V}_{\mu}^8/\sqrt{3}  ) (\mathcal{V}_{\nu}^3+ \mathcal{V}_{\nu}^8/\sqrt{3}  )e^{i\mathcal{L}_{\mathrm{QCD}} }  | 0  \right \rangle\nonumber\\
    &&=\varepsilon_{\mu \nu \rho \sigma } q_{1}^{\rho } q_{2}^{\sigma }\mathcal{F} _{P \gamma^* \gamma^{*}  }(q_1^2,q_2^2),
\end{eqnarray}
The singly-virtual TFFs can be obtained through the doubly-virtual TFFs by setting one of the photons on-shell, 
    \begin{equation}
        \mathcal{F}_{P\gamma^{*}  \gamma } (q^2)=\mathcal{F} _{P \gamma^* \gamma^{*}  }(q^2,0).
    \end{equation}
The Feynman diagrams of $P\to \gamma^*\gamma^*$ are shown in Fig.\ref{Fig:Feynman}. 
The first graph is about the WZW term, which dominates in the low energy region. Hence, it mostly affects the double-photon decay and single-Dalitz decay processes. The other two diagrams are related to the vector resonances and dominate in the higher energy region. They have a greater effect on electron-positron and two-photon annihilation processes.

In the ideal mixing case, the $\rho^0\!-\!\omega$ mixing angle $\delta$, $\gamma V$ higher order term $\alpha_{V}$, and heavier vectors $V',V''$ are ignored, the $\omega\!-\!\phi$ mixing angles are set as $\theta_V=35.26^\circ$ (${\rm sin}\theta_V=\frac{1}{\sqrt{3}}$), $\theta _{8}= \theta _{0}=-54.74^\circ$ (${\rm sin}\theta_0=-\sqrt{\frac{2}{3} } )$, and the $\eta\!-\!\eta'$ mixing parameters are set as $F_{8}=F_{0}=F$. The TFFs of the lightest pseudoscalars are given as 
\begin{eqnarray} 
\mathcal{F}^{\rm ideal}_{P \gamma^* \gamma^{*}  }\!(q_1^2\!,\!q_2^2) \!=\!\mathcal{F}^{\mathrm{local} } _{P  }\!+\!\mathcal{F}^{\mathrm{1R} } _{P   }(q_1^2\!,\!q_2^2)\!+\!\mathcal{F}^{\mathrm{2R} } _{P  }(q_1^2\!,\!q_2^2), 
\end{eqnarray}
where one has 
\begin{align*}
\mathcal{F}^{\mathrm{local} } _{\pi^{0}  }=&-\frac{N_C}{12 \pi ^2 F},\\
\mathcal{F}^{\mathrm{1R} } _{\pi^{0}  }(q_1^2,q_2^2)=&-\frac{2 \sqrt{2} F_{V}}{3FM_V}\left(\frac{1}{\Delta _{\rho ^{0} }(q_1^2)}+\frac{1}{\Delta _{\omega }(q_1^2)} \right)\\&\times (\tilde{c}_{125}q_2^2-\tilde{c}_{1256}q_1^2+\tilde{c}_{1235}m_\pi^2) \\
            &+\bigg\{q_1\leftrightarrow q_2\bigg\},\\
\mathcal{F}^{\mathrm{2R} } _{\pi^{0}  }(q_1^2,q_2^2)&= \frac{4 F_{V}^2}{3F}\left[\left(q_1^2+q_2^2\right)\tilde{d}_3+m_\pi^2\tilde{d}_{123}\right]\\
        &\times\left(\frac{1}{\Delta _{\rho ^{0} }(q_1^2)\Delta _{\omega }(q_2^2)}+\frac{1}{\Delta _{\rho ^{0} }(q_2^2)\Delta _{\omega }(q_1^2)}\right).
        \end{align*}
        
        \begin{align*}
            \mathcal{F}^{\mathrm{local} } _{\eta }&=-\frac{N_C(5C_q-\sqrt{2}C_s)}{36 \pi ^2 F},\\
            \mathcal{F}^{\mathrm{1R} } _{\eta }(q_1^2,q_2^2)&= \frac{8F_{V} }{3FM_{V} } \left\{9 \left[2 \sqrt{3} M_V^2 \tilde{c} _{8} \left(\sqrt{2} C_s-2 C_q\right)\right.\right.\\
            &-3 \sqrt{2} C_q \left(m_{\eta}^2 \left(\tilde{c} _{1235}  -8 \tilde{c} _{3}  \right)+8 m_{\pi }^2 \tilde{c} _{3} ^{*}\right.\\
            &\left.\left.+ \tilde{c} _{125}q_2^2-\tilde{c} _{1256}q_1^2  \right)\right]\left(\frac{1}{\Delta _{\rho ^{0} }(q_1^2)}+\frac{1}{9\Delta _{\omega }(q_1^2)} \right)\\
            &+\left[12 C_s \left(m_\eta^2 \tilde{c}_{1235}+q_2^2 \tilde{c}_{125}-q_1^2 \tilde{c}_{1256}\right.\right.\\
            &\left.-8 \tilde{c}_3 \left(-2 m_K^2+m_{\pi }^2+m_\eta^2\right)\right)\\
            &\left.+4 \sqrt{3} M_V^2 \tilde{c}_8 \left(\sqrt{2} C_s-2 C_q\right)\right]\frac{1}{\Delta_\phi(q_1^2)}\\
            &+\bigg\{q_1\leftrightarrow q_2\bigg\},\\   
            \mathcal{F}^{\mathrm{2R} } _{\eta }(q_1^2,q_2^2)&= \frac{16F_{V}^2 }{3F } \left\{\left[27 C_q \left(m_\eta^2 \left(\tilde{d}_{123}-8 \tilde{d}_2\right)+8 m_{\pi }^2 \tilde{d}_2\right.\right.\right.\\
            &\left.\left.\left.+\tilde{d}_3 \left(q_1^2+q_2^2\right)\right)  +18 \sqrt{3} M_V^2\tilde{d}_5 \left(\sqrt{2} C_q-C_s\right)\right]\right.\\
            &\left.\times\left(\frac{1}{\Delta _{\rho ^{0} }(q_1^2)\Delta _{\rho ^{0} }(q_2^2)}+\frac{1}{9\Delta _{\omega }(q_1^2)\Delta _{\omega }(q_2^2)}\right)\right.\\
            &\left.+\frac{1}{\Delta_\phi(q_1^2)\Delta_\phi(q_2^2)}\left[4 \sqrt{3} M_V^2 \tilde{d}_5 \left(\sqrt{2} C_q-C_s\right)\right.\right.\\
            &\left.\left.-6 \sqrt{2} C_s \left(m_\eta^2 \tilde{d}_{123}-8 \tilde{d}_2 \left(-2 m_K^2+m_{\pi }^2+m_\eta^2\right)\right.\right.\right.\\            &\left.\left.\left.+\tilde{d}_3\left(q_1^2+q_2^2\right)\right)\right]
\right\}. 
        \end{align*}

        \begin{align*}
            \mathcal{F}^{\mathrm{local} } _{\eta' }&=-\frac{N_C(5C'_q+\sqrt{2}C'_s)}{36 \pi ^2 F},\\
            \mathcal{F}^{\mathrm{1R} } _{\eta'  }(q_1^2,q_2^2)&= \frac{8F_{V} }{3FM_{V} }  \left\{9 \left[2 \sqrt{3} M_V^2 \tilde{c} _{8} \left(-\sqrt{2} C'_s-2 C'_q\right)\right.\right.\\
            &\left.\left.-3 \sqrt{2} C'_q \left(m_{\eta'}^2 \left(\tilde{c} _{1235}  -8 \tilde{c} _{3}  \right)+8 m_{\pi }^2 \tilde{c} _{3} ^{*}\right.\right.\right.\\
            &\left.\left.\left.+ \tilde{c} _{125}q_2^2-\tilde{c} _{1256}q_1^2  \right)\right]\left(\frac{1}{\Delta _{\rho ^{0} }(q_1^2)}+\frac{1}{9\Delta _{\omega }(q_1^2)} \right)\right.\\
            &\left.+\left[-12 C'_s \left(m_{\eta'}^2 \tilde{c}_{1235}+q_2^2 \tilde{c}_{125}-q_1^2 \tilde{c}_{1256}\right.\right.\right.\\
            &\left.\left.\left.-8 \tilde{c}_3 \left(-2 m_K^2+m_{\pi }^2+m_{\eta'}^2\right)\right)\right.\right.\\
            &\left.\left.+4 \sqrt{3} M_V^2 \tilde{c}_8 \left(-\sqrt{2} C'_s-2 C'_q\right)\right]\frac{1}{\Delta_\phi(q_1^2)}\right\}\\
            &+\bigg\{q_1\leftrightarrow q_2\bigg\},\\   
            \mathcal{F}^{\mathrm{2R} } _{\eta'  }(q_1^2,q_2^2)&= \frac{16F_{V}^2 }{3F } \left\{\left[27 C'_q \left(m_\eta^2 \left(\tilde{d}_{123}-8 \tilde{d}_2\right)+8 m_{\pi }^2 \tilde{d}_2\right.\right.\right.\\
            &\left.\left.\left.+\tilde{d}_3 \left(q_1^2+q_2^2\right)\right)  +18 \sqrt{3} M_V^2\tilde{d}_5 \left(\sqrt{2} C'_q+C'_s\right)\right]\right.\\
            &\left.\times\left(\frac{1}{\Delta _{\rho ^{0} }(q_1^2)\Delta _{\rho ^{0} }(q_2^2)}+\frac{1}{9\Delta _{\omega }(q_1^2)\Delta _{\omega }(q_2^2)}\right)\right.\\
            &\left.+\frac{1}{\Delta_\phi(q_1^2)\Delta_\phi(q_2^2)}\left[4 \sqrt{3} M_V^2 \tilde{d}_5 \left(\sqrt{2} C'_q+C'_s\right)\right.\right.\\
            &\left.\left.+6 \sqrt{2} C'_s \left(m_{\eta'}^2 \tilde{d}_{123}-8 \tilde{d}_2 \left(-2 m_K^2+m_{\pi }^2+m_{\eta'}^2\right)\right.\right.\right.\\            &\left.\left.\left.+\tilde{d}_3\left(q_1^2+q_2^2\right)\right)\right]
            \right\}. 
        \end{align*}
The definition of combinations of the unknown couplings is given as \cite{Dai:2013joa, Guevara:2018rhj} 
\begin{eqnarray}
    \tilde{c}_{1235} & = & \tilde{c}_1+\tilde{c}_2+8 \tilde{c}_3-\tilde{c}_5, \nonumber \\
    \tilde{c}_{1256} & = & \tilde{c}_1-\tilde{c}_2-\tilde{c}_5+2 \tilde{c}_6, \nonumber \\
    \tilde{c}_{125} & = & \tilde{c}_1-\tilde{c}_2+\tilde{c}_5, \nonumber \\
    \tilde{d}_{123} & = & \tilde{d}_1+8 \tilde{d}_2-\tilde{d}_3 .
\end{eqnarray}
The complete form of TFFs with all the mixing angles, heavier resonances, and higher order $\gamma V$ terms is shown in the Appendix~\ref{app:TFF}.

With these TFFs, one can calculate out the double-photon and single-Dalitz decay widths \cite{Landsberg:1985gaz,Danilkin:2019mhd}. The double-photon decay width is given as   
\begin{equation}
\Gamma_{P\gamma\gamma }=\frac{1}{4 }\pi\alpha^2 m_{P}^3 |\mathcal{F}_{P \gamma^* \gamma } (0)| ^2,
\end{equation}
and the normalized invariant mass spectrum of single-Dalitz decay is given by 
\begin{eqnarray}
    \frac{d\Gamma_{P\to l^{+}\! l^{-}\!\gamma } }{dq^{2}\Gamma\!_{P\to \gamma \gamma }  }\!&=&\!\frac{2\alpha }{3\pi q^2 }\!\sqrt{1\!\!-\!\!\frac{4m_{l}^2 }{q^2} } \!\left(\!1\!\!+\!\!\frac{2m_{l}^2 }{q^2}\!\right)\left(\!1\!\!-\!\!\frac{q^2}{m_{P }^2 } \!\right)^3\!\!|F_{P}(q\!^2)|^{2}\nonumber \\
    &&\times (1+\delta^{\mathrm{NLO}})\,,
\end{eqnarray}
where one has 
\begin{equation}
F_{P}(q^2)=\frac{\mathcal{F}_{P \gamma^* \gamma } (q^2)}{\mathcal{F}_{P \gamma^* \gamma } (0)}. \nonumber
\end{equation}
The studies of single-Dalitz decay \cite{Husek:2015sma,Husek:2017vmo,Afanasev:2023gev} indicate that the NLO radiative corrections for $\eta,\eta'$ are significant. Hence, we will include them in our analysis. See Appendix \ref{app:RNLO}. Unlike $\eta$ and $\pi$, there is only a limited amount of available data for the $\eta'$ time-like TFF \cite{Gan:2020aco}, which is in the very low energy region, $\sqrt{q^2}<100$~MeV. Therefore, we include the $e^+e^-$ invariant mass spectrum of $\eta'\to \omega e^{+}e^{-}$ to give an extra constraint, which shares the same parameters as $\mathcal{F}_{\eta' \gamma^* \gamma } (q^2)$ as both of them are calculated within the same framwork of RChT. The normalized $e^+e^-$ invariant mass spectrum is given by \cite{Landsberg:1985gaz}
\begin{eqnarray}
\frac{d\Gamma\!_{\eta'\to \omega e^{+}\! e^{-}  } }{dq^{2}\Gamma _{\eta'\to \omega \gamma }  }\!&=&\!\frac{\alpha }{3 \pi q^2}\!\sqrt{1\!-\!\frac{4 m_e^2}{q^2}}\! \left(\!1\!+\!\frac{2 m_e^2}{q^2}\!\right) \left\lvert F_{ \omega\eta' \gamma^* }(q^2) \right\rvert ^2 \nonumber\\[3mm]
        &&\! \left[\!\left(\!1\!-\!\frac{q^2}{m_{\eta'}^2\!\!-\!M_{\omega}^2}\right)^2\!\!\!-\!\!\frac{4 q^2 M_{\omega}^2}{(m_{\eta'}^2\!-\!M_{\omega}^2)^2}\!\right]^{3/2}. \nonumber\\
\end{eqnarray}
where one has 
\begin{equation}
F_{\omega\eta' \gamma^* }(q^2)=\frac{\mathcal{F}_{\omega\eta'\gamma^* } (q^2)}{\mathcal{F}_{\omega\eta' \gamma^* } (0)}. \nonumber
\end{equation}
The details of the TFF $\mathcal{F}_{\omega\eta'\gamma^* }(q^2)$ is given in the Appendix~\ref{app:A2}.

The total cross-section of the electron-positron annihilation into a photon and a pseudoscalar is also helpful to study the TFFs, and we take them into our analysis. The expression of the cross-section is given as  
\begin{equation}
\sigma_{e^{+}e^{-}\to P\gamma}(s)=\frac{2}{3} \pi ^2 \alpha ^3 |F_{P \gamma^\ast \gamma } (s)|^{2} \left(1-\frac{m_{P }^2}{s}\right)^3\,.
\end{equation}
Notice that our analysis of the cross-section is limited from $\sqrt{s} \approx m_{P}$ up to approximately $\sqrt{s} \approx 2.3$~GeV,  as only $V', V"$ are included, and so on for the time-like TFFs.

To reduce the unknown couplings, we apply the high energy constraints given in Refs.~\cite{Ruiz-Femenia:2003jdx,Dai:2013joa,Wang:2023njt}, where  
the matching on VVP Green functions between RChT and QCD at leading order \cite{Ruiz-Femenia:2003jdx, Dai:2013joa, Wang:2023njt,Kampf:2011ty,Chen:2012vw} is performed. One has 
\begin{eqnarray}
\tilde{c} _{125} &=&\tilde{c} _{1235} =0 \,,\nonumber\\
\tilde{c} _{1256}&=&-\frac{N_CM_V}{32\sqrt{2} \pi^2F_V}  \,,\nonumber\\
\tilde{c} _{8}&=&-\frac{\sqrt{2}M_0^2}{\sqrt{3} M_V^2}\tilde{c}_1=\frac{4\sqrt{2}M_0^2}{\sqrt{3} M_V^2}\tilde{c}_3  \,,\nonumber\\
\tilde{d}_{123}&=&\frac{F^2}{8F_V^2}   \,.
\end{eqnarray}
The Brodsky-Lepage (B-L) limit of the singly-virtual TFFs \cite{Lepage:1979zb,Hoferichter:2020lap}, which requires the TFFs to have asymptotic behavior of $1/Q^2$ for large $Q^2$, can be included.  Matching with it at leading order gives  
\begin{equation}
\tilde{d} _{3}=-\frac{N_CM_V^2}{64 \pi^2F_V^2}. \label{Eq:B-L}
\end{equation}
The time-like TFFs of the lightest pseudoscalars $\mathcal{F}_{P \gamma^* \gamma^{*}  }$ discussed above are calculated within RChT. They work well in the energy region of $0\leq q^2\leq (2.3~{\rm GeV})^2$ and can be extended to express the space-like TFFs in the corresponding energy region, $ -(2.3~{\rm GeV})^2\leq q^2 < 0 $.

\subsection{TFFs in the space-like region}  
As mentioned above, 
the space-like TFFs $\mathcal{F}_{P \gamma^* \gamma^{*}  }$ in the energy region of $Q_{1,2}^2=-q_{1,2}^2\geq (2.3{\rm GeV})^2$ has been given in the previous section through RChT. For the one in the higher energy region, pQCD can describe it well  \cite{Braaten:1982yp,Greiner:2007zz,Hoferichter:2020lap,Chai:2025xuz}. Thus, we will cut off our TFFs at roughly 2.3~GeV and 
align them with the pQCD results in the high energy region, as well as the VMD results in the middle energy region \cite{Hoferichter:2018kwz}. The space-like TFFs are given as 
\begin{equation}
\mathcal{F}^{\mathrm{SL} }_{P \gamma^*\gamma^{*}   }(q_1^2,q_2^2)=\mathcal{F}^{\mathrm{had} }_{P \gamma^* \gamma^{*}  }(q_1^2,q_2^2)+\mathcal{F}^{\mathrm{asym} }_{P \gamma^* \gamma^{*}   }(q_1^2,q_2^2)\,, \label{Eq:TFF}
\end{equation} 
with $q_{1,2}^2<0$. For convenience, people often rewrite it into 
\begin{eqnarray}
\mathcal{F}^{\mathrm{SL} }_{P \gamma^*\gamma^{*}   }(-Q_1^2,-Q_2^2)&=&\mathcal{F}^{\mathrm{had} }_{P \gamma^* \gamma^{*}  }(-Q_1^2,-Q_2^2)\nonumber\\
&+&\mathcal{F}^{\mathrm{asym} }_{P \gamma^* \gamma^{*}   }(-Q_1^2,-Q_2^2)\,,\nonumber
\end{eqnarray} 
where $Q$ is positive. We will use this formalism in the following plots. 
The hadronic part $\mathcal{F}^{\mathrm{had} }_{P \gamma^* \!\gamma^{*}  }$ is given as 
    \begin{widetext}
        \begin{equation}
            \mathcal{F}^{\mathrm{had} }_{P \gamma^* \gamma^{*}  }(q_1^2,q_2^2)=
            \begin{cases}\mathcal{F} _{P \gamma^* \gamma^{*}  }(q_1^2,q_2^2) & \mathrm{if}\, q_1^2\ge -s_{1}\, \& \,q_2^2\ge -s_{1}
 \\ \mathcal{F} _{P \gamma^* \gamma^{*}  }(-s_{1},q_2^2)\times \left( \frac{(1- \epsilon )(M_1^2+s_{1})}{M_1^2-q_1^2} + \frac{\epsilon (M_2^2+s_{1})}{M_2^2-q_1^2} \right) & \mathrm{if}\, q_1^2< -s_{1}\, \& \,q_2^2\ge -s_{1}
 \\ \mathcal{F}_{P \gamma^* \gamma^{*}  }(q_1^2,-s_{1})\times \left( \frac{(1-\epsilon )(M_1^2+s_{1})}{M_1^2-q_2^2} + \frac{\epsilon (M_2^2+s_{1})}{M_2^2-q_2^2} \right) & \mathrm{if}\, q_1^2\ge -s_{1}\, \& \, q_2^2< -s_{1}
 \\ \mathcal{F}_{P \gamma^* \gamma^{*}  }(-s_{1},-s_{1})\times \left( \frac{(1- \epsilon )(M_1^2+s_{1})^2}{(M_1^2-q_1^2)(M_1^2-q_2^2)} + \frac{\epsilon (M_2^2+s_{1})^2}{(M_2^2-q_1^2)(M_2^2-q_2^2)} \right) & \mathrm{if}\, q_1^2< -s_{1}\, \& \,q_2^2< -s_{1}        
            \end{cases}.
            \label{Eq:TFF;1}
        \end{equation}
    \end{widetext}
Notice that the present formalism ensures that the doubly-virtual TFFs and their first derivative are continuous at the energy point $s_1$. 
$s_1$ is either fixed to be $(2.3~{\rm GeV})^2$ or set as a free parameter. See discussions in the next section.
The values of $M_1^2$ and $M_2^2$ are determined by the following equations
\begin{equation}
\left\{\begin{matrix}(1-\epsilon) (M_1^2+s_1)+\epsilon (M_2^2+s_1)=z_P/a
            \\\frac{(1-\epsilon )}{M_{1}^2+s_1 } +\frac{\epsilon }{M_{2}^2+s_1 } =-b/a
\end{matrix}\right. \,,
\label{cut-off equ}
\end{equation}
where
\begin{eqnarray}
z_P&=&\lim_{q^2 \to -\infty} -q^2\mathcal{F}^{\mathrm{QCD} } _{P\gamma^{*}  \gamma } (q^2),   \nonumber\\
a&=&\begin{cases}
    \mathcal{F}_{P\gamma^{*}  \gamma } (-s_1,q_2^2)& \mathrm{if}\, q_1^2< -s_{1}\, \& \,q_2^2\ge -s_{1} \\
    \mathcal{F}_{P\gamma^{*}  \gamma } (q_1^2,-s_1)&  \mathrm{if}\, q_1^2\ge -s_{1}\, \& \,q_2^2< -s_{1} \\
    \mathcal{F}_{P\gamma^{*}  \gamma } (-s_1,-s_1)& \mathrm{if}\, q_1^2< -s_{1}\, \& \,q_2^2< -s_{1}        
\end{cases},  \nonumber\\
b&=&\begin{cases}
 -\frac{\partial \mathcal{F}_{P\gamma^{*}  \gamma } (q^2,q_2^2)}{\partial q^2}\bigg|_{q^2=-s_1}& \mathrm{if}\, q_1^2< -s_{1}\, \& \,q_2^2\ge -s_{1} \\
 -\frac{\partial \mathcal{F}_{P\gamma^{*}  \gamma } (q_1^2,q^2)}{\partial q^2}\bigg|_{q^2=-s_1}&  \mathrm{if}\, q_1^2\ge -s_{1}\, \& \,q_2^2< -s_{1} \\
 -\frac{\partial \mathcal{F}_{P\gamma^{*}  \gamma } (q^2,-s_1)}{\partial q^2}\bigg|_{q^2=-s_1}& \mathrm{if}\, q_1^2< -s_{1}\, \& \,q_2^2< -s_{1}     
\end{cases}. \nonumber
\end{eqnarray}
By imposing the condition $ M_1^2 \le M_2^2$, the final solution of $M_{1,2}$ are given as follows
    \begin{equation}
        \left\{\begin{matrix} M_1^2=\frac{\sqrt{\left(a^2+b z_P\right) \left(a^2 (1-2 \epsilon )^2+b z_P\right)}+a^2 (2 \epsilon -1)+b z_P}{2 a b (1-\epsilon)}-s_1
            \\M_2^2=\frac{-\sqrt{\left(a^2+b z_P\right) \left(a^2 (1-2 \epsilon )^2+b z_P\right)}+a^2 (1-2 \epsilon )+b z_P}{2 a b \epsilon }-s_1
        \end{matrix}\right.. \nonumber
    \label{eq:Mi}
    \end{equation}
The existence conditions of the solution are 
\begin{equation}
z_P>0\,\&\, b<0\,\&\,0< \epsilon < 1 \,\&\,a^2+bz_P<0. \nonumber
\end{equation}
The first condition $(z_P>0)$ is naturally satisfied by the asymptotic behavior of QCD; The second condition $(b<0)$ is inherently fulfilled by the VMD-like extension; The third condition $(0< \epsilon < 1)$ ensures that $M_1^2\le M_2^2$; The last condition $(a^2+bz_P<0)$ needs to be discussed in detail. First, it ensures that there is no imaginary part of $M_2^2$. This is natural as in the space-like region the decay width of a resonance in the calculation should be ignored. Second, by considering $b>-a/s_1$, which could be examined by VMD model, $a^2+bz_P<0$ gives us a weaker condition $s_1a<z_P$, which implies $-q^2\mathcal{F} _{P\gamma^{*} \gamma^{*} }(-s_1,q^2)$ can not cross the asymptotic line of QCD if both $-q^2$ and $s_1$ are sufficiently large. The parameter $\epsilon$ can be determined by the fit as suggested by Ref.~\cite{Hoferichter:2018kwz}. Nevertheless, different values of $\epsilon$ actually lead to little difference in our analysis, as the TFFs in the high energy region are basically determined by its first order derivative and the asymptotic behavior. Further, it is noteworthy that if one sets $\epsilon=1/2$, the expressions are much simplified. Specifically, when substituting this value into the Eq. \eqref{Eq:TFF;1}, all the radicals in Eq. \eqref{eq:Mi} cancel out in the final expression, and one does not need to worry about the last condition $a^2+bz_P<0$. Because this condition is derived from the requirement that there should not be an imaginary part of $\sqrt{\left(a^2+b z_P\right) \left(a^2 (1-2 \epsilon )^2+b z_P\right)}$. Once the radical disappears in the final expression, the imaginary part will not exist. As a result, the working range of the model will significantly increase. Consequently, we will adopt this particular value, $\epsilon=1/2$. 

The VMD-like model with only the lightest vectors may cause incorrect asymptotic behaviors of the doubly off-shell TFFs, resulting in large uncertainties on estimation of the $(g-2)_{\mu}$. However, as discussed in Refs.~\cite{Guevara:2018rhj,Estrada:2024cfy}, including the heavier resonances, $V'$, will modify the momentum structure of the doubly-virtual TFFs and provide the correct asymptotic behavior. In our approach, not only $V'$ but also $V''$ are included, ensuring a comprehensive analysis on the TFFs in the higher energy region.

To elaborate the asymptotic contribution of the high energy behaviour, we also incorporate an asymptotic contribution to the doubly-virtual TFFs \cite{Hoferichter:2018kwz,Holz:2024lom},
\begin{eqnarray}
    &\mathcal{F}^{\mathrm{asym} } _{P \gamma^* \gamma^{*}  }(q_1^2,q_2^2)=\frac{-z_P }{m_P^4}\int_{2s_{m}^P }^{\infty } dv\Big[\frac{q_2^2}{v-q_1^2}f^{\mathrm{asym} } _{P  }(v,q_1^2)\nonumber\\
    &\times\left (\frac{1}{v-q_1^2-q_2^2}-\frac{1}{q_1^2-q_2^2}  \right)+\left\{ q_1\leftrightarrow q_2\right\}\Big],
\end{eqnarray}
\begin{equation}
 f^{\mathrm{asym} } _{P  }(v,x)=\frac{(v-2x)^2-m_P^2v}{\sqrt{(v-2x)^2-2m_P^2v+m_P^4}}+2x-v.
\end{equation}
The $s_m^{\pi,\eta,\eta'}$ are fixed by fitting to the space-like doubly-virtual TFF data that is measured through the two-photon annihilation process. However, such kinds of data of $\pi^0$ and $\eta$ are still lacking. To solve this problem, we include the doubly-virtual TFFs provided by LQCD as another kind of data. It is found that to fit the data well, $s_m^{\eta}$ and $s_m^{\eta'}$ can be set as the same, but $s_m^{\pi}$ should be different. Hence, we will use two parameters, $s_m^{\pi}$ and $s_m^{\eta/\eta'}$, in our analysis.

\section{Fit results and discussions}\label{sec3}
The strategy of the analysis is to include as many constraints as possible, both from experiment and theory. 
We perform a combined analysis on the single-Dalitz decays of $P\to \gamma e^{+}e^{-}$ and $\eta'\to \omega e^{+}e^{-}$, the cross-sections of $e^+e^-\to P\gamma$, the experimental data of space-like singly-virtual and doubly-virtual TFFs, LQCD  data of space-like diagonal-virtual TFFs, and $P\to\gamma\gamma$ decay widths. 
The masses and widths of the resonances, $V$, $V'$, $V''$ are fixed by PDG \cite{ParticleDataGroup:2024cfk}.  They are shown in Table \ref{tab:RMG}. Note that some of them are not the central values of the PDG, but rather fall within the uncertainties, due to the improved fit quality. 
\begin{table}[htb!]
\centering
{\footnotesize 
\renewcommand{\arraystretch}{2}
\newcommand{\tabincell}[2]{\begin{tabular}{@{}#1@{}}#2\end{tabular}}
\begin{ruledtabular}
\begin{tabular}{ccc}
\tabincell{c}{Parameter}&This work  &PDG \cite{ParticleDataGroup:2024cfk}\\
\colrule
$M_{\rho}$ &$775.26$ &$775.26\!\pm \!0.23$ \\
$M_{\omega}$ &$782.66$ &$782.66\!\pm\! 0.13$\\
$\Gamma_{\omega}$ &$8.90$ &$8.68\!\pm\! 0.13 $\\
$M_{\phi}$ &$1019.36 $ &$1019.461\!\pm\! 0.016$\\
$\Gamma_{\phi}$ & $4.23 $ &$4.249\!\pm\! 0.013$\\
$M_{\rho'}$ &$1480$  &$1465\!\pm\! 25$ \\
$\Gamma_{\rho'}$ &$340$ &$400\!\pm\! 60 $\\
$M_{\omega'}$ &$1430 $ &$1410\!\pm\! 60$\\
$\Gamma_{\omega'}$ &$290 $ &$290\!\pm\! 190 $\\
 $M_{\phi'}$ &$1680 $ &$1680\!\pm\! 20$\\
 $\Gamma_{\phi'}$ & $150$ &$150\!\pm\! 50$\\
 $M_{\rho''}$ &$1720$ &$1720\!\pm\! 20$\\
 $\Gamma_{\rho''}$ &$250$ &$250\!\pm\! 100$\\
 $M_{\omega''}$ & $1670$ &$1670\!\pm\!30$\\
 $\Gamma_{\omega''}$ &$315$ &$315\!\pm\!35$\\
 $M_{\phi''}$ &$2162$ &$2162\!\pm\!70$\\
 $\Gamma_{\phi''}$ & $100$ &$100\!\pm \!27$\\         \end{tabular}
\end{ruledtabular}
\caption{Masses and widths of $V,\,V'\,,V''$ used in our TFFs, these values are given in unit of MeV. They are fixed in both Fits. }
\label{tab:RMG}
}
\end{table}
The unknown couplings in the TFFs are $\tilde{c} _{3} $, $\tilde{d} _{2} $, $\tilde{d} _{5}  $, $\alpha _{V} $, $\beta _{P \gamma }'$, $\beta _{P \gamma }'' $, the mixing angles and decay constants, i.e., $f_0$, $f_8$ and $\theta_{0}$, $\theta_{8}$ for $\eta-\eta'$ mixing, $\delta$ and $\theta_{V}$ for $\rho-\omega$ and $\omega-\phi$ mixing, $s_m^{P}$ for the asymptotic TFFs. 
The parameter $F_{V}$ always appears together with other parameters such as $\tilde{d}_{i}$ and $\tilde{c}_{i}$. Hence, one can either redefine the latter parameters \cite{Guevara:2018rhj} or just fix them.
Here, we fix $F_V=\sqrt{3}F$ following Ref.~\cite{Roig:2014uja,Roig:2014,Chen:2012vw}. The cutoff energy point, $s_1$, as shown in Eq.~(\ref{Eq:TFF;1}), can be set as $(2.3~{\rm GeV})^2$, and we use the same value of $s_1$ for all three pseudoscalar mesons to reduce the parameters. Also, this assumption is compatible with the fact that all the TFFs of the three pseudoscalars are calculated in the same theoretical framework, RChT. Nevertheless, once $s_1$ is larger, one can describe better the space-like singly-virtual TFF data of $\pi^0$ \cite{BaBar:2009rrj} from BaBar. Notice that the BaBar data are quite different from other experiments \cite{Brodsky:2011yv,Guevara:2018rhj,Chai:2025xuz} at the energy region $q^2<-10$~GeV$^2$. Therefore, we perform two fits. One is to set $s_1=(2.3~\mathrm{GeV})^2$ and exclude BaBar's space-like singly-virtual TFF data of $\pi^0$, named as Fit A. As a comparison, we set $s_1$ as a free parameter and fit all the data to fix it, called Fit B. To obtain the statistical uncertainties of the parameters and the physical observables, we employ the bootstrap method \cite{Efron:1979bxm}, where the data points are varied within their uncertainties by multiplying a normal distribution function. The fit parameters are shown in Table \ref{tab:para}. 
\begin{table}[htb]
\centering
{\footnotesize 
\renewcommand{\arraystretch}{2}
\newcommand{\tabincell}[2]{\begin{tabular}{@{}#1@{}}#2\end{tabular}}
\begin{ruledtabular}
\begin{tabular}{cccc}
\tabincell{c}{Parameter}& Fit A & Fit B & Ref. \cite{Chen:2012vw}\\
\colrule
$\tilde{d}_{2}(10^{-1}) $&$1.94\pm 0.05  $& $1.70\pm 0.03$ &$0.86\pm 0.85 $\\
 $\tilde{d} _{5}(10^{-1}) $&$9.83\pm 0.22$  & $8.78\pm 0.12$  &$3.6\pm 4.0$ \\
 $\tilde{c}_{3}(10^{-3}) $ &$-6.08\pm 0.25 $ & $-6.92\pm 0.30 $  &$11\pm 16$\\
 $\alpha _{V}(10^{-3}) $ &$-6.4\pm 0.7 $& $-5.6\pm 0.3 $  &-\\
 $\beta _{\pi \gamma } '(10^{-2})$ &$1.8\pm 2.6 $& $7.2\pm 2.2$  &-\\
 $\beta _{\pi \gamma }''(10^{-2})$ &$-2.1 \pm 3.0$ & $-8.5\pm 2.4$  &-\\
 $\beta _{\eta \gamma } '(10^{-1})$ &$0.98\pm 0.21 $& $1.03\pm 0.34 $  &-\\
 $\beta _{\eta \gamma }''(10^{-1})$ &$-1.08\pm 0.29  $ & $-0.83\pm 0.36$ &-\\
 $\beta _{\eta' \gamma } '(10^{-1})$ & $2.95\pm 0.73$& $1.71\pm 0.20 $  &-\\
 $\beta _{\eta' \gamma }''(10^{-1})$ &$-3.05\pm 0.83 $& $-1.66\pm 0.21$ &-\\
 $f_0$ &$1.207\pm 0.030$& $1.115 \pm 0.016$  &$1.11\pm 0.18$\\
 $f_8$ &$1.438\pm 0.014  $& $1.417\pm 0.011  $  &$1.37\pm 0.07$\\
 $\theta_0(\degree)$ &$-10.49\pm0.29 $ & $-10.46\pm0.23$ &$-2.5\pm 8.2$\\
 $\theta_8(\degree)$ &$-12.77\pm 0.30 $ & $-11.62\pm 0.19$  &$-21.1\pm 6.0$\\
 $\delta(\degree)$ &$-1.80$~(fixed) & $-1.80$~(fixed) &- \\
 $\theta_V(\degree)$ &$38.62$~(fixed) & $38.62$~(fixed) &- \\
$s_1(\mathrm{GeV}^2 )$ &$5.29 $~(fixed)  & $16.76\pm 1.62$  &-  \\
$s_m^{\pi}(\mathrm{GeV}^2 )$ &$0.893\pm 0.034$  & $0.951\pm 0.038$   &-  \\
$s_m^{\eta/\eta'}(\mathrm{GeV}^2 )$ & $0.712 \pm 0.043$ & $0.642 \pm 0.037$  & - \\
\end{tabular}
\end{ruledtabular}
\caption{Parameters of Fits A and B.  Their $\chi^2_{\mathrm{d.o.f.} }$ are $1.41$ and $1.49$ respectively. The uncertainties of the parameters are taken from bootstrap method \cite{Efron:1979bxm}. }
\label{tab:para}
}
\end{table}
Their $\chi^2_{\mathrm{d.o.f.} }$ are $1.41$ and $1.49$ respectively.
Notice that the dataset of space-like singly-virtual pion TFF from BaBar \cite{BaBar:2009rrj} has been excluded in Fit A. Once this data is included, the $\chi^2_{\mathrm{d.o.f.} }$ of Fit A will be $1.56$.
\begin{figure}[!htb]
\centering
\includegraphics[width=1\linewidth,height=0.32\textheight]{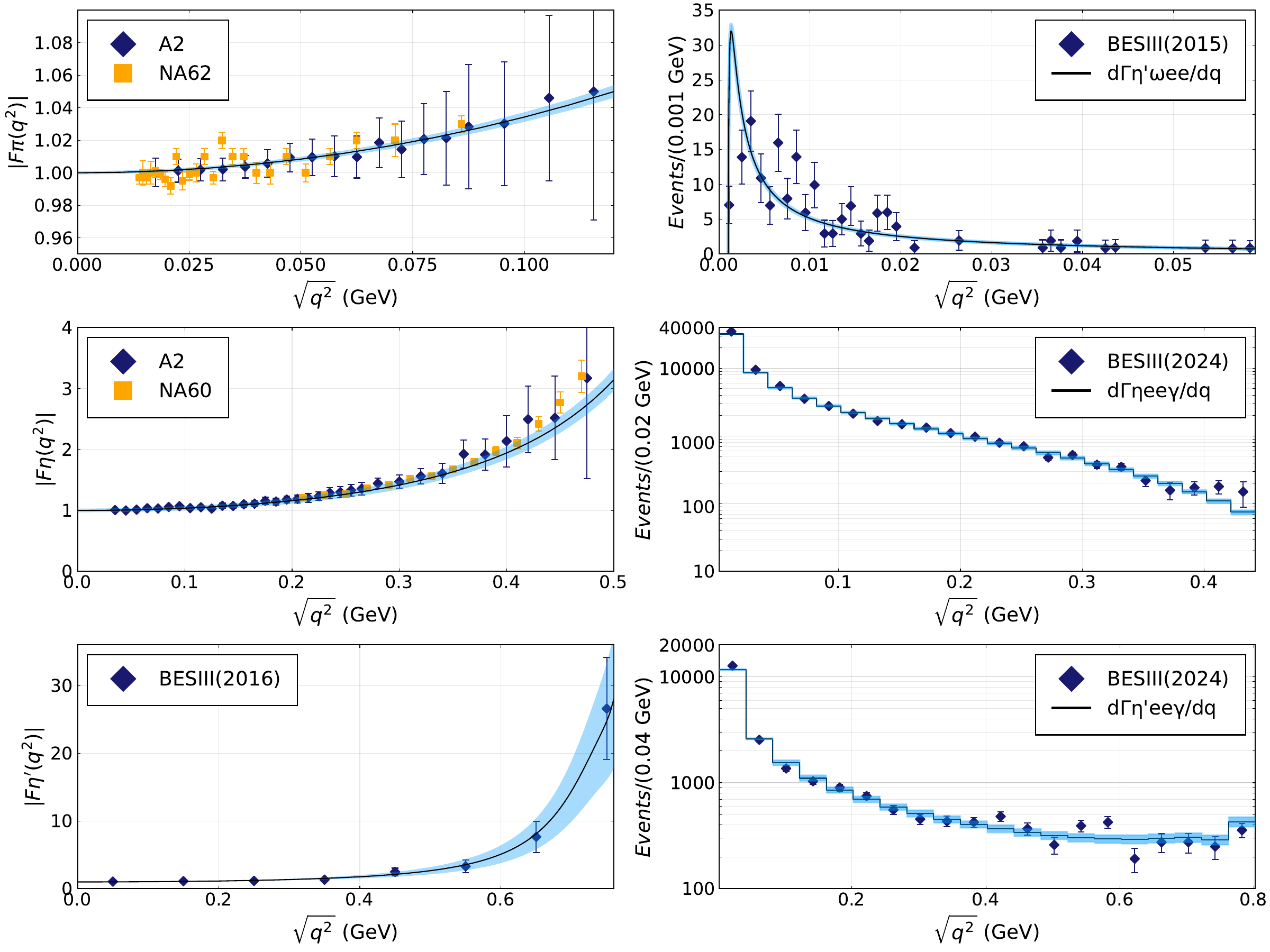}
\caption{Normalized form factors $|F_{P}(q^2)|$ and $e^+e^-$ invariant mass spectra of $P\to \gamma e^{+}e^{-}$ and $\eta'\to\omega e^+e^-$. Fits A and B get almost the same results.}
\label{fig:IMS}
\end{figure}
In Fig.~\ref{fig:IMS}, there are the fit results of the normalized TFFs in the left column and the $e^+e^-$ invariant mass spectra of $P\to \gamma e^{+}e^{-}$ and $\eta'\to \omega e^+e^-$ in the right column. 
For the invariant mass spectra, $\sqrt{q^2}$ is the momentum of the virtual photon, which transits into electron-positron pair. 
The experimental datasets are taken from A2 \cite{A2:2016sjm,Adlarson:2016hpp}, NA60 \cite{NA60:2016nad}, NA62 \cite{NA62:2016zfg}, BESIII \cite{BESIII:2015zpz,BESIII:2024pxo,BESIII:2015jiz}.  
For the invariant mass spectra of $\eta,\eta'\to\gamma e^+e^-$, two significant effects should be taken into account. First, the NLO radiative corrections must be considered, as shown by the theoretical analysis \cite{Husek:2017vmo}. In practice,  a reasonable approach to include the complicated NLO radiative corrections is to use the interpolation method with the existing numerical results, e.g., Ref.~\cite{Husek:2017vmo}. One can obtain these radiative corrections from Refs.~\cite{Husek:2015sma,Husek:2017vmo,Afanasev:2023gev}. See Appendix \ref{app:RNLO} for details of our numerical results using the cubic spline interpolation method.  Second, the widths of the lightest vector mesons play an essential role on the decay process of $\eta'\to\gamma e^+e^-$, as the the masses of the lightest vector mesons ($\rho$, $\omega$, $\phi$) are close to that of $\eta'$. As can be found, our results are consistent with the data. The fits are of high quality  except for the energy region of 0.45-0.5~GeV  for the TFF and invariant mass spectrum of $\eta$, as shown in the second row graphs. Nevertheless, ours are still compatible with the data within uncertainties. Indeed, the behaviour of the $\eta$ TFF in this energy region is sensitive to the mass of the $\rho$, but it has been fixed by PDG. Also, our analysis combines all the datasets. The cross section is also sensitive to the mass of the $\rho$ and give a strong contraint on it.  
As has been checked, the NLO radiative correction of $\eta'$ reduces the magnitude by about twelve percent in the energy region around $\rho,\omega$ resonances, which is crucial for fitting the data and determining the TFF. 
The results of branching ratios for $V\to P\gamma$ and $P\to V\gamma$ decays of our model are shown in Tab.~\ref{tab:t5}. In general, our results agree well with PDG values~\cite{ParticleDataGroup:2024cfk} and previous theoretical predictions~\cite{Chen:2012vw,Qin:2020udp,Wang:2023njt}. 
\begin{table}
\centering
{\footnotesize 
\renewcommand{\arraystretch}{2}
\newcommand{\tabincell}[2]{\begin{tabular}{@{}#1@{}}#2\end{tabular}}
\begin{ruledtabular}
\begin{tabular}{cccc}
 Process & This work (Fit A) & \tabincell{c}{Experimental value\\ \cite{ParticleDataGroup:2024cfk}} \\
\colrule
 $\rho^0\to \pi^0\gamma$ & $(5.1\pm 1.2)\times10^{-4}$ & $(4.7\pm0.8)\times10^{-4}$ \\
$\omega \to \pi^0\gamma$ & $(6.14\pm 2.12)\times10^{-2}$ & $(8.33\pm0.25)\times10^{-2}$ \\
$\phi \to \pi^0\gamma$ & $(1.73\pm 0.55 )\times10^{-3}$ & $(1.33\pm0.05)\times10^{-3}$ \\
$\rho^0\to \eta\gamma$ & $(2.35\pm 0.87)\times10^{-4}$ & $(3.00\pm0.21)\times10^{-4}$ \\
$\omega \to \eta\gamma$ & $(4.2\pm 1.1 )\times10^{-4}$ & $(4.5\pm0.4)\times10^{-4}$ \\
$\phi \to \eta\gamma$ & $(1.653\pm 0.521 )\times10^{-2}$ & $(1.306\pm0.024)\times10^{-2}$ \\
$\eta'\to \rho^0\gamma$ & $(26.78\pm 9.21 )\times10^{-2}$  & $(29.48\pm 0.35)\times10^{-2}$\\
$\eta'\to \omega\gamma$ & $(3.64\pm 1.24)\times10^{-2}$  & $(2.52\pm 0.07)\times10^{-2}$\\
$\phi \to \eta'\gamma$ & $(5.77 \pm 1.68 )\times10^{-5}$ & $(6.23\pm0.21)\times10^{-5}$ \\
\end{tabular}
\end{ruledtabular}
\caption{Fit results of $V\to P\gamma$ and $P\to V\gamma$ decays. Fits A and B get almost same results.}
\label{tab:t5}
}       
\end{table}

\begin{figure}[htb]
\centering
\includegraphics[width=1\linewidth,height=0.32\textheight]{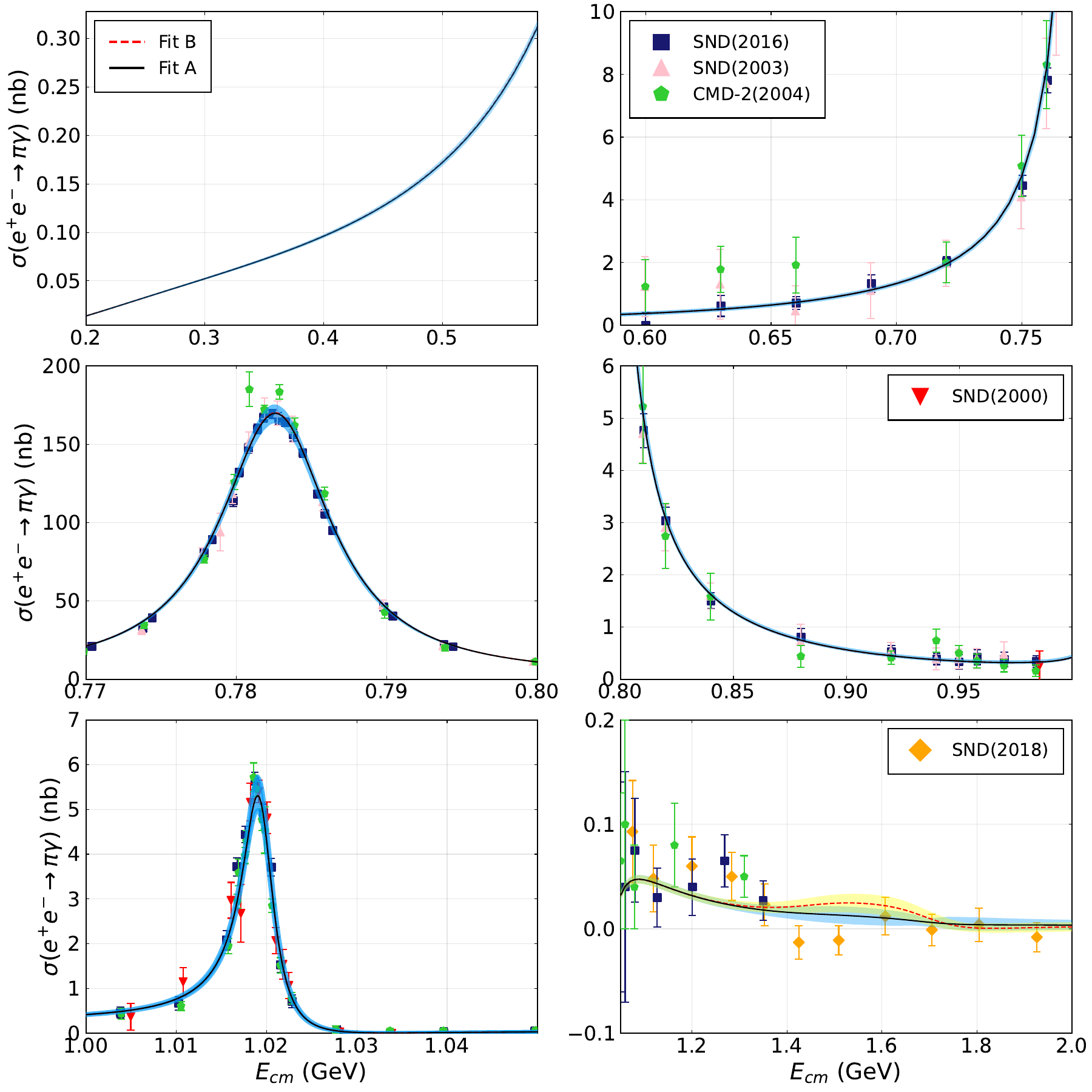}
\caption{Fit to the $e^+e^-\to \pi^0\gamma$ cross-section. The experimental datasets are from SND \cite{Achasov:2000zd,Achasov:2003ed,SND:2016drm,Achasov:2018ujw}, CMD-2 \cite{CMD-2:2004ahv}. }
\label{fig:cs;pi}
\end{figure}

\begin{figure}[htb]
\centering
\includegraphics[width=1\linewidth,height=0.32\textheight]{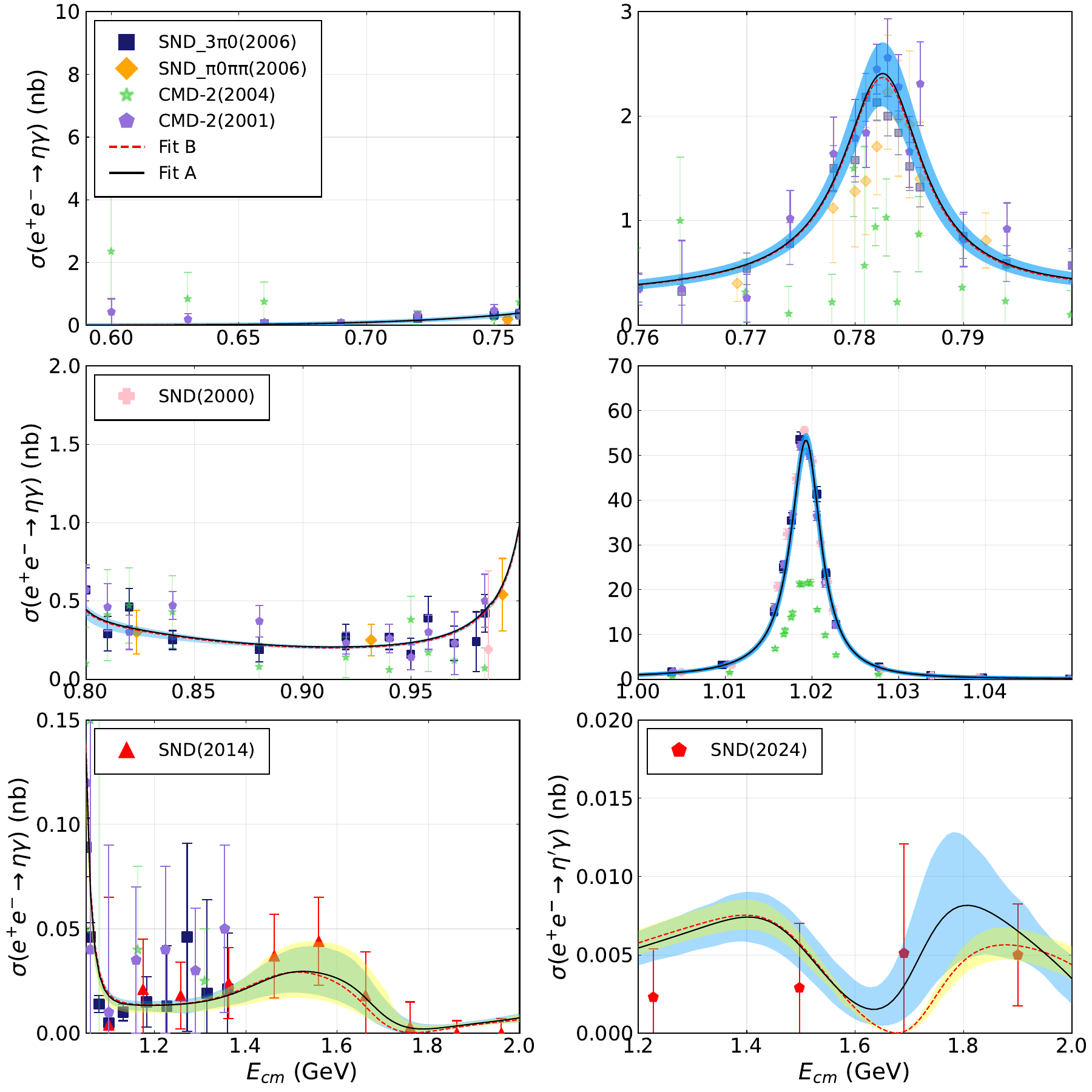}
\caption{Fit to the cross-sections of  $e^+e^-\to \eta\gamma, \eta'\gamma$. The datasets are taken from SND \cite{Achasov:2000zd,Achasov:2006dv,Achasov:2013eli,SND:2024qaq}, CMD-2 \cite{CMD-2:2004ahv,CMD-2:2001dnv}.}
\label{fig:cs;eta}
\end{figure}

The results of $e^+e^-\to P\gamma$ cross-sections are shown in Fig.~\ref{fig:cs;pi} for the $\pi^0$ case and Fig.~\ref{fig:cs;eta} for the $\eta,\eta'$ cases. 
For  $e^+e^-\to \pi^0\gamma$, the peak around 1.020~GeV is caused by the $\phi$ resonance, which is mostly determined by $\theta_{V}$ and would not appear in the ideal mixing case. For simplicity, we will fix the  mixing angle $\theta_{V}=38.62\degree$ as that given in Ref.~\cite{Wang:2023njt}, where it has been fixed by analysis on $e^+e^-\to \pi\gamma$ as well as other processes, e.g., $e^+e^-\to\bar{K}K$. For $\rho-\omega$ mixing angle $\delta$, we also use the value given in Ref.~\cite{Wang:2023njt} due to a similar reason. The values of $\tilde{d}_{2,5}$ are highly correlated to these cross-sections~\cite{Chen:2012vw}. Our results are consistent with those of Ref.~\cite{Chen:2012vw}. See Table~\ref{tab:para}. The value of $\alpha_{V}$ is given by the fit. It is consistent with that of Ref.~\cite{Wang:2023njt}, too. As shown in the last graph of Fig.~\ref{fig:cs;pi} and the last two graphs of Fig.~\ref{fig:cs;eta}, Fit A fits the cross-section data in the energy region of 1.4-1.9~GeV a bit better than Fit B. Therefore, we will take Fit A as the optimal solution.
The parameters $\beta_{P\gamma}^{\prime}$ and $\beta_{P\gamma}^{\prime\prime}$ are determined by a combined fit from not only the cross-sections of electron positron annihilation, but also the space-like datasets  of singly- and doubly-virtual virtual TFFs. 
The cross-section data has more significant effects on these parameters. The cross-section data for $e^+e^-\to\eta'\gamma$ below 2 GeV is quite poor; only one dataset is available \cite{SND:2024qaq}. This time, the singly- and doubly-virtual virtual TFFs have more effects on the fit.  Future measurements on the cross-section of $e^+e^-\to\eta'\gamma$ would be crucial for refining the TFFs. 
    
\begin{figure}[htb]
\centering
\includegraphics[width=1\linewidth]{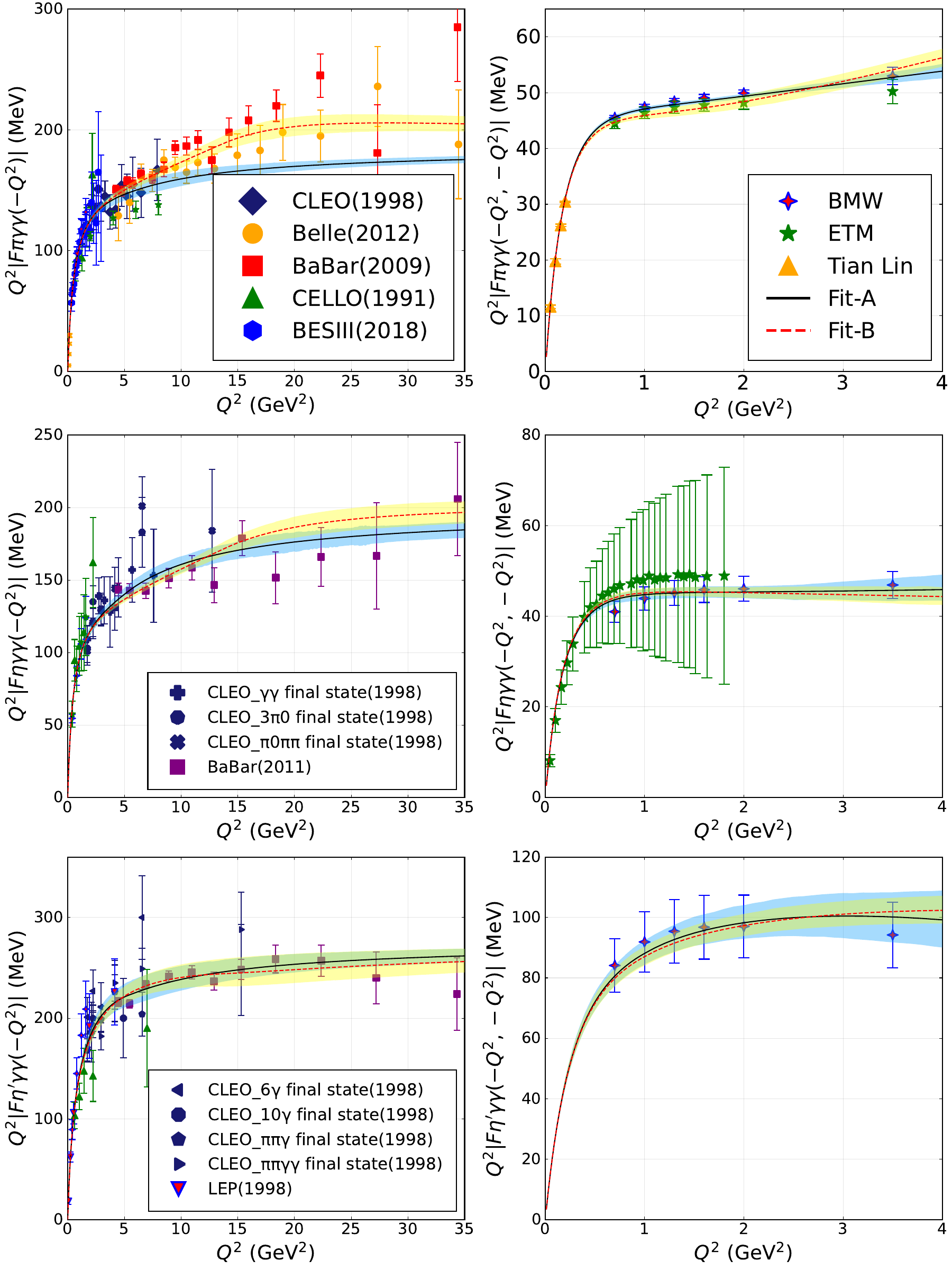}
\caption{Fit to the Space-like singly-virtual and doubly-virtual TFFs. The experimental datasets are sourced from BaBar \cite{BaBar:2009rrj,BaBar:2011nrp}, Belle \cite{Belle:2012wwz}, LEP \cite{L3:1997ocz}, CLEO \cite{CLEO:1997fho}, CELLO \cite{CELLO:1990klc} and BESIII \cite{Redmer:2019zzr}. The LQCD datasets are sourced from BMW \cite{Gerardin:2023naa}, ETM \cite{ExtendedTwistedMass:2022ofm,ExtendedTwistedMass:2023hin} and Lin's work \cite{Lin:2024khg}. Here we use points with same color and different shapes for different data of CLEO \cite{CLEO:1997fho}.}
\label{Fig:TFFs}
\end{figure}

The results of space-like singly-virtual TFFs are presented in Fig.~\ref{Fig:TFFs}. Here we used the notion $Q^2=-q^2$ for space-like $q^2$.
As can be found, our solutions can describe the datasets well. Besides the experimental datasets, we also take LQCD results as one kind of dataset. For $\pi^0$, we take four points ($Q^2=[0.4, 0.8, 1.2, 1.6]~\mathrm{GeV}^2$) from BMW \cite{Gerardin:2023naa}, four points ($Q^2=[2,4,6,8]~\mathrm{GeV}^2$) from ETM \cite{ExtendedTwistedMass:2023hin} and four points ($Q^2=[0.02, 0.06, 0.10, 0.14]~\mathrm{GeV}^2$) from Ref.~\cite{Lin:2024khg}. For $\eta$, we take four points ($Q^2=[0.4, 0.8, 1.2, 1.6]~\mathrm{GeV}^2$) from BMW \cite{Gerardin:2023naa} and four points ($Q^2=[0.4, 0.8, 1.2, 1.6]~\mathrm{GeV}^2$) from ETM \cite{ExtendedTwistedMass:2023hin}. For $\eta'$, we take four points ($Q^2=[0.4, 0.8, 1.2, 1.6]~\mathrm{GeV}^2$) from BMW \cite{Gerardin:2023naa}. 
Most of the data points of space-like singly-virtual TFFs, as shown in the left column of Fig.~\ref{Fig:TFFs}, are consistent with the asymptotic behavior of QCD, except for the pion TFF from BaBar \cite{BaBar:2009rrj}. The BaBar one exhibits linear behavior in the high energy region and crosses the asymptotic line ($Q^2F_{\pi\gamma^*\gamma}(-Q^2)=2F_{\pi}$). However, a later measurement given by Belle \cite{Belle:2012wwz} did not support such linear growth of the TFFs. Also, the space-like $\eta$ and $\eta'$ TFFs measured by BaBar \cite{BaBar:2011nrp} did not show such linear growth behavior. From the theoretical perspective, Ref.~\cite{Brodsky:2011yv} suggested that it is challenging to interpret the pion TFF from BaBar consistently within the existing theoretical framework in the high energy region. For instance, a recent study based on pQCD  \cite{Chai:2025xuz} supports the measurements of Belle. In both Fit A and Fit B, we apply the asymptotic limit strictly following QCD. We set it as \cite{Feldmann:1999uf} $\lim_{Q^2 \to \infty} Q^2\mathcal{F} _{P\gamma^*\gamma} (-Q^2)=6\sqrt{2} \sum_{a}C_{a} F^{P}_{a}$, and the corrections due to the anomalous dimension of $F_0$ is also considered by replacing $F_0\to F_0(1+\delta_{\infty } )$~\cite{Escribano:2015nra,Escribano:2015yup}, with $\delta_{\infty }=-0.17$ given by Ref.~\cite{Escribano:2015nra}. For $\eta$ and $\eta'$, their asymptotic limits are determined by the two-angle mixing parameters. The coefficients obtained from our best fits are in accordance with Ref.~\cite{Chen:2012vw}.

\begin{figure}[htb]
\centering
\includegraphics[width=1\linewidth,height=0.24\textheight]{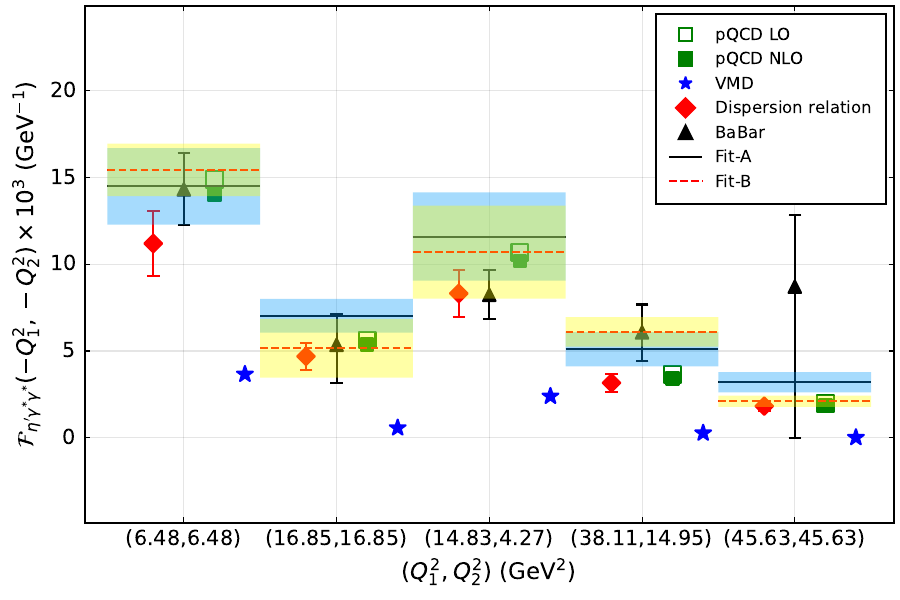}
\caption{Space-like doubly-virtual form factor of $\eta'$. The data is taken from BaBar \cite{BaBar:2018zpn}. The blue and yellow bands are the uncertainties for Fits A and B, respectively. The green bands are overlapping regions of them.}
\label{Fig:etapdou}
\end{figure}

The results of space-like doubly-virtual TFFs are shown in the right column of Fig.~\ref{Fig:TFFs} and Fig.~\ref{Fig:etapdou} (non-diagonal), where the former is for diagonal TFF, $Q_1=Q_2=Q$, and the latter for diagonal (the first two and the last data points) and non-diagonal one (the middle two data points), $Q_1\neq Q_2$. 
 For each pseudoscalar meson, we take six points ($Q^2=[0.7, 1, 1.3, 1.6, 2, 3.5]\mathrm{GeV}^2$) from BMW as the dataset \cite{Gerardin:2023naa}. For $\eta$ and pion TFFs, we also take the results of ETM \cite{ExtendedTwistedMass:2022ofm,ExtendedTwistedMass:2023hin} at the same six energy points as the data. Besides, the results of Ref.~\cite{Lin:2024khg} ($Q^2=[0.05,0.10,0.15,0.20]\mathrm{GeV}^2$) are included as data for the pion TFFs. As can be found, our results are compatible with those of LQCD. 
\begin{table}
\centering
{\footnotesize 
\renewcommand{\arraystretch}{2}
\newcommand{\tabincell}[2]{\begin{tabular}{@{}#1@{}}#2\end{tabular}}
\begin{ruledtabular}
\begin{tabular}{cccc}
 Process & This work (Fit A) & \tabincell{c}{Experimental value\\ \cite{ParticleDataGroup:2024cfk,BESIII:2022cul,BESIII:2024ddb}} \\
\colrule
 $\pi^0\to e^+e^-\gamma$ & $(1.17\pm 0.01)\times 10^{-2}$ & $(1.174\pm0.035)\times10^{-2}$ \\
 $\eta\to e^+e^-\gamma$ & {$(6.3\pm0.5)\times 10^{-3}$} & $(6.9\pm0.4)\times 10^{-3}$ \\
 $\eta\to \mu^+\mu^-\gamma$ & $(3.0\pm0.2)\times 10^{-4}$ & $(3.1\pm0.4)\times 10^{-4}$ \\
 $\eta'\to e^+e^-\gamma$ &$(4.85\pm0.21)\times 10^{-4}$ & $(4.91\pm0.27)\times 10^{-4}$ \\
 $\eta'\to \mu^+\mu^-\gamma$ & $(1.01\pm0.20)\times 10^{-4}$ & $(1.13\pm0.28)\times 10^{-4}$ \\
 $\pi^0\to 2e^+e^-$ & $(3.37\pm 0.02)\times 10^{-5}$ & $(3.34\pm0.16)\times 10^{-5}$ \\
 $\eta\to 2e^+e^-$ & $(2.60\pm0.17)\times 10^{-5}$ & $(2.40\pm0.22)\times 10^{-5}$ \\
 $\eta\to 2\mu^+\mu^-$ & $(3.6\pm0.2)\times 10^{-9}$ & $<3.6\times 10^{-4}$ \\
 $\eta\to e^+e^-\mu^+\mu^-$ & $(2.0 \pm0.2)\times 10^{-6}$ & $<6.88\times 10^{-6}$ \\
 $\eta'\to 2e^+e^-$ & $(2.4\pm0.2)\times 10^{-6}$ & $(4.5\pm1.0\pm0.5)\times 10^{-6}$ \\
 $\eta'\to 2\mu^+\mu^-$ & $(2.2\pm0.4)\times 10^{-8}$ & $<5.28\times 10^{-7}$ \\
 $\eta'\to e^+e^-\mu^+\mu^-$ & $(6.4\pm1.2)\times 10^{-7}$ & $<1.75\times 10^{-6}$ \\
\end{tabular}
\end{ruledtabular}
\caption{Predictions of branching ratios for $P\to l^+l^- \gamma$ and $P\to 2l^+l^-$ decays. Fits A and B get almost same results.}
\label{tab:t6}
}       
\end{table}
In Fig.~\ref{Fig:etapdou}, the data is taken from BaBar \cite{BaBar:2018zpn}, where the diagonal and non-diagonal TFF at the energy point $(Q_1^2,Q_2^2)$ is obtained by the pQCD results \cite{Braaten:1982yp,BaBar:2018zpn} multiplying with the weighted averaging ratio of the measured cross-section over that of the Monte Carlo simulation in the corresponding energy region of $(Q_1^2,Q_2^2)$. Notice that our results, though shown as the bands, correspond to the energy point $(Q_1^2,Q_2^2)$. As we have discussed above, the asymptotic TFF dominates in the high energy region. Hence, both Fits A and B can describe the datasets well, except for the data point at $(45.63,45.63)$. Nevertheless, taking into account the large uncertainty of the data point, ours are still compatible with the data. 
We also listed the results given by pQCD \cite{Braaten:1982yp,BaBar:2018zpn}, dispersion relation \cite{Holz:2024lom}, and VMD method \cite{Landsberg:1985gaz,BaBar:2018zpn} to have a comparison. As can be found, ours describes the data well, and is compatible with pQCD and dispersion relation. Because we adopted the asymptotic TFFs used in dispersion relation, which conform with pQCD for large $Q_i^2$. Although the asymptotic TFF dominates, RChT part also makes nonnegligible contribution, which makes our results perform better as a whole. For the last point $(45.63,45.63)$, its central value seems too large to be explained in the existing theoretical framework.

Our prediction of branching fractions for single- and double-Dalitz decays are shown in Table \ref{tab:t6}. They are consistent with that of the experiments (as shown in the Table) and previous theoretical predictions~\cite{Escribano:2015vjz,Guevara:2018rhj}. This implies that our TFFs are well constrained and have strong predictive ability. 
Indeed,	our pion TFF is successfully applied in analyzing the process of $J/\psi \to \pi^0 e^+e^-$~\cite{Zhang:2025}, where the first invariant mass spectrum is measured by BESIII \cite{BESIII:2025xjh} recently. 

\section{pole contribution to \texorpdfstring{$a_{\mu }^{\mathrm{HLbL}}$}{aHLbL} }
\label{sec4}
 The definition of Hadronic Light-by-Light tensor is given as \cite{Colangelo:2014dfa}
    \begin{eqnarray}
    \Pi_{\mu \nu \rho \sigma } (q_{1}\!,q_{2}\!,q_{3})\! & = & \!\!-\!i\!\!\int \!\!\mathrm{d}^{4}  x_{1}  \mathrm{d}^{4} x_{2}  \mathrm{d}^{4} x_{3}e^{i(q_{1}\cdot x_1+q_{2}\cdot x_2 +q_{3}\cdot x_3)} \nonumber \\
    & & \!\!\!\!\left \langle 0| T\left \{ j_{\mu }(x_{1}) j_{\nu }(x_{2}) j_{\rho }(x_{3}) j_{\sigma }(0)  \right \} |0\right \rangle\!,
\end{eqnarray}
 where $j_{\mu }=(\mathcal{V}_{\mu}^3+ \mathcal{V}_{\mu}^8/\sqrt{3})$ is the electromagnetic current. Applying the Cutkosky rules and dispersion relation, the pole contribution to HLbL tensor is \cite{Colangelo:2014dfa}
    \begin{eqnarray}
    &&\Pi_{\mu \nu \rho \sigma}^{P }(q_{1},q_{2},q_{3})\nonumber\\
    &= &  \frac{\mathcal{F}_{P \gamma^* \gamma^*}\left(q_1^2, q_2^2\right) \mathcal{F}_{P \gamma^* \gamma^*}\left(q_3^2, 0\right)}{s-m_{P}^2} \epsilon_{\mu \nu \alpha \beta} q_1^\alpha q_2^\beta \epsilon_{\rho \sigma \gamma \delta} q_3^\gamma k^\delta \nonumber \\
    & + & \frac{\mathcal{F}_{P \gamma^* \gamma^*}\left(q_1^2, q_3^2\right) \mathcal{F}_{P \gamma^* \gamma^*}\left(q_2^2, 0\right)}{t-m_{P}^2} \epsilon_{\mu \rho \alpha \beta} q_1^\alpha q_3^\beta \epsilon_{\nu \sigma \gamma \delta} q_2^\gamma k^\delta \nonumber \\
    & + & \frac{\mathcal{F}_{P \gamma^* \gamma^*}\left(q_2^2, q_3^2\right) \mathcal{F}_{P \gamma^* \gamma^*}\left(q_1^2, 0\right)}{u-m_{P}^2} \epsilon_{\nu \rho \alpha \beta} q_2^\alpha q_3^\beta \epsilon_{\mu \sigma \gamma \delta} q_1^\gamma k^\delta \nonumber,\\
\end{eqnarray}
with $k$ the momentum of the on-shell photon for HLbL and $s=(q_1+q_2)^2$, $t=(q_1-q_3)^2$ and $u=(q_1-k)^2$ the Mandelstam variables.
 Using the projection formula, the anomalous magnetic moment is then given by \cite{Colangelo:2014dfa,Roskies:1990ki,Jegerlehner:2009ry}
    \begin{equation}
 a_{\mu}\!=\!-\!\frac{1}{48m_{\mu}}\!\lim_{k \to 0}\!  \mathrm{Tr}\! \left \{ \!(\slashed{p}\!+\!m_{\mu}) [\gamma ^{\sigma }\!,\!\gamma ^{\nu } ] (\slashed{p}\!+\!m_{\mu})\frac{\partial \Gamma_{\nu }^{\mathrm{HLbL} }\! (\!k^2\!)}{\partial k_{\sigma } }\! \right \},
    \end{equation}
 where $\Gamma_{\nu }^{\mathrm{HLbL} } (k^2)$ is the electromagnetic vertex. The HLbL contribution to the electromagnetic vertex is
    \begin{eqnarray}
    \Gamma_{\nu }^{\mathrm{HLbL} }(k^2) & = & e^{6} \int \frac{\mathrm{d}^{4}q_{1} \mathrm{d}^{4}q_{2}   }{(2\pi )^{8} } \frac{1}{q_{1}^{2} }\frac{1}{q_{2}^{2}  }\frac{1}{(k-q_{1}-q_{2} )^{2}  } \nonumber \\
    & \times & \Pi _{ \nu\delta \rho \sigma } (q_{1},q_{2},k-q_{1}-q_{2}) \nonumber \\
    & &\! \gamma ^{\delta }\!\frac{(\slashed{p}+\slashed{q}_{1} +m_{\mu}  )}{(p+q_{1}  )^{2} -m_{\mu}^{2} }\! \gamma ^{\rho }\!  \frac{(\slashed{p}+\slashed{q}_{1}  +\slashed{q}_{2}+m_{\mu}  )}{(p+q_{1}+q_{2}  )^{2} -m_{\mu}^{2} }\!  \gamma ^{\sigma }.\nonumber \\
\end{eqnarray}
Here, $p$ is the momentum of the muon. 
 After performing the Wick rotation and  averaging over the directions, the pole contribution to $a_{\mu } ^{\mathrm{HLbL} }$ is then given by the master formula \cite{Hoferichter:2018kwz}
    \begin{eqnarray}
    &&a_\mu^{P \text {-HLbL }} =  \left(\frac{\alpha}{\pi}\right)^3 \int_0^{\infty} \mathrm{d} Q_1 \int_0^{\infty} \mathrm{d} Q_2 \int_{-1}^1 \mathrm{d} \tau \nonumber \\
    &&\times \left[w_1^{P}\left(Q_1, Q_2, \tau\right) \mathcal{F}_{P \gamma^* \gamma^*}\left(-Q_1^2,-Q_3^2\right) \mathcal{F}_{P \gamma^* \gamma^*}\left(-Q_2^2, 0\right)\right. \nonumber\\
    &&+ \left.w_2^{P}\left(Q_1, Q_2, \tau\right) \mathcal{F}_{P \gamma^* \gamma^*}\left(-Q_1^2,-Q_2^2\right) \mathcal{F}_{P \gamma^* \gamma^*}\left(-Q_3^2, 0\right)\right],\nonumber\\
\end{eqnarray}
 where $Q_3^2=Q_1^2+Q_2^2+2\tau Q_1 Q_2$, $\tau=\cos\theta$, $\theta$ is the angle between $Q_1$ and $Q_2$. And $w_{1,2}^{P}$ are the weight functions
    \begin{eqnarray}
    w_1^{P}\left(Q_1, Q_2, \tau\right) & = & -\frac{2 \pi}{3} \sqrt{1-\tau^2} \frac{Q_1^3 Q_2^3}{Q_2^2+m_{P }^2} T_1\left(Q_1, Q_2, \tau\right), \nonumber \\
    w_2^{P}\left(Q_1, Q_2, \tau\right) & = & -\frac{2 \pi}{3} \sqrt{1-\tau^2} \frac{Q_1^3 Q_2^3}{Q_3^2+m_{P}^2} T_2\left(Q_1, Q_2, \tau\right),\nonumber \\
\end{eqnarray}
 where the kernel functions $T_i$ are given in Ref.~\cite{Hoferichter:2018kwz}
\begin{eqnarray}
T_1(Q_1, Q_2, \tau) &=&X \frac{8 (\tau^2 - 1) (2m_\mu^2 - Q_2^2)}{Q_3^2 m_\mu^2}+ \frac{1}{Q_1 Q_2 Q_3^2 m_\mu^2}\nonumber \\&& \times\bigg[ Q_1 \big(\sigma_\mu(-Q_1^2) - 1\big) \Big( Q_1 \tau \big(\sigma_\mu(-Q_1^2)\nonumber\\&& + 1\big) + 4Q_2 (\tau^2 - 1) \Big) - 4\tau m_\mu^2 \bigg] \\
T_2(Q_1, Q_2, \tau) &=& \frac{1}{2Q_1 Q_2 Q_3^2 m_\mu^2}\nonumber\\
&&\bigg[ Q_1^2 \tau \big(\sigma_\mu(-Q_1^2) - 1\big) \big(\sigma_\mu(-Q_1^2) + 5\big) \nonumber \\
&& + Q_2^2 \tau \big(\sigma_\mu(-Q_2^2) - 1\big) \big(\sigma_\mu(-Q_2^2) + 5\big) \nonumber \\
&& + 4Q_1 Q_2 \big(\sigma_\mu(-Q_1^2) + \sigma_\mu(-Q_2^2) - 2\big) \nonumber \\
&&- 8\tau m_\mu^2 \bigg]  + X \bigg( \frac{8 (\tau^2 - 1)}{Q_3^2} - \frac{4}{m_\mu^2} \bigg),
\end{eqnarray}
with 
\begin{eqnarray}
X &=& \frac{1}{Q_1 Q_2 x} \arctan \left( \frac{zx}{1 - z\tau} \right), \\
x &=& \sqrt{1 - \tau^2}, \\
z &=& \frac{Q_1 Q_2}{4m_\mu^2} \big(1 - \sigma_\mu(-Q_1^2)\big)\big(1 - \sigma_\mu(-Q_2^2)\big), \\
\sigma_\mu(x) &=& \sqrt{1 - \frac{4m_\mu^2}{x}}.
\end{eqnarray}

Following Eq.~(\ref{Eq:TFF}), the pole contribution of the pseudoscalar can be divided into two parts: the hadronic one calculated from RChT and the one from the asymptotic QCD form factor, 
\begin{equation}
a_\mu^{P}=a_\mu^{P\mathrm{-h}}+a_\mu^{P\mathrm{-a}}.
\end{equation}  
The pion pole contribution is given as  
\begin{eqnarray}
a_\mu^{\pi^0}|_{\mathrm{Fit~A}}&=&(61.6\!\pm\! 0.5_{\mathrm{sta}}\! \pm\! 1.7_{N_C}\!\pm \!0.2_{\mathrm{R}} )\!\times\! 10^{-11}  \,,\nonumber\\
a_\mu^{\pi^0\text {-h }}|_{\mathrm{Fit~A}}&=&(56.5\!\pm\! 0.5_{\mathrm{sta}}\! \pm\! 1.7_{N_C}\!\pm \!0.2_{\mathrm{R}} )\!\times\! 10^{-11}  \,,\nonumber\\
 a_\mu^{\pi^0 \text {-a }}|_{\mathrm{Fit~A}}&=&(5.1\!\pm\! 0.3_{\mathrm{sta}}\!\pm\! 0.1_{N_C}\!\pm\! 0.1_{\mathrm{R}})\times 10^{-11}  \,.\nonumber\\
a_\mu^{\pi^0 }|_{\mathrm{Fit~B}}&=&(62.2\!\pm\! 0.4_{\mathrm{sta}}\!\pm\! 1.7_{N_C}\!\pm\!0.2_{\mathrm{R}} )\!\times\! 10^{-11}  \,,\nonumber\\
a_\mu^{\pi^0 \text {-h }}|_{\mathrm{Fit~B}}&=&(57.7\!\pm\! 0.4_{\mathrm{sta}}\!\pm\! 1.7_{N_C}\!\pm\!0.2_{\mathrm{R}} )\!\times\! 10^{-11}  \,,\nonumber\\
a_\mu^{\pi^0 \text {-a}}|_{\mathrm{Fit~B}}&=&(4.5\!\pm\! 0.3_{\mathrm{sta}}\!\pm\! 0.1_{N_C}\!\pm\!0.1_{\mathrm{R}})\!\times\! 10^{-11}  \,.\nonumber\\
\end{eqnarray}
For $\eta$ and $\eta'$, our results read
\begin{eqnarray}
a_\mu^{\eta }|_{\mathrm{Fit~A}}&=&(15.2\!\pm \!1.1_{\mathrm{sta}} \!\pm\! 1.2_{N_C}\!\pm\! 0.3_{\mathrm{R}})\!\times\! 10^{-11}  \,,\nonumber\\
a_\mu^{\eta \text {-h } }|_{\mathrm{Fit~A}}&=&(12.5\!\pm \!0.9_{\mathrm{sta}} \!\pm\! 1.2_{N_C}\!\pm\! 0.3_{\mathrm{R}})\!\times\! 10^{-11}  \,,\nonumber\\
a_\mu^{\eta \text {-a }}|_{\mathrm{Fit~A}}&=&(2.7\!\pm\! 0.4_{\mathrm{sta}} \!\pm\! 0.1_{N_C}\!\pm\! 0.2_{\mathrm{R}})\!\times 10^{-11}  \,.\nonumber\\
a_\mu^{\eta}|_{\mathrm{Fit~ B}}&=&(15.5\!\pm\! 1.1_{\mathrm{sta}}\! \pm \!1.3_{N_C}\!\pm\! 0.3_{\mathrm{R}}  )\times 10^{-11}  \,,\nonumber\\
a_\mu^{\eta \text {-h }}|_{\mathrm{Fit~ B}}&=&(12.7\!\pm\! 0.9_{\mathrm{sta}}\! \pm \!1.3_{N_C}\!\pm\! 0.3_{\mathrm{R}}  )\times 10^{-11}  \,,\nonumber\\
a_\mu^{\eta \text {-a }}|_{\mathrm{Fit~ B}}&=&(2.8\!\pm\! 0.4_{\mathrm{sta}} \!\pm\! 0.1_{N_C}\!\pm\! 0.2_{\mathrm{R}} )\times 10^{-11}  \,.\nonumber\\
\end{eqnarray} 
and
\begin{eqnarray}
a_\mu^{\eta' }|_{\mathrm{Fit~ A}}&=&(16.0\!\pm\! 1.0_{\mathrm{sta}}\!\pm\! 0.7_{\mathrm{N_C}}\!\pm\! 0.2_{\mathrm{R}})\!\times\! 10^{-11}  \,,\nonumber\\
a_\mu^{\eta' \text {-h }}|_{\mathrm{Fit~ A}}&=&(12.5\!\pm\! 0.9_{\mathrm{sta}}\!\pm\! 0.7_{\mathrm{N_C}}\!\pm\! 0.2_{\mathrm{R}})\!\times\! 10^{-11}  \,,\nonumber\\
a_\mu^{\eta' \text {-a }}|_{\mathrm{Fit~ A}}&=&(3.5\!\pm\! 0.4_{\mathrm{sta}}\! \pm \! 0.1_{N_C}\!\pm\! 0.1_{\mathrm{R}})\!\times\! 10^{-11}  \,.\nonumber\\
a_\mu^{\eta' }|_{\mathrm{Fit~B}}&=&(15.7\!\pm \!0.9_{\mathrm{sta}}\!\pm\! 0.7_{\mathrm{N_C}}\!\pm\!0.2_{\mathrm{R}})\!\times\! 10^{-11}  \,,\nonumber\\
a_\mu^{\eta'\text {-h } }|_{\mathrm{Fit~B}}&=&(11.9\!\pm \!0.8_{\mathrm{sta}}\!\pm\! 0.7_{\mathrm{N_C}}\!\pm\!0.2_{\mathrm{R}})\!\times\! 10^{-11}  \,,\nonumber\\
a_\mu^{\eta' \text {-a }}|_{\mathrm{Fit~B}}&=&(3.8\!\pm\! 0.4_{\mathrm{sta}} \!\pm\! 0.1_{N_C}\!\pm\! 0.1_{\mathrm{R}})\!\times\! 10^{-11}  \,.\nonumber\\
\end{eqnarray}
where the subscript \lq stat, ${N_C}$, R' denote for statistical uncertainty, systematical uncertainties from $1/N_C$ corrections, and systematical ones from the finite number of vector resonances, respectively. 
The statistical uncertainties are taken from the bootstrap method~\cite{Efron:1979bxm}. 
The large-$N_C$ correction uncertainties are taken by considering the effects from the momentum-dependent Breit-Wigner propagator~\cite{Guevara:2018rhj,Estrada:2024cfy}, where the NLO $N_C$ correction is implemented by the $\pi\pi$ loop contributions \cite{GomezDumm:2000fz}. It is composed of two components: The first part is obtained by replacing the $\rho$ mass with the real part of the $\pi\pi$-loop contribution.  This approach has proven practical for estimating HLbL uncertainties~\cite{Guevara:2018rhj,Estrada:2024cfy}; The second part is given by difference between the decay width forms, either a constant width or a momentum-dependent form like Eq.~(\ref{Eq:BW}).  
In addition, the mixing angles are also related to NLO $N_C$ corrections, we therefore perform the fit by fixing $f_{0/8}$ and $\theta_{0/8}$ to the central values of Ref.~\cite{Guo:2015xva} (NNLO Fit-B, $f_0=1.18$, $f_8=1.37$, $\theta_0=-6.8^\circ$, $\theta_8=-27.9^\circ$), and compare the resulting $(g-2)_\mu$ values with our central values. The final large-$N_C$ uncertainties are square root of quadratic sum of the three ones discussed above.
The uncertainties of finite number of vector resonances are estimated by performing the fit without $V'$ and $V''$, i.e., $\beta'_{P\gamma}=\beta''_{P\gamma}=0$. 
The total uncertainties can be obtained by the sum of the three uncertainties.
As can be seen, the asymptotic behaviour of space-like doubly-virtual TFF makes a significant contribution to $a_\mu^{P}$, roughly one-eighth of the pure hadronic contribution.
Finally, the total pole contributions to HLBL can be expressed as $a_\mu^{\pi^0+\eta+\eta'}|_{\mathrm{Fit~A}}=(92.8\pm 2.9 )\times 10^{-11}$, $a_\mu^{\pi^0+\eta+\eta'}|_{\mathrm{Fit~B}}=(93.4\pm 2.9 )\times 10^{-11}$.

\begin{figure}
    \centering
    \includegraphics[width=1\linewidth]{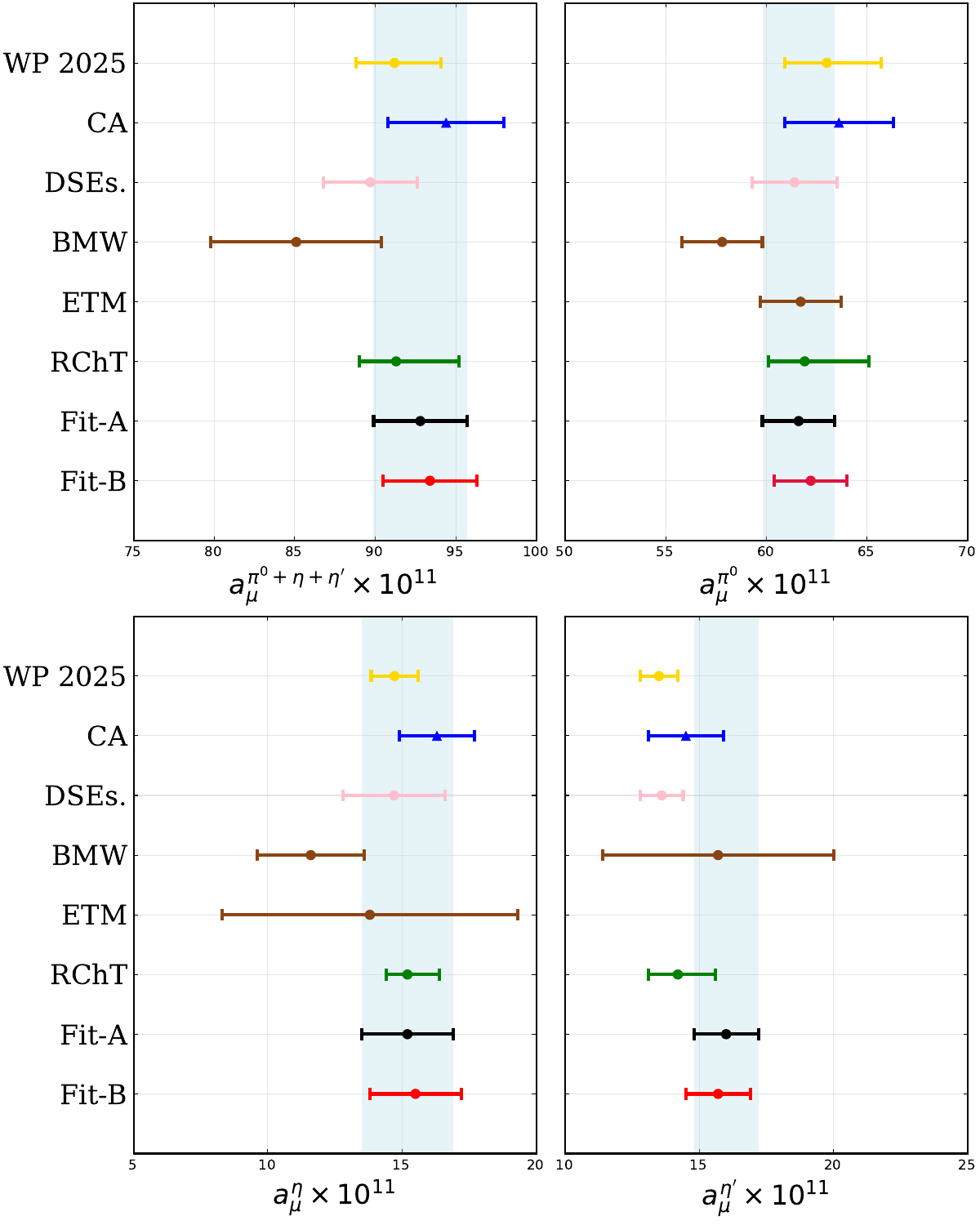}
    \caption{Recent results of pole contributions to $a_{\mu } ^{\mathrm{HLbL} }$. The blue error bands are given by our Fit-A. \lq WP' represents the white paper~\cite{Aliberti:2025beg}. The results of \lq CA' are taken from Ref.~\cite{Masjuan:2017tvw}, the results of \lq DSEs.' are taken from Ref.~\cite{Raya:2019dnh}, the results of LQCD are taken from BMW~\cite{Gerardin:2023naa} and ETM~\cite{ExtendedTwistedMass:2022ofm,ExtendedTwistedMass:2023hin}, the results of RChT are taken from Ref.~\cite{Estrada:2024cfy}. }
    \label{fig:pole}
\end{figure}

The contribution of each pseudoscalar and their total contribution are shown in Fig. \ref{fig:pole}. 
As can be found, Fits A and B are pretty close to each other. Nevertheless, we choose Fit-A as our final results according to the reasons discussed in the last subsection. 
It is shown that the pion pole contribution is much larger than that of the other pseudoscalars. This is not a surprise as the mass of the pion is much smaller than that of the $\eta$ and $\eta'$.
For the reader's convenience, the numerical results of the pseudoscalar pole contributions to HLbL are shown in Fig.~\ref{fig:pole}, together with other recent results~\cite{Aliberti:2025beg,Masjuan:2017tvw,Raya:2019dnh,Gerardin:2023naa,ExtendedTwistedMass:2022ofm,ExtendedTwistedMass:2023hin,Estrada:2024cfy}. 

As shown in Fig.~\ref{fig:pole}, a slightly large discrepancy of $\eta$-pole contribution between ours and LQCD is observed. It is caused by the different behaviour of the TFFs of $\eta$ in the energy region of $Q_i<0.5$~GeV as discussed in Ref.~\cite{Gerardin:2023naa}. Besides, ours is compatible with the result of another LQCD group ETM~\cite{ExtendedTwistedMass:2022ofm}, which gave $(13.8\pm5.5)\times 10^{-11}$. For $\eta'$ case, ours is closer to that of LQCD than to the white paper \cite{Aliberti:2025beg}. The reason is that we include the results (diagonal $\eta'$ TFF) of LQCD as a kind of data, which deviates from that of the white paper \cite{Aliberti:2025beg}. Interestingly, our results for the sum of the lightest pseudoscalars are in agreement with that of the white paper \cite{Aliberti:2025beg} but not with that of LQCD. Also, ours are close to other results based on phenomenological analysis, e.g., Refs.~\cite{Masjuan:2017tvw,Estrada:2024cfy,Holz:2024diw}.  
Our total contribution (of Fit A) to HLbL will be $(104.9 \pm 6.9)\times 10^{-11}$, with the other contributions taken from WP~\cite{Aliberti:2025beg}. Our leading order HLbL contributions are close to that of the data-driven method, $(103.3 \pm 8.8)\times 10^{-11}$~\cite{Aliberti:2025beg}, but much smaller than that of LQCD, ($125.5 \pm 11.6 \pm 0.4)\times 10^{-11}$~\cite{Fodor:2024jyn}. The total LQCD contribution of HLbL is obtained from direct computation of the HLbL diagram \cite{Fodor:2024jyn}, same as HVP. The reason of these discrepancies is still unknown, but it should be noticed that the pole contribution of the lightest pseudoscalars given by LQCD is compatible with ours within the uncertainties. See Fig.~\ref{fig:pole}. Indeed, unlike the total contribution, our pole contribution is even larger than that of LQCD.

    \section{Conclusion}\label{sec5}
 In this paper, we calculated the doubly-virtual TFFs of $\pi^0(\eta,\eta')\gamma^*\gamma^*$ within the framework of resonance chiral theory. A comprehensive analysis of time-like singly-virtual and space-like singly-virtual and doubly-virtual TFFs is performed. The experimental datasets as well as the results from LQCD are well described, and the unknown couplings are fixed. The TFFs of the $\pi^0,\eta,\eta'$ are obtained, and with these TFFs, we evaluate the contributions of these pseudoscalars to the HLbL, $a_\mu^{P}$, shown in Fig.~\ref{fig:pole}. Our Fit-A gives, $a_\mu^{\pi^0 }=(61.6\pm 1.8)\times 10^{-11} $, $a_\mu^{\eta }=(15.2\pm 1.7)\times 10^{-11} $, and $a_\mu^{\eta'}=(16.0\pm 1.2)\times 10^{-11}$ . The total contribution of these pseudoscalar meson poles to the HLbL is  $a_\mu^{\pi^0+\eta+\eta'}=(92.8\pm 2.9)\times 10^{-11}$.  
The scattering and decay processes studied in this analysis are inherently linked. Consequently, future measurements of any process (e.g., $ee\to P\gamma$, $P\to ee\gamma$, $V\to P\gamma$ and $\eta'\to V\gamma$) will thereby provide crucial input for refining the pseudoscalar transition form factors.
    
    \section*{Acknowledgements}
 We thank Profs. Yun-Hua Chen, Zhi-Hui Guo, and Shan Cheng for helpful discussions.
 This work is supported by the National Natural Science Foundation of China (NSFC) with Grants No.12322502, 12335002, Joint Large Scale Scientific Facility Funds of the NSFC and Chinese Academy of Sciences (CAS) under contract No.U1932110, Hunan Provincial Natural Science Foundation with Grant No.2024JJ3044, and Fundamental Research Funds for the central universities.

\appendix
\section{Transition form factors and useful expressions}\label{app1}
\subsection{TFFs of the lightest pseudoscalars}\label{app:TFF}
The TFFs calculated with all the assumptions, including $\omega-\phi$ and $\rho-\omega$ mixing, are given as 
\begin{widetext}   
\begin{equation}   
\mathcal{F}_{P\gamma^*\gamma^* }(q_1^2,q_2^2)=\mathcal{F}_{P\gamma^*\gamma^* }^{\mathrm{local}}+\mathcal{F}_{P\gamma^*\gamma^* }^{1R}(q_1^2,q_2^2)+\mathcal{F}_{P\gamma^*\gamma^* }^{2R}(q_1^2,q_2^2). 
\end{equation}
The one resonance part of the pion form factor is given as 
\begin{eqnarray}
\mathcal{F}_{\pi ^{0}\gamma^*\gamma^* }^{1R}\!(q_1^2,\!q_2^2)\!&=&\!\frac{-2}{3 F M_V}\!\bigg[\!F\!_\rho(q_{2}^{2}) \text{BW}(\rho,q_{2}^{2} ) \big(\sqrt{2} \cos\!\delta\!+\!\sqrt{3} \sin\!\delta_\rho(q_{2}^{2} ) (\sqrt{2} \sin\!\theta_V\!+\!2 \cos\! \theta_V\!)\big)(\tilde{c}_{1235} m_{\pi }^2\!+\!\tilde{c}_{125} q_{1}^{2} \!-\!\tilde{c}_{1256}q_{2}^{2} )\!\bigg] \nonumber\\   
    \!&+&\!\frac{-2}{3 F M_V}\!\bigg[\!F\!_\omega(q_{2}^{2} ) \text{BW}(\omega ,q_{2}^{2} ) \big(2 \sqrt{3} \cos\!\delta \cos\! \theta_V\!\!+\!\!\sqrt{6} \cos\!\delta \sin\!\theta_V\!\!-\!\!\sqrt{2} \sin\!\delta_\omega(q_{2}^{2})\big)(\tilde{c}_{1235} m_{\pi }^2\!+\!\tilde{c}_{125} q_{1}^{2} \!\!-\!\tilde{c}_{1256}q_{2}^{2} )\!\bigg]  \nonumber\\ 
    \!&+&\!\frac{-2}{3 F M_V}\!\bigg[\!F\!_\phi(q_{2}^{2} )\text{BW}(\phi,q_{2}^{2}) (\sqrt{6} \cos\!\theta_V\!-\!2\sqrt{3} \sin\!\theta_V ) (\tilde{c}_{1235} m_{\pi }^2\!+\!\tilde{c}_{125} q_{1}^{2}\!-\!\tilde{c}_{1256}q_{2}^{2} )\bigg]\!+\!\bigg\{q_1\leftrightarrow q_2\bigg\}\,,   
\end{eqnarray}
where one has 
\begin{eqnarray}
F\!_\rho(q^2)\!&=&\!\frac{F_V}{9} \bigg[9 \cos\!\delta \big(1\!+\! \frac{8\sqrt{2}\alpha _{V} }{M_V^2}  m_{\pi }^2\big) 
\!+\!\sqrt{3} \sin\!\delta_\rho(q^2) \big(3  \sin\!\theta_V \!-\! \frac{16\sqrt{2}\alpha _{V} }{M_V^2}  m_K^2 (\!\sqrt{2} \cos\! \theta_V\!-\!2 \sin\!\theta_V \!) \nonumber\\
\!&+&\! \frac{8\sqrt{2}\alpha_{V} }{M_V^2}  m_{\pi }^2 (2 \sqrt{2} \cos\!\theta_V\!-\!\sin\!\theta_V \!)\big)\!\bigg],\nonumber\\   
F\!_\omega(q^2)\!&=&\!\frac{F_V}{9} \bigg[\sqrt{3} \cos\!\delta \Big(\!3  \sin\!\theta_V \!-\! \frac{16\sqrt{2}\alpha _{V} }{M_V^2}  m_K^2 \big(\!\sqrt{2} \cos\! \theta_V\!-\!2 \sin\!\theta_V \!\big) \!+\! \frac{8\sqrt{2}\alpha _{V} }{M_V^2}  m_{\pi }^2 \big(\!2 \sqrt{2} \cos\!\theta_V\!-\!\sin\!\theta_V \!\big)\!\Big) \nonumber\\
\!& &\! -9 \sin\!\delta_\omega(q^2) \big(1\!+\! \frac{8\sqrt{2}\alpha _{V} }{M_V^2}  m_{\pi }^2\big)\bigg],\nonumber\\  
F\!_\phi(q^2)\!&=&\! \frac{F_V}{3\sqrt{3}} \bigg[3 \cos\!\theta_V \!+\! \frac{16\sqrt{2}\alpha_V}{M_V^2} m_K^2 \big(\!\sqrt{2} \sin\!\theta_V\!+\!2 \cos\!\theta_V \!\big) \! - \! \frac{8\sqrt{2}\alpha_V}{M_V^2} m_{\pi}^2 \big(\!2\sqrt{2} \sin\!\theta_V \!+\! \cos\!\theta_V \!\big) \bigg].\nonumber
\end{eqnarray}
The two resonances part is given as 
\begin{eqnarray}
\mathcal{F}_{\pi^{0}\gamma^*\gamma^*}^{2R}\!(q_1^2,\!q_2^2)\!&=&\!-\frac{1}{\sqrt{3} F}\!\big(\!\tilde{d}_{123} m_{\pi}^2\!+\!\tilde{d}_{3}(q_1^2\!+\!q_2^2)\big) \nonumber\\
\!& &\!\times \bigg\{\!-\!4\cos\!\delta\, F\!_\rho(q_{1}^{2}) F\!_\rho(q_{2}^{2}) \big(\!\sin\!\theta_V\!+\!\sqrt{2} \cos\!\theta_V\!\big) \big(\!\sin\!\delta_\rho(q_{1}^{2})\!+\!\sin\!\delta_\rho(q_{2}^{2})\big) \text{BW}(\rho,q_{1}^{2}) \text{BW}(\rho,q_{2}^{2})\nonumber\\
\!& &\!-\!2(\sin\!\theta_V\!+\!\sqrt{2} \cos\!\theta_V\!) \Big[F\!_\rho(q_{2}^{2}) F\!_\omega(q_{1}^{2}) \text{BW}(\omega,q_{1}^{2}) \text{BW}(\rho,q_{2}^{2}) \big(\!\cos\!2\delta\!-\!2 \sin\!\delta_\omega(q_{1}^{2}) \sin\!\delta_\rho(q_{2}^{2})\!+\!1\big) \nonumber\\
\!& &\;\;+F\!_\rho(q_{1}^{2}) F\!_\omega(q_{2}^{2}) \text{BW}(\rho,q_{1}^{2}) \text{BW}(\omega,q_{2}^{2}) \big(\!\cos\!2\delta\!-\!2 \sin\!\delta_\rho(q_{1}^{2}) \sin\!\delta_\omega(q_{2}^{2})\!+\!1\big)\Big]\nonumber\\
\!& &\!+\!4\cos\!\delta \big(\!\sqrt{2} \sin\!\theta_V\!\!-\!\cos\!\theta_V\!\big) \Big[F\!_\rho(q_{2}^{2})F\!_\phi(q_{1}^{2}) \text{BW}(\phi,q_{1}^{2}) \text{BW}(\rho,q_{2}^{2}) \!+\!F\!_\rho(q_{1}^{2}) F\!_\phi(q_{2}^{2}) \text{BW}(\rho,q_{1}^{2}) \text{BW}(\phi,q_{2}^{2})\Big]\nonumber\\
\!& &\!+\!4\cos\!\delta F\!_\omega(q_{1}^{2}) F\!_\omega(q_{2}^{2}) \big(\!\sin\!\theta_V\!\!+\!\!\sqrt{2} \cos\!\theta_V\!\big) \Big(\!\sin\!\delta_\omega(q_{1}^{2})\!+\!\sin\!\delta_\omega(q_{2}^{2})\!\Big) \text{BW}(\omega,q_{1}^{2}) \text{BW}(\omega,q_{2}^{2})\nonumber\\
\!& &\!-\!4\big(\!\sqrt{2} \sin\!\theta_V\!-\!\cos\!\theta_V\!\big) \Big[F\!_\omega(q_{1}^{2}) F\!_\phi(q_{2}^{2}) \sin\!\delta_\omega(q_{1}^{2}) \text{BW}(\omega,q_{1}^{2}) \text{BW}(\phi,q_{2}^{2}) \nonumber\\
\!& &\!\;\;+F\!_\omega(q_{2}^{2}) F\!_\phi(q_{1}^{2}) \sin\!\delta_\omega(q_{2}^{2}) \text{BW}(\phi,q_{1}^{2}) \text{BW}(\omega,q_{2}^{2})\Big]\!\bigg\}.
\end{eqnarray}
For the TFFs of $\eta$, the one resonance part it is given as 
\begin{eqnarray}
\mathcal{F}_{\eta\gamma^*\gamma^*}^{1R}\!(q_1^2,\!q_2^2)\!\!&=&\!-\frac{2}{9 F M_V}\! F\!_\rho(q_2^2) \text{BW}(\rho, q_2^2) \!\bigg\{\! 2 C_s \Big[\! -\sin\!\delta_\rho(q_2^2) \Big(\! 3 \sqrt{2} \tilde{c}_{8} M_V^2 \sin\!\theta_V \!-\!\sqrt{3} (\!\sqrt{2} \cos\!\theta_V\!-\!2 \sin\!\theta_V \!) \nonumber\\
\!& &\!\times\!\big(\! m_{\eta}^2 (\tilde{c}_{1235} \!-\! 8 \tilde{c}_{3}) \!+\! \tilde{c}_{125} q_{1}^{2} \!-\! \tilde{c}_{1256} q_{2}^{2} \!+\! 8 \tilde{c}_{3} (2 m_K^2 \!-\! m_{\pi}^2) \!\big)\!\Big) \!-\! 3 \sqrt{6} \tilde{c}_{8} M_V^2 
\!\cos\!\delta \!\Big] \nonumber\\
\!&&\!+\! C_q \Big[\! m_{\eta}^2 (\tilde{c}_{1235} \!-\! 8 \tilde{c}_{3}) \big(\! 9 \sqrt{2} \cos\!\delta \!+\! \sqrt{3} \sin\!\delta_\rho(q_2^2) (\!\sqrt{2} \sin\!\theta_V \!+\! 2 \cos\!\theta_V )\!\big) \nonumber\\
\!& &\!+\!\tilde{c}_{125} q_{1}^{2} \big(\! 9 \sqrt{2} \cos\!\delta \!+\! \sqrt{3} \sin\!\delta_\rho(q_2^2) (\sqrt{2} \sin\!\theta_V \!+\! 2 \cos\!\theta_V )\!\big) \!-\!(\!\tilde{c}_{1256} q_{2}^{2} \!-\! 8 \tilde{c}_{3} m_{\pi}^2 ) \nonumber\\
\!& &\!\times\!\big( 9 \sqrt{2} \cos\!\delta \!+\! \sqrt{3} \sin\!\delta_\rho(q_2^2) (\sqrt{2} \sin\!\theta_V \!+\! 2 \cos\!\theta_V )\big) \!+\!12 \tilde{c}_{8} M_V^2 \big(\sqrt{3} \cos\!\delta \!+\! \sin\!\theta_V \!\sin\!\delta_\rho(q_2^2)\big)\!\Big]\bigg\}  \nonumber\\
\!&-&\!\frac{2}{9 F M_V}\! F\!_\omega(q_2^2) \text{BW}(\omega, q_2^2)\!\bigg\{\! m_{\eta}^2 (\tilde{c}_{1235} \!-\! 8 \tilde{c}_{3}) \Big[\! -\!9\! \sqrt{2} C_q \sin\!\delta_\omega(q_2^2) \!+\!\sqrt{3} \cos\!\delta \big( C_q (\sqrt{2} \sin\!\theta_V \!+\! 2 \cos\!\theta_V) \nonumber\\
\!& &\!+\! 2 C_s (\sqrt{2} \cos\!\theta_V \!-\! 2 \sin\!\theta_V)\!\big)\!\Big] \!+\!\cos\!\delta \sin\!\theta_V \Big[\! C_q \Big(\!\sqrt{6} \big(\tilde{c}_{125} q_{1}^{2} \!-\! \tilde{c}_{1256} q_{2}^{2} \!+\! 8 \tilde{c}_{3} m_{\pi}^2\big) 
+\! 12 \tilde{c}_{8} M_V^2 \!\Big) \nonumber\\
\!& &\!-\!2 C_s \Big(\! 2\sqrt{3} \big(\!\tilde{c}_{125} q_{1}^{2} \!-\! \tilde{c}_{1256} q_{2}^{2} \!+\! 8 \tilde{c}_{3} (2 m_K^2 \!-\! m_{\pi}^2)\!\big) \!+\! 3\sqrt{2} \tilde{c}_{8} M_V^2 \!\Big)\!\Big] \nonumber\\
\!& &\!-\!3 \sin\!\delta_\omega(q_2^2) \Big[ C_q \big(\! 3\sqrt{2} (\!\tilde{c}_{125} q_{1}^{2} \!-\! \tilde{c}_{1256} q_{2}^{2} \!+\! 8 \tilde{c}_{3} m_{\pi}^2) \!+\! 4\sqrt{3} \tilde{c}_{8} M_V^2 \big) \!-\!2\sqrt{6} \tilde{c}_{8} \!M_V^2 C_s \!\Big] \nonumber\\ 
\!&&\!+\!2\sqrt{3} \cos\!\delta \cos\!\theta_V \Big[ \tilde{c}_{125} q_{1}^{2} (C_q \!+\! \sqrt{2} C_s) \!-\! \tilde{c}_{1256} q_{2}^{2} (C_q \!+\! \sqrt{2} C_s) \!+\!8 \tilde{c}_{3} \big(\sqrt{2} C_s 
(2 m_K^2 \!-\! m_{\pi}^2) \!+\! m_{\pi}^2 C_q \big) \Big] \!\bigg\}  \nonumber\\
\!&-&\!\frac{2}{9 F M_V}\! F\!_\phi(q_2^2) \text{BW}(\phi, q_2^2) \bigg\{\! C_q \Big[ \sqrt{3} (\sqrt{2} \cos\!\theta_V \!-\! 2 \sin\!\theta_V ) \big( m_{\eta}^2 (\tilde{c}_{1235} \!-\! 8 \tilde{c}_{3}) \!+\!\tilde{c}_{125} q_{1}^{2} \!-\! \tilde{c}_{1256} q_{2}^{2}  \nonumber\\
\!&&\!+\! 8 \tilde{c}_{3} m_{\pi}^2 \big) \!+\! 12 \tilde{c}_{8} M_V^2 \cos\!\theta_V \!\Big] \!-\!2 C_s \Big[ \sqrt{3} (\sqrt{2} \sin\!\theta_V \!+\! 2 \cos\!\theta_V ) \big( m_{\eta}^2 (\tilde{c}_{1235} \!-\! 8 \tilde{c}_{3}) \nonumber\\
\!& &\!+\!\tilde{c}_{125} q_{1}^{2} \!-\! \tilde{c}_{1256} q_{2}^{2} \!+\! 8 \tilde{c}_{3} (2 m_K^2 \!-\! m_{\pi}^2) \big) \!+\! 3\sqrt{2} \tilde{c}_{8} M_V^2 \cos\!\theta_V \!\Big] \!\bigg\} \!+\!\bigg\{q_1\leftrightarrow q_2\bigg\}.
\end{eqnarray}
The two resonances part is given as 
\begin{eqnarray}
\mathcal{F}_{\eta\gamma^*\gamma^* }  ^{2R}\!&=&\!-\frac{2}{3 F}\! F\!_\rho(q_1^2) F\!_\rho(q_2^2) \text{BW}(\rho,q_1^2) \text{BW}(\rho,q_2^2) \nonumber\\
\!& &\!\times\!\bigg\{\! C_s \Big[\! \sin\delta_\rho(q_1^2) \sin\delta_\rho(q_2^2) \Big(\! 4\sqrt{3} \tilde{d}_{5} M_V^2 \!-\!(4 \sin2\theta_V\!+\!\sqrt{2}\cos\!2\theta_V\!-\!3\sqrt{2}) \nonumber\\
\!& &\!\times\!\big(\tilde{d}_{123} m_{\eta}^2 \!+\! 8\tilde{d}_{2} (2m_K^2\!-\!m_{\eta}^2\!-\!m_{\pi}^2) \!+\!\tilde{d}_{3} (q_1^2\!+\!q_2^2)\big)\!\Big) \!+\! 4\sqrt{3} \tilde{d}_{5} M_V^2 \cos^{2}\!\delta \Big] \nonumber\\
\!& &\!+ C_q \Big[ -\big(\!6\cos^{2}\!\delta \!+\! \sin\delta_\rho(q_1^2) \sin\delta_\rho(q_2^2) (\!2\sqrt{2}\sin2\theta_V\!+\!\cos\!2\theta_V\!+\!3)\big) \nonumber\\
\!& &\!\times\!\big( m_{\eta}^2 (\tilde{d}_{123}\!-\!8\tilde{d}_{2}) \!+\! 8m_{\pi}^2 \tilde{d}_{2} \!+\!\tilde{d}_{3} (q_1^2\!+\!q_2^2)\!\big) \!-\! 2\sqrt{6} \tilde{d}_{5} M_V^2 \big(\!\cos\!2\delta \!+\! 2\sin\delta_\rho(q_1^2) \!\sin\delta_\rho(q_2^2)\!+\!1\big)\Big]\!\bigg\}\nonumber\\
\!&-&\!\frac{2}{3 F}\! \bigg\{ \! F\!_\rho(q_1^2) F\!_\omega(q_2^2) \text{BW}(\rho,q_1^2) \text{BW}(\omega,q_2^2) \!\times\! \bigg[\! -\sin\delta_\rho(q_1^2) \Big[ C_s\!\Big( (4\sin2\theta_V\!+\!\sqrt{2}\cos\!2\theta_V\!-\!3\sqrt{2}) \nonumber\\
\!& &\!\times\!\big(\!\tilde{d}_{123} m_{\eta}^2 \!+\! 8\tilde{d}_{2} (2m_K^2\!-\!m_{\eta}^2\!-\!m_{\pi}^2) \!+\!\tilde{d}_{3} (q_1^2\!+\!q_2^2)\!\big) \!-\! 4\sqrt{3} \tilde{d}_{5} M_V^2\!\Big) \!+\! C_q \Big(\! (\!2\sqrt{2}\sin2\theta_V\!+\!\cos\!2\theta_V\!+\!3) \nonumber\\
\!& &\!\times\!\big(\! m_{\eta}^2 (\tilde{d}_{123}\!-\!8\tilde{d}_{2}) \!+\!8m_{\pi}^2 \tilde{d}_{2} \!+\!\tilde{d}_{3} (q_1^2\!+\!q_2^2)\!\big) \!+\! 4\sqrt{6} \tilde{d}_{5} M_V^2\!\Big)\Big] \!+\! 2 \sin\delta_\omega(q_2^2) \Big(\!3 C_q \big(\! m_{\eta}^2 (\tilde{d}_{123}\!-\!8\tilde{d}_{2}) \nonumber\\
\!& &\!+\!8m_{\pi}^2 \tilde{d}_{2} \!+\!\tilde{d}_{3} (q_1^2\!+\!q_2^2)\!\big) \!+\!2\sqrt{3} \tilde{d}_{5} M_V^2 (\sqrt{2} C_q\!-\! C_s)\!\Big)\!\bigg] \!+\! \bigg[\!q_1\!\leftrightarrow\! q_2\!\bigg]\!\bigg\}\nonumber\\
\!&-&\!\frac{2}{3 F}\! \bigg\{ \bigg[\! F\!_\rho(q_1^2) F\!_\phi(q_2^2) \sin\delta_\rho(q_1^2) \text{BW}(\rho,q_1^2) \text{BW}(\phi,q_2^2)  \!\times\! \Big(\! m_{\eta}^2 (\tilde{d}_{123}\!-\!8\tilde{d}_{2}) \big(2\cos\!2\theta_V (\sqrt{2} C_q\!+\!2 C_s)  \nonumber\\
\!& &\!-\!\sin2\theta_V (C_q\!+\!\sqrt{2} C_s)\big)+\!8\tilde{d}_{2} \big(\! C_s (2m_K^2\!-\!m_{\pi}^2) (4\cos\!2\theta_V\!-\!\sqrt{2}\sin2\theta_V) \!+\! m_{\pi}^2 C_q (\!2\sqrt{2}\cos\!2\theta_V\!-\!\sin2\theta_V)\!\big) \nonumber\\
\!& &\!+\!\tilde{d}_{3} (q_1^2\!+\!q_2^2) \big(2\cos\!2\theta_V (\sqrt{2} C_q\!+\!2 C_s) \!-\!\sin\!2\theta_V (C_q\!+\!\sqrt{2} C_s)\!\big)\!\Big)\bigg]\!+\! \bigg[\!q_1\!\leftrightarrow\! q_2\!\bigg]\!  \bigg\}\nonumber\\
\!&+&\!\frac{2}{3 F}\! F\!_\omega(q_1^2) F\!_\omega(q_2^2) \text{BW}(\omega,q_1^2) \text{BW}(\omega,q_2^2) \!\times\! \bigg\{\! C_s \Big(\! \cos^{2}\!\delta (4\sin2\theta_V\!+\!\sqrt{2}\cos\!2\theta_V\!-\!3\sqrt{2}) \nonumber\\
\!& &\!\times\!\big(\!\tilde{d}_{123} m_{\eta}^2 \!+\!8\tilde{d}_{2} (2m_K^2\!-\!m_{\eta}^2\!-\!m_{\pi}^2) \!+\!\tilde{d}_{3} (q_1^2\!+\!q_2^2)\big) \!-\!2\sqrt{3} \tilde{d}_{5} M_V^2 \big(\!\cos\!2\delta \!+\! 2 \sin\delta_\omega(q_1^2)\!\sin\delta_\omega(q_2^2) \!+\!1\!\big)\!\Big)  \nonumber\\
\!& &\!+\! C_q \Big(\! \big(\cos^{2}\!\delta (\!2\sqrt{2}\sin2\theta_V\!+\!\cos\!2\theta_V\!+\!3) \!+\!6 \sin\delta_\omega(q_1^2)\sin\delta_\omega(q_2^2)\!\big) \nonumber\\
\!& &\! \big(m_{\eta}^2 (\tilde{d}_{123}\!-\!8\tilde{d}_{2}) \!+\!8m_{\pi}^2 \tilde{d}_{2} \!+\!\tilde{d}_{3} (q_1^2\!+\!q_2^2)\big) \!+\! 2\sqrt{6} \tilde{d}_{5} M_V^2 \big(\cos\!2\delta \!+\!2 \sin\delta_\omega(q_1^2) \sin\delta_\omega(q_2^2) \!+\!1\!\big)\!\Big)\!\bigg\}\nonumber\\
\!&-&\!\frac{2}{3 F}\!\bigg\{ F\!_\omega(q_1^2) F\!_\phi(q_2^2) \text{BW}(\omega,q_1^2) \text{BW}(\phi,q_2^2) \cos\!\delta  \nonumber\\
\!& &\!\times\!\bigg[\!-\!m_{\eta}^2 (\tilde{d}_{123}\!-\!8\tilde{d}_{2}) \big(2\cos\!2\theta_V (\sqrt{2} C_q\!+\!2 C_s) \!-\!\sin2\theta_V (C_q\!+\!\sqrt{2} C_s)\big) \!+\!8\tilde{d}_{2} \big( C_s (2m_K^2\!-\!m_{\pi}^2) \nonumber\\
\!& &\!\times\!(\sqrt{2}\sin2\theta_V\!-\!4\cos\!2\theta_V) \!+\! m_{\pi}^2 C_q (\sin2\theta_V\!-\!2\sqrt{2}\cos\!2\theta_V\!)\big) \!-\!\tilde{d}_{3} (q_1^2\!+\!q_2^2) \nonumber\\
\!& &\!\times\!\big(2\cos\!2\theta_V (\sqrt{2} C_q\!+\!2 C_s) \!-\!\sin2\theta_V (C_q\!+\!\sqrt{2} C_s)\big)\!\bigg]\!+\! \bigg[\!q_1\!\leftrightarrow\! q_2\!\bigg]\! \bigg\}\nonumber\\
\!&-&\!\!\frac{2}{3 F}\! F\!_\phi(q_1^2) F\!_\phi(q_2^2) \text{BW}(\phi,q_1^2) \text{BW}(\phi,q_2^2)  \bigg\{\! C_s \Big( (4\sin2\theta_V\!+\!\sqrt{2}\cos\!2\theta_V\!+\!3\sqrt{2}) \big(\tilde{d}_{123} m_{\eta}^2 \nonumber\\
\!& &\!+\!8\tilde{d}_{2} (2m_K^2\!-\!m_{\eta}^2\!-\!m_{\pi}^2) \!+\!\tilde{d}_{3} (q_1^2\!+\!q_2^2)\!\big) \!+\! 4\sqrt{3} \tilde{d}_{5} M_V^2\!\Big) \!+\! C_q \Big( (\!2\sqrt{2}\sin2\theta_V\!+\!\cos\!2\theta_V\!-\!3\!) \nonumber\\
\!& &\!\times\!\big( m_{\eta}^2 (\tilde{d}_{123}\!-\!8\tilde{d}_{2}) \!+\!8m_{\pi}^2 \tilde{d}_{2} \!+\!\tilde{d}_{3} (q_1^2\!+\!q_2^2)\!\big) \!-\! 4\sqrt{6} \tilde{d}_{5} M_V^2\!\Big)\!\bigg\}.
\end{eqnarray}
For the TFF of $\eta'$, $\mathcal{F}_{\eta'\gamma^*\gamma^* }(q_1^2,q_2^2)$, one only needs to perform the following changes on $\mathcal{F}_{\eta\gamma^*\gamma^* }(q_1^2,q_2^2)$, $C_q\to C'_q\,\,,\,\,C_s\to -C'_s\,\,,\,\,m_{\eta }\to m_{\eta' }$.

\subsection{\texorpdfstring{$VP\gamma^{*}$}{V-P-gamma*} form factor}
\label{app:A2}
The relevant Feynman diagrams for the TFF of $\mathcal{F} _{VP\gamma^{*} }(q^2)$ is shown in Fig.~\ref{Fig:etapwg}. The LO diagrams are shown in the first row, and the electromagnetic corrections are listed in the other rows. Notice that the $V-\gamma^*-V$ transition in Fig.~\ref{Fig:etapwg} (d) and (f) would be absorbed into the redefinition of the field through wave-function renormalization, thus only different type of the internal vector resonances contribute to the $VP\gamma^{*}$ TFF.
\begin{figure}[!htb]
\centering
\includegraphics[width=0.5\linewidth]{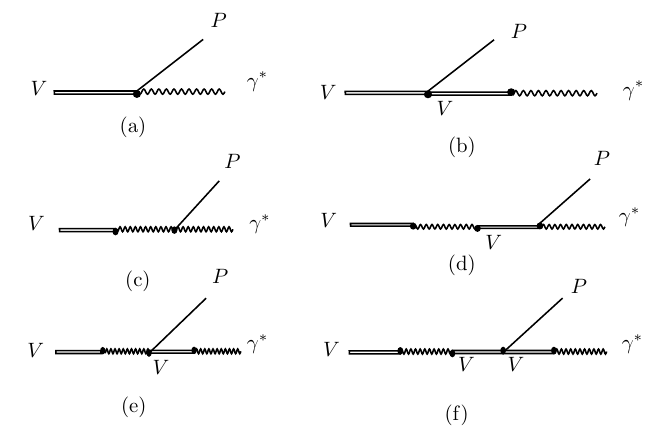}
\caption{Feynman diagrams for TFF of $V\to P\gamma^*$. \label{Fig:etapwg}}
\end{figure}
The form factor is given as       
        \begin{eqnarray}
             \mathcal{F} _{VP\gamma^{*} }(q^2) &=& \mathcal{F} _{VP\gamma^{*} } ^{\mathrm{LO} }(q^2)+\mathcal{F} _{VP\gamma^{*} } ^{\mathrm{EM} }(q^2)\,.
        \end{eqnarray}
The $\mathcal{F} _{VP\gamma^{*} }^{\mathrm{LO} }(q^2)$ is
\begin{eqnarray}
            \mathcal{F} _{VP\gamma^{*} }^{\mathrm{LO} }(q^2) &=& \mathcal{F} _{VP\gamma^{*} } ^{\mathrm{1R} }(q^2)+\mathcal{F} _{VP\gamma^{*} } ^{\mathrm{2R} }(q^2) \,,
\end{eqnarray}
where one has 
\begin{eqnarray}
    F_{\rho^0\pi^0\gamma^*}^{\mathrm{1R}}(q^2)\!&=&\! \frac{-2}{3FM_V M_\rho}\!\!\big[\!\sqrt{2}\!\cos{\!\delta}\!\!+\!\!\sqrt{3}(2\!\cos{\!\theta_V}\!\!+\!\!\sqrt{2}\!\sin{\!\theta_V})\sin\!{\delta\!_\rho}\!(M\!_\rho\!\!^2) \big](\tilde{c}_{125}q^2+\tilde{c}_{1235}m_\pi^2-\tilde{c}_{1256}M\!_\rho\!\!^2),\nonumber\\
    \mathcal{F}_{\rho^0\pi^0\gamma^{*}}^{\mathrm{2R}}(q^2) \!&=&\! \frac{2}{3F M_\rho} \bigg[ 2 \text{BW}(\phi, q^2) \cos\!\delta F_\phi(q^2) \big( \cos\!\theta_V \!-\! \sqrt{2} \sin\!\theta_V \big)
\!\! +\! \big( \sqrt{2} \cos\!\theta_V \!+\! \sin\!\theta_V \big)\nonumber\\&&
\times\big( 2 \text{BW}(\rho, q^2) \cos\!\delta F_\rho(q^2) \big( \sin\!\delta_\rho(q^2) 
\!\! +\! \sin\!\delta_\rho(M_\rho^2) \big) \big) \!+\! \text{BW}(\omega, q^2) F_\omega(q^2)\nonumber\\
& & \times \big( 1 \!+\! \cos\!2\delta \!-\! 2 \sin\!\delta_\rho(M_\rho^2)  \sin\!\delta_\omega(q^2) \big) \bigg] \big( \tilde{d}_{123} m_\pi^2 \!+\! \tilde{d}_3 (q^2 \!+\! M_\rho^2) \big). \\
\mathcal{F}_{\omega\pi^0\gamma^{*}}^{\mathrm{1R}}(q^2) \!&=&\! \frac{-2}{3F M_V M_\omega} \bigg[ \sqrt{3} \cos\!\delta (2 \cos\!\theta_V \!+\! \sqrt{2} \sin\!\theta_V) \!-\! \sin\!\delta_\omega(M_\omega^2) \bigg] 
\!\!  \big( \tilde{c}_{125} q^2 \!+\! \tilde{c}_{1235} m_\pi^2 \!-\! \tilde{c}_{1256} M_\omega^2 \big),\nonumber\\
\mathcal{F}_{\omega\pi^0\gamma^{*}}^{\mathrm{2R}}(q^2) \!&=&\! \frac{2}{\sqrt{3}F M_\omega} \bigg[ 2 \text{BW}(\phi, q^2) \sin\!\delta_\omega(M_\omega^2) F_\phi(q^2) \big( \sqrt{2} \sin\!\theta_V \!-\! \cos\!\theta_V \big)
\!\! -\! \big( 2 \text{BW}(\omega, q^2) \cos\delta F_\omega(q^2)\nonumber\\
&&\times  \big( \sin\!\delta_\omega(q^2) \!+\! \sin\!\delta_\omega(M_\omega^2) \big) \big) 
\!\! +\! \text{BW}(\rho, q^2) F_\rho(q^2) \big( 1 \!+\! \cos\!2\delta \!-\! 2 \sin\!\delta_\rho(q^2) \sin\!\delta_\omega(M_\omega^2) \big) \nonumber\\
& & \times \big( \sqrt{2} \cos\!\theta_V \!+\! \sin\!\theta_V \big) \bigg] \big( \tilde{d}_{123} m_\pi^2 \!+\! \tilde{d}_3 (q^2 \!+\! M_\omega^2) \big).\\
\mathcal{F}_{\phi\pi^0\gamma^{*}}^{\mathrm{1R}}(q^2) \!&=&\! \frac{-2}{\sqrt{3}F M_V M_\phi} \big( \sqrt{2} \cos\!\theta_V - 2 \sin\!\theta_V \big) \big( \tilde{c}_{125} q^2 
\!\! +\! \tilde{c}_{1235} m_\pi^2 \!-\! \tilde{c}_{1256} M_\phi^2 \big),\nonumber\\
\mathcal{F}_{\phi\pi^0\gamma^{*}}^{\mathrm{2R}}(q^2) \!&=&\! \frac{4}{\sqrt{3}F M_\phi} \bigg[ \text{BW}(\rho, q^2) \cos\!\delta F_\rho(q^2) 
\!\! -\! \text{BW}(\omega, q^2) \sin\!\delta_\omega(q^2) F_\omega(q^2) \bigg] \nonumber\\
& & \times \big( \cos\!\theta_V \!-\! \sqrt{2} \sin\!\theta_V \big) \big( \tilde{d}_{123} m_\pi^2 \!+\! \tilde{d}_3 (q^2 \!+\! M_\phi^2) \big).
\end{eqnarray}

\begin{eqnarray}
\mathcal{F}_{\rho\eta\gamma^{*}}^{\mathrm{1R}}(q^2) \!&=&\! \frac{4}{9 F M_V M_{\rho}} \bigg\{ C_s \Big[ \sin\!\delta_\rho(M_{\rho}^2) \Big( 3 \sqrt{2} \tilde{c}_{8} M_V^2 \sin\!\theta_V \nonumber\\
\!& &\! -\! \sqrt{3} \big(\sqrt{2} \cos\!\theta_V \!-\! 2 \sin\!\theta_V\big) \big( (\tilde{c}_{1235}\!-\!8 \tilde{c}_{3}) m_{\eta}^2 \!+\! \tilde{c}_{125} q^2 \!-\! \tilde{c}_{1256} M_{\rho}^2 \!+\! 8 \tilde{c}_{3} (2 m_K^2 \!-\! m_{\pi}^2) \big) \Big) \nonumber\\
\!& &\! +\! 3 \sqrt{6} \tilde{c}_{8} M_V^2 \cos\!\delta \Big] \!+\! \frac{1}{2} C_q \Big[ \!-\!(\tilde{c}_{1235}\!-\!8 \tilde{c}_{3}) m_{\eta}^2 \big( 9 \sqrt{2} \cos\!\delta \!+\! \sqrt{3} (\sqrt{2} \sin\!\theta_V \!+\! 2 \cos\!\theta_V) \sin\!\delta_\rho(M_{\rho}^2) \big) \nonumber\\
\!& &\! -\! \big(\tilde{c}_{125} q^2 \!+\! 8 \tilde{c}_{3} m_{\pi}^2\big) \big( 9 \sqrt{2} \cos\!\delta \!+\! \sqrt{3} (\sqrt{2} \sin\!\theta_V \!+\! 2 \cos\!\theta_V) \sin\!\delta_\rho(M_{\rho}^2) \big) \nonumber\\
\!& &\! +\! \tilde{c}_{1256} M_{\rho}^2 \big( 9 \sqrt{2} \cos\!\delta \!+\! \sqrt{3} (\sqrt{2} \sin\!\theta_V \!+\! 2 \cos\!\theta_V) \sin\!\delta_\rho(M_{\rho}^2) \big) \nonumber\\
\!& &\! -\! 12 \tilde{c}_{8} M_V^2 \big( \sqrt{3} \cos\!\delta \!+\! \sin\!\theta_V \sin\!\delta_\rho(M_{\rho}^2) \big) \Big] \bigg\},\nonumber\\
\mathcal{F}_{\rho\eta\gamma^{*}}^{\mathrm{2R}}(q^2) \!&=&\! \frac{2}{3 F M_{\rho}} \bigg\{ F_\rho(q) \text{BW}(\rho, q^2) \Big[ C_s \big( \sin\!\delta_\rho(q^2) \sin\!\delta_\rho(M_{\rho}^2) \big( (4 \sin\!2\theta_V \!+\! \sqrt{2} \cos\!2\theta_V \!-\! 3 \sqrt{2}) \nonumber\\
\!& &\! \times \big( (\tilde{d}_{123} \!-\! 8\tilde{d}_{2}) m_{\eta}^2 \!+\! 16\tilde{d}_{2} m_K^2 \!-\! 8\tilde{d}_{2} m_{\pi}^2 \!+\! \tilde{d}_{3} M_{\rho}^2 \!+\! \tilde{d}_{3} q^2 \big) \!-\! 4 \sqrt{3} \tilde{d}_{5} M_V^2 \big) \!-\! 4 \sqrt{3} \tilde{d}_{5} M_V^2 \cos^2\!\delta \big) \nonumber\\
\!& &\! +\! C_q \big( \big( (\tilde{d}_{123} \!-\! 8\tilde{d}_{2}) m_{\eta}^2 \!+\! 8\tilde{d}_{2} m_{\pi}^2 \!+\! \tilde{d}_{3} (M_{\rho}^2 \!+\! q^2) \big) \big( 6 \cos^2\!\delta \!+\! \sin\!\delta_\rho(q^2) \nonumber\\
\!& &\! \times (2 \sqrt{2} \sin\!2\theta_V \!+\! \cos\!2\theta_V \!+\! 3) \sin\!\delta_\rho(M_{\rho}^2) \big) \!+\! 2 \sqrt{6} \tilde{d}_{5} M_V^2 \big( \cos\!2\delta \!+\! 2 \sin\!\delta_\rho(q^2) \sin\!\delta_\rho(M_{\rho}^2) \!+\! 1 \big) \big) \Big] \nonumber\\
\!& &\! +\! \cos\!\delta F_\omega(q) \text{BW}(\omega, q^2) \Big[ \sin\!\delta_\rho(M_{\rho}^2) \big( C_s \big( (4 \sin\!2\theta_V \!+\! \sqrt{2} \cos\!2\theta_V \!-\! 3 \sqrt{2}) \nonumber\\
\!& &\! \times \big( (\tilde{d}_{123} \!-\! 8\tilde{d}_{2}) m_{\eta}^2 \!+\! 16\tilde{d}_{2} m_K^2 \!-\! 8\tilde{d}_{2} m_{\pi}^2 \!+\! \tilde{d}_{3} M_{\rho}^2 \!+\! \tilde{d}_{3} q^2 \big) \!-\! 4 \sqrt{3} \tilde{d}_{5} M_V^2 \big) \nonumber\\
\!& &\! +\! C_q \big( (2 \sqrt{2} \sin\!2\theta_V \!+\! \cos\!2\theta_V \!+\! 3) \big( (\tilde{d}_{123} \!-\! 8\tilde{d}_{2}) m_{\eta}^2 \!+\! 8\tilde{d}_{2} m_{\pi}^2 \!+\! \tilde{d}_{3} (M_{\rho}^2 \!+\! q^2) \big) \!+\! 4 \sqrt{6} \tilde{d}_{5} M_V^2 \big) \big) \nonumber\\
\!& &\! -\! 2 \sin\!\delta_\omega(q^2) \big( 3 C_q \big( (\tilde{d}_{123} \!-\! 8\tilde{d}_{2}) m_{\eta}^2 \!+\! 8\tilde{d}_{2} m_{\pi}^2 \!+\! \tilde{d}_{3} (M_{\rho}^2 \!+\! q^2) \big) \!+\! 2 \sqrt{3} \tilde{d}_{5} M_V^2 (\sqrt{2} C_q \!-\! C_s) \big) \Big] \nonumber\\
\!& &\! +\! F_\phi(q) \sin\!\delta_\rho(M_{\rho}^2) \text{BW}(\phi, q^2) \Big[ (\tilde{d}_{123} \!-\! 8\tilde{d}_{2}) m_{\eta}^2 \big( 2 \cos\!2\theta_V (\sqrt{2} C_q \!+\! 2 C_s) \nonumber\\
\!& &\! -\! \sin\!2\theta_V (C_q \!+\! \sqrt{2} C_s) \big) \!+\! 8\tilde{d}_{2} \big( C_s (2 m_K^2 \!-\! m_{\pi}^2) (4 \cos\!2\theta_V \!-\! \sqrt{2} \sin\!2\theta_V) \nonumber\\
\!& &\! +\! m_{\pi}^2 C_q (2 \sqrt{2} \cos\!2\theta_V \!-\! \sin\!2\theta_V) \big) \!+\! \tilde{d}_{3} (M_{\rho}^2 \!+\! q^2) \big( 2 \cos\!2\theta_V (\sqrt{2} C_q \!+\! 2 C_s) \nonumber\\
\!& &\! -\! \sin\!2\theta_V (C_q \!+\! \sqrt{2} C_s) \big) \Big] \bigg\}.
\end{eqnarray}

\begin{eqnarray}
\mathcal{F}_{\omega\eta\gamma^{*}}^{\mathrm{1R}}(q^2) \!&=&\! \frac{2}{9 F M_V M_{\omega}} \bigg\{ \cos\!\delta \sin\!\theta_V \Big[ 2 C_s \Big( 2 \sqrt{3} \big( (\tilde{c}_{1235} \!-\! 8\tilde{c}_{3}) m_{\eta}^2 \!+\! \tilde{c}_{125} q^2 \!-\! \tilde{c}_{1256} M_{\omega}^2 \nonumber\\
\!& &\! +\! 8\tilde{c}_{3} (2 m_K^2 \!-\! m_{\pi}^2) \big) \!+\! 3 \sqrt{2} \tilde{c}_{8} M_V^2 \Big) \!-\! C_q \Big( \sqrt{6} \big( (\tilde{c}_{1235} \!-\! 8\tilde{c}_{3}) m_{\eta}^2 \!+\! \tilde{c}_{125} q^2 \!-\! \tilde{c}_{1256} M_{\omega}^2 \nonumber\\
\!& &\! +\! 8\tilde{c}_{3} m_{\pi}^2 \big) \!+\! 12 \tilde{c}_{8} M_V^2 \Big) \Big] \!+\! 3 \sin\!\delta_\omega(M_{\omega}^2) \Big[ C_q \Big( 3 \sqrt{2} \big( (\tilde{c}_{1235} \!-\! 8\tilde{c}_{3}) m_{\eta}^2 \!+\! \tilde{c}_{125} q^2 \nonumber\\
\!& &\! -\! \tilde{c}_{1256} M_{\omega}^2 \!+\! 8\tilde{c}_{3} m_{\pi}^2 \big) \!+\! 4 \sqrt{3} \tilde{c}_{8} M_V^2 \Big) \!-\! 2 \sqrt{6} \tilde{c}_{8} M_V^2 C_s \Big] \!-\! 2 \sqrt{3} \cos\!\delta \cos\!\theta_V \Big[ \nonumber\\
\!& &\! (\tilde{c}_{1235} \!-\! 8\tilde{c}_{3}) m_{\eta}^2 \big( C_q \!+\! \sqrt{2} C_s \big) \!+\! \sqrt{2} C_s \big( \tilde{c}_{125} q^2 \!+\! 16\tilde{c}_{3} m_K^2 \!-\! 8\tilde{c}_{3} m_{\pi}^2 \big) \!+\! C_q \big( \tilde{c}_{125} q^2 \nonumber\\
\!& &\! +\! 8\tilde{c}_{3} m_{\pi}^2 \big) \!-\! \tilde{c}_{1256} M_{\omega}^2 \big( C_q \!+\! \sqrt{2} C_s \big) \Big] \bigg\},\nonumber\\
\mathcal{F}_{\omega\eta\gamma^{*}}^{\mathrm{2R}}(q^2) \!&=&\! \frac{2}{3 F M_{\omega}} \bigg\{ \cos\!\delta \Big[ F_\rho(q) \text{BW}(\rho, q^2) \Big( \sin\!\delta_\rho(q^2) \big( C_s \big( (4 \sin\!2\theta_V \!+\! \sqrt{2} \cos\!2\theta_V \!-\! 3 \sqrt{2}) \nonumber\\
\!& &\! \times \big( (\tilde{d}_{123} \!-\! 8\tilde{d}_{2}) m_{\eta}^2 \!+\! 16\tilde{d}_{2} m_K^2 \!-\! 8\tilde{d}_{2} m_{\pi}^2 \!+\! \tilde{d}_{3} M_{\omega}^2 \!+\! \tilde{d}_{3} q^2 \big) \!-\! 4 \sqrt{3} \tilde{d}_{5} M_V^2 \big) \nonumber\\
\!& &\! +\! C_q \big( (2 \sqrt{2} \sin\!2\theta_V \!+\! \cos\!2\theta_V \!+\! 3) \big( (\tilde{d}_{123} \!-\! 8\tilde{d}_{2}) m_{\eta}^2 \!+\! 8\tilde{d}_{2} m_{\pi}^2 \!+\! \tilde{d}_{3} (M_{\omega}^2 \!+\! q^2) \big) \nonumber\\
\!& &\! +\! 4 \sqrt{6} \tilde{d}_{5} M_V^2 \big) \big) \!-\! 2 \sin\!\delta_\omega(M_{\omega}^2) \big( 3 C_q \big( (\tilde{d}_{123} \!-\! 8\tilde{d}_{2}) m_{\eta}^2 \!+\! 8\tilde{d}_{2} m_{\pi}^2 \!+\! \tilde{d}_{3} (M_{\omega}^2 \!+\! q^2) \big) \nonumber\\
\!& &\! +\! 2 \sqrt{3} \tilde{d}_{5} M_V^2 (\sqrt{2} C_q \!-\! C_s) \big) \Big) \!+\! F_\phi(q) \text{BW}(\phi, q^2) \Big( (\tilde{d}_{123} \!-\! 8\tilde{d}_{2}) m_{\eta}^2 \big( 2 \cos\!2\theta_V (\sqrt{2} C_q \!+\! 2 C_s) \nonumber\\
\!& &\! -\! \sin\!2\theta_V (C_q \!+\! \sqrt{2} C_s) \big) \!+\! 8\tilde{d}_{2} \big( C_s (2 m_K^2 \!-\! m_{\pi}^2) (4 \cos\!2\theta_V \!-\! \sqrt{2} \sin\!2\theta_V) \nonumber\\
\!& &\! +\! m_{\pi}^2 C_q (2 \sqrt{2} \cos\!2\theta_V \!-\! \sin\!2\theta_V) \big) \!+\! \tilde{d}_{3} (M_{\omega}^2 \!+\! q^2) \big( 2 \cos\!2\theta_V (\sqrt{2} C_q \!+\! 2 C_s) \nonumber\\
\!& &\! -\! \sin\!2\theta_V (C_q \!+\! \sqrt{2} C_s) \big) \Big) \Big] \!+\! F_\omega(q) \text{BW}(\omega, q^2) \Big[ C_s \big( \cos^2\!\delta (4 \sin\!2\theta_V \!+\! \sqrt{2} \cos\!2\theta_V \!-\! 3 \sqrt{2}) \nonumber\\
\!& &\! \times \big( (\tilde{d}_{123} \!-\! 8\tilde{d}_{2}) m_{\eta}^2 \!+\! 16\tilde{d}_{2} m_K^2 \!-\! 8\tilde{d}_{2} m_{\pi}^2 \!+\! \tilde{d}_{3} M_{\omega}^2 \!+\! \tilde{d}_{3} q^2 \big) \!-\! 2 \sqrt{3} \tilde{d}_{5} M_V^2 \big( \cos\!2\delta \nonumber\\
\!& &\! +\! 2 \sin\!\delta_\omega(q^2) \sin\!\delta_\omega(M_{\omega}^2) \!+\! 1 \big) \big) \!+\! C_q \big( \big( (\tilde{d}_{123} \!-\! 8\tilde{d}_{2}) m_{\eta}^2 \!+\! 8\tilde{d}_{2} m_{\pi}^2 \!+\! \tilde{d}_{3} (M_{\omega}^2 \!+\! q^2) \big) \nonumber\\
\!& &\! \times \big( \cos^2\!\delta (2 \sqrt{2} \sin\!2\theta_V \!+\! \cos\!2\theta_V \!+\! 3) \!+\! 6 \sin\!\delta_\omega(q^2) \sin\!\delta_\omega(M_{\omega}^2) \big) \!+\! 2 \sqrt{6} \tilde{d}_{5} M_V^2 \nonumber\\
\!& &\! \times \big( \cos\!2\delta \!+\! 2 \sin\!\delta_\omega(q^2) \sin\!\delta_\omega(M_{\omega}^2) \!+\! 1 \big) \big) \Big] \bigg\}.
\end{eqnarray}

\begin{eqnarray}
\mathcal{F}_{\phi\eta\gamma^{*}}^{\mathrm{1R}}(q^2) \!&=&\! \frac{4}{9 F M_V M_{\phi}} \bigg\{ C_s \Big[ \sqrt{3} \big(\sqrt{2} \sin\!\theta_V \!+\! 2 \cos\!\theta_V\big) \big( (\tilde{c}_{1235} \!-\! 8\tilde{c}_{3}) m_{\eta}^2 \!+\! \tilde{c}_{125} q^2 \nonumber\\
\!& &\! -\! \tilde{c}_{1256} M_{\phi}^2 \!+\! 8\tilde{c}_{3} (2 m_K^2 \!-\! m_{\pi}^2) \big) \!+\! 3 \sqrt{2} \tilde{c}_{8} M_V^2 \cos\!\theta_V \Big] \!-\! \frac{1}{2} C_q \cos\!\theta_V \Big( \sqrt{6} \big( (\tilde{c}_{1235} \!-\! 8\tilde{c}_{3}) m_{\eta}^2 \nonumber\\
\!& &\! +\! \tilde{c}_{125} q^2 \!-\! \tilde{c}_{1256} M_{\phi}^2 \!+\! 8\tilde{c}_{3} m_{\pi}^2 \big) \!+\! 12 \tilde{c}_{8} M_V^2 \Big) \!+\! \sqrt{3} C_q \sin\!\theta_V \big( (\tilde{c}_{1235} \!-\! 8\tilde{c}_{3}) m_{\eta}^2 \nonumber\\
\!& &\! +\! \tilde{c}_{125} q^2 \!-\! \tilde{c}_{1256} M_{\phi}^2 \!+\! 8\tilde{c}_{3} m_{\pi}^2 \big) \bigg\},\nonumber\\
\mathcal{F}_{\phi\eta\gamma^{*}}^{\mathrm{2R}}(q^2) \!&=&\! \frac{2}{3 F M_{\phi}} \bigg\{ F_\phi(q) \text{BW}(\phi, q^2) \Big[ C_q \big( 4 \sqrt{6} \tilde{d}_{5} M_V^2 \!-\! (2 \sqrt{2} \sin\!2\theta_V \!+\! \cos\!2\theta_V \!-\! 3) \nonumber\\
\!& &\! \times \big( (\tilde{d}_{123} \!-\! 8\tilde{d}_{2}) m_{\eta}^2 \!+\! 8\tilde{d}_{2} m_{\pi}^2 \!+\! \tilde{d}_{3} (M_{\phi}^2 \!+\! q^2) \big) \big) \!-\! C_s \big( (4 \sin\!2\theta_V \!+\! \sqrt{2} \cos\!2\theta_V \!+\! 3 \sqrt{2}) \nonumber\\
\!& &\! \times \big( (\tilde{d}_{123} \!-\! 8\tilde{d}_{2}) m_{\eta}^2 \!+\! 16\tilde{d}_{2} m_K^2 \!-\! 8\tilde{d}_{2} m_{\pi}^2 \!+\! \tilde{d}_{3} M_{\phi}^2 \!+\! \tilde{d}_{3} q^2 \big) \!+\! 4 \sqrt{3} \tilde{d}_{5} M_V^2 \big) \Big] \nonumber\\
\!& &\! +\! \big( F_\rho(q) \sin\!\delta_\rho(q^2) \text{BW}(\rho, q^2) \!+\! \cos\!\delta F_\omega(q) \text{BW}(\omega, q^2) \big) \Big[ (\tilde{d}_{123} \!-\! 8\tilde{d}_{2}) m_{\eta}^2 \big( 2 \cos\!2\theta_V \nonumber\\
\!& &\! \times (\sqrt{2} C_q \!+\! 2 C_s) \!-\! \sin\!2\theta_V (C_q \!+\! \sqrt{2} C_s) \big) \!+\! 8\tilde{d}_{2} \big( C_s (2 m_K^2 \!-\! m_{\pi}^2) (4 \cos\!2\theta_V \!-\! \sqrt{2} \sin\!2\theta_V) \nonumber\\
\!& &\! +\! m_{\pi}^2 C_q (2 \sqrt{2} \cos\!2\theta_V \!-\! \sin\!2\theta_V) \big) \!+\! \tilde{d}_{3} (M_{\phi}^2 \!+\! q^2) \big( 2 \cos\!2\theta_V (\sqrt{2} C_q \!+\! 2 C_s) \nonumber\\
\!& &\! -\! \sin\!2\theta_V (C_q \!+\! \sqrt{2} C_s) \big) \Big] \bigg\}.
\end{eqnarray}
And $C_q\to C'_q\,\,,\,\,C_s\to -C'_s\,\,,\,\,m_{\eta }\to m_{\eta' }$ for $\mathcal{F}_{V\eta'\gamma^{*}}$ as before.

The $\mathcal{F} _{VP\gamma^{*} } ^{\mathrm{EM} }(q^2)$ is given as
\begin{equation}
    \mathcal{F} _{VP\gamma^{*} } ^{\mathrm{EM} }(q^2)=\frac{4\pi\alpha F_V(q^2)}{M_V}\mathcal{\bar{F}}_{P\gamma^*\gamma^*}^{-V}(M_V^2,q^2).
\end{equation}
Here the $\mathcal{\bar{F}}_{P\gamma^*\gamma^*}^{-V}(M_V^2,q^2)$ means setting $\mathrm{BW}(V,M_V^2)=0$ with other terms unchanged in $\mathcal{F}_{P\gamma^*\gamma^*}(M_V^2,q^2)$.

With these TFFs, the $V\to P\gamma$ and $P\to V\gamma$ decay widths are given as
\begin{equation}
    \Gamma_{V\to P\gamma}=\frac{\alpha(M_V^2-m_P^2)^3}{24M_V^3}|\mathcal{F} _{VP\gamma^{*} }(0)|^2,
\end{equation}
\begin{equation}
    \Gamma_{P\to V\gamma}=\frac{\alpha(m_P^2-M_V^2)^3}{8m_P^3}|\mathcal{F} _{VP\gamma^{*} }(0)|^2.
\end{equation}

\subsection{Off-shell widths of vector resonances}\label{app:BW}
The off-shell widths of the resonance $\rho$ is taken from Ref.~\cite{GomezDumm:2000fz,Dai:2013joa,Wang:2023njt}
\begin{equation}
 \Gamma_\rho\left(q^2\right)=\frac{M_\rho q^2}{96 \pi F^2}\left[\sigma_\pi^3\left(q^2\right) \theta\left(q^2-4 m_\pi^2\right)+\frac{1}{2} \sigma_K^3\left(q^2\right) \theta\left(q^2-4 m_K^2\right)\right] ,
\end{equation}
Where $\theta(x)$ is the step function, and $\sigma_P(x)=\sqrt{1-4m_P^2/x}$ is the phase space factor. Since $\omega$ and $\phi$ are quite narrow, we use their constant values in our TFFs. One has 
\begin{eqnarray}
    \Gamma_\omega(q^2)&=&\Gamma_\omega\,\theta(q^2-9m_\pi^2), \nonumber \\
    \Gamma_\phi(q^2)&=&\Gamma_\phi\,\theta(q^2-4m_K^2).
\end{eqnarray}
Here and after, the $\Gamma_{V,V',V''}$ are the physical decay widths of the vectors. 
The off-shell widths of heavier vector resonances, $V'$ and $V''$, are parameterized by momentum dependent forms \cite{Dai:2013joa,Wang:2023njt},
\begin{eqnarray}
    \Gamma_{\rho'}(q^2)&=&\Gamma_{\rho'}\frac{\sqrt{q^2}}{M_{\rho'}}\frac{\sigma^3_\pi(q^2)}{\sigma^3_\pi(M_{\rho'}^2)}\theta(q^2-4m_\pi^2), \nonumber\\
    \Gamma_{\rho''}(q^2)&=&\Gamma_{\rho''}\frac{\sqrt{q^2}}{M_{\rho''}}\frac{\sigma^3_\pi(q^2)}{\sigma^3_\pi(M_{\rho''}^2)}\theta(q^2-4m_\pi^2)\,, \nonumber\\
    \Gamma_{\phi'}(q^2)&=&\Gamma_{\phi'}\frac{\sqrt{q^2}}{M_{\phi'}}\frac{\sigma^3_K(q^2)}{\sigma^3_K(M_{\phi'}^2)}\theta(q^2-4m_K^2)\,, \nonumber\\
    \Gamma_{\phi''}(q^2)&=&\Gamma_{\phi''}\frac{\sqrt{q^2}}{M_{\phi''}}\frac{\sigma^3_K(q^2)}{\sigma^3_K(M_{\phi''}^2)}\theta(q^2-4m_K^2)\,, \nonumber\\
    \Gamma_{\omega'}(q^2)&=&\Gamma_{\omega'}\frac{M_{\omega'}^3}{(q^2)^{\frac{3}{2}}}\frac{\lambda^{\frac{3}{2}}(q^2,M_\rho^2,m_\pi^2)}{\lambda^{\frac{3}{2}}(M_{\omega'}^2,M_\rho^2,m_\pi^2)}\theta(q^2-(M_\rho+m_\pi)^2)\,, \nonumber\\
    \Gamma_{\omega''}(q^2)&=&\Gamma_{\omega''}\frac{M_{\omega''}^3}{(q^2)^{\frac{3}{2}}}\frac{\lambda^{\frac{3}{2}}(q^2,M_\rho^2,m_\pi^2)}{\lambda^{\frac{3}{2}}(M_{\omega''}^2,M_\rho^2,m_\pi^2)}\theta(q^2-(M_\rho+m_\pi)^2)\,,
\end{eqnarray}
where $\lambda(x,y,z)=x^2+y^2+z^2-2xy-2xz-2yz$. 

\subsection{NLO radiative corrections of Single-Dalitz decay}\label{app:RNLO}
The NLO corrections include three parts,
\begin{equation}
\delta^{\mathrm{NLO}}(x,y)=\delta^{\mathrm{virt}}(x,y)+\delta^{\mathrm{1\gamma IR}}(x,y)+\delta^{\mathrm{BS}}(x,y)\,,
\end{equation}
where $\delta^{\mathrm{virt}}$, $\delta^{\mathrm{1\gamma IR}}$, and $\delta^{\mathrm{BS}}$ correspond to the virtual radiative corrections, one-photon-irreducible contribution, and the bremsstrahlung correction, respectively. The one-fold NLO radiative corrections used in the Single-Dalitz decay is given by
\begin{equation}
    \delta^{\mathrm{NLO}}(x)=\frac{3}{8\beta_l}\frac{1}{(1+\frac{\nu_l^2}{2x})}\int_{-\beta_l}^{\beta_l} \mathrm{d} y ~\delta^{\mathrm{NLO}}(x,y)\Big[1+y^2+\frac{\nu_l^2}{x} \Big].
\end{equation}
\begin{figure}[htb]
\centering
\includegraphics[width=1\linewidth,height=0.24\textheight]{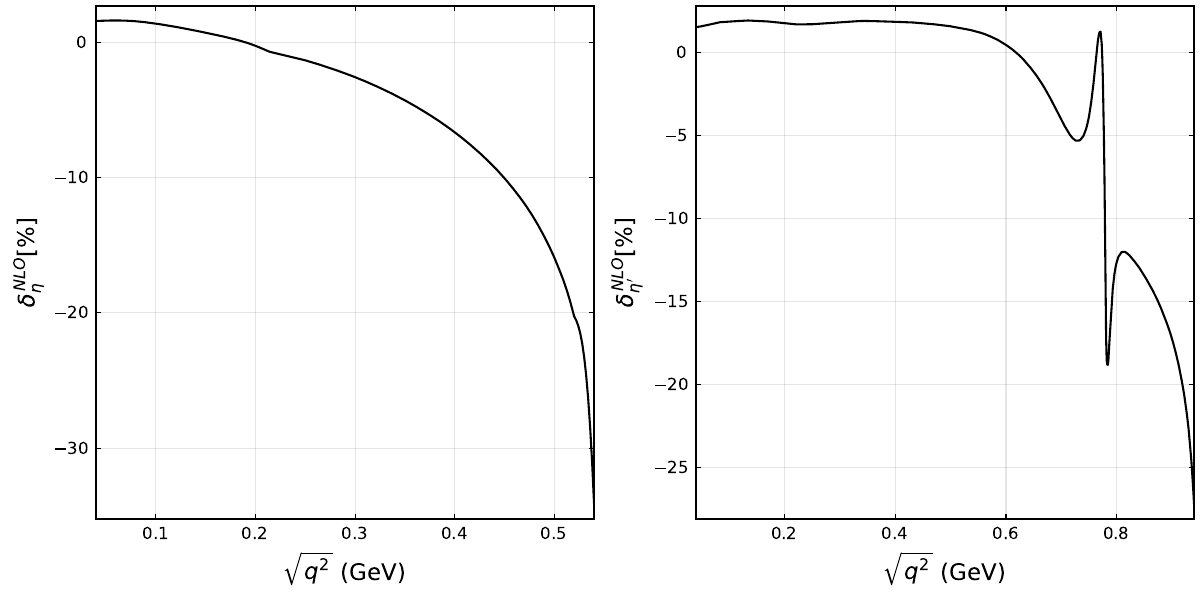}
\caption{NLO radiative corrections of Single-Dalitz decays $\eta\to \gamma e^+e^-$ and $\eta'\to \gamma e^+e^-$.} 
\label{Fig:NLO}
\end{figure}
$\delta^{\mathrm{virt}}$, the virtual radiative corrections, can be calculated analytically or given by dispersion integral, given in Ref.~\cite{Husek:2017vmo}
\begin{equation}
    \delta^{\mathrm{virt}}(x,y)=\frac{1}{|1+\Pi (m_P^2x)|}-1+2\mathrm{Re}\Bigg \{F_1(x)+\frac{2F_2(x)}{1+y^2+\frac{\nu_l^2}{x}} \Bigg \}, 
\end{equation}
where the kinematical variables $x$ and $y$ for Single-Dalitz decay $P(p)\to \gamma(k)\gamma^*(q) \to \gamma(k)l^+(p_1)l^-(p_2)$ are given by Ref.~\cite{Husek:2017vmo}, 
\begin{equation}
    x=\frac{q^2}{m_P^2}=\frac{(p_1+p_2)^2}{m_P^2},~y=-\frac{2}{m_P^2}\frac{p\cdot(p_1-p_2)}{(1-x)}.
\end{equation}
And $\gamma_{l}=(1-\beta_l)/(1+\beta_l)$, $\beta_l=\sqrt{1-\nu_l^2/x}$, $\nu_l=2m_l/m_P$.

The vacuum polarization function $\Pi (m_P^2x)$ includes contributions of leptonic and hadronic ones $\Pi (m_P^2x)=\Pi_{\mathrm{H}} (m_P^2x)+\Pi_{\mathrm{L}} (m_P^2x)$. The definitions of $\Pi_{\mathrm{H}} (m_P^2x)$ and $\Pi_{\mathrm{L}} (m_P^2x)$ are
\begin{eqnarray}
    \Pi_{\mathrm{H}}(m_P^2x)&=&-\frac{m_P^2x}{4 \pi^2 \alpha} \int_{4 m_\pi^2}^{\infty} \frac{\sigma_{\mathrm{H}}\left(s^{\prime}\right) \mathrm{d} s^{\prime}}{m_P^2x-s^{\prime}+i \epsilon},\nonumber\\
    \Pi_{\mathrm{L}} (m_P^2x)&=&\frac{\alpha}{\pi} \sum_{l^{\prime}=e, \mu}\left\{\frac{8}{9}-\frac{\beta_{l^{\prime}}^2}{3}+\left(1-\frac{\beta_{l^{\prime}}^2}{3}\right) \frac{\beta_{l^{\prime}}}{2} \log \left[-\gamma_{l^{\prime}}+i \epsilon\right]\right\}.
\end{eqnarray}
Here $\sigma_{\mathrm{H}}(s)$ is the total cross-section of $e^+e^-\to \mathrm{Hadrons}$. The definitions of $F_1(x)$ and $F_2(x)$ are given in Ref.~\cite{Husek:2015sma}. 
$\delta^{\mathrm{1\gamma IR}}$ and $\delta^{\mathrm{BS}}$ are given  in Ref.~\cite{Husek:2017vmo}. The former involves one-loop integral of doubly off-shell TFF, Ref.~\cite{Husek:2017vmo} calculated this correction using the VMD model. The bremsstrahlung correction $\delta^{\mathrm{BS}}$ is important for canceling infrared divergences, Ref.~\cite{Husek:2017vmo} calculated this correction using dispersive approach.
We do not list the formulas of these two terms as the final expression is very complicated and lengthy, but they can be found in  Ref.~\cite{Husek:2017vmo}.
The numerical results of  $\delta^{\mathrm{NLO}}(q^2)$ used in this work are shown in Fig.~\ref{Fig:NLO}.

\end{widetext}

\clearpage
\bibliographystyle{unsrt}
\bibliography{ref}

\begin{thebibliography}{100}

\bibitem{Greiner:2007zz}
Walter Greiner, Stefan Schramm, and Eckart Stein.
\newblock {\em Quantum Chromodynamics}.
\newblock Sprigner, 2007.

\bibitem{Weinberg:1978kz}
Steven Weinberg.
\newblock {Phenomenological Lagrangians}.
\newblock {\em Physica A}, 96(1-2):327--340, 1979.

\bibitem{Gasser:1983yg}
J.~Gasser and H.~Leutwyler.
\newblock {Chiral Perturbation Theory to One Loop}.
\newblock {\em Annals Phys.}, 158:142, 1984.

\bibitem{Eichmann:2019tjk}
Gernot Eichmann, Christian~S. Fischer, Esther Weil, and Richard Williams.
\newblock {Single pseudoscalar meson pole and pion box contributions to the
  anomalous magnetic moment of the muon}.
\newblock {\em Phys. Lett. B}, 797:134855, 2019.
\newblock [Erratum: Phys.Lett.B 799, 135029 (2019)].

\bibitem{Raya:2019dnh}
Kh\'epani Raya, Adnan Bashir, and Pablo Roig.
\newblock {Contribution of neutral pseudoscalar mesons to
  $a_\mu^{\mathrm{HLbL}}$ within a Schwinger-Dyson equations approach to QCD}.
\newblock {\em Phys. Rev. D}, 101(7):074021, 2020.

\bibitem{Nambu:1961tp}
Yoichiro Nambu and G.~Jona-Lasinio.
\newblock {Dynamical Model of Elementary Particles Based on an Analogy with
  Superconductivity. 1.}
\newblock {\em Phys. Rev.}, 122:345--358, 1961.

\bibitem{Klevansky:1992qe}
S.~P. Klevansky.
\newblock {The Nambu-Jona-Lasinio model of quantum chromodynamics}.
\newblock {\em Rev. Mod. Phys.}, 64:649--708, 1992.

\bibitem{Sakurai:1969ss}
J.~J. Sakurai.
\newblock {Vector meson dominance and high-energy electron proton inelastic
  scattering}.
\newblock {\em Phys. Rev. Lett.}, 22:981--984, 1969.

\bibitem{Schildknecht:2005xr}
Dieter Schildknecht.
\newblock {Vector meson dominance}.
\newblock {\em Acta Phys. Polon. B}, 37:595--608, 2006.

\bibitem{Witten:1998qj}
Edward Witten.
\newblock {Anti de Sitter space and holography}.
\newblock {\em Adv. Theor. Math. Phys.}, 2:253--291, 1998.

\bibitem{Cappiello:2010uy}
Luigi Cappiello, Oscar Cata, and Giancarlo D'Ambrosio.
\newblock The hadronic light by light contribution to the $(g-2)_{\mu}$ with
  holographic models of {QCD}.
\newblock {\em Phys. Rev. D}, 83:093006, 2011.

\bibitem{Gupta:1997nd}
Rajan Gupta.
\newblock {Introduction to lattice QCD: Course}.
\newblock In {\em {Les Houches Summer School in Theoretical Physics, Session
  68: Probing the Standard Model of Particle Interactions}}, pages 83--219, 7
  1997.

\bibitem{Gattringer:2010zz}
Christof Gattringer and Christian~B. Lang.
\newblock {\em Quantum chromodynamics on the lattice}, volume 788.
\newblock Springer, Berlin, 2010.

\bibitem{Colangelo:2001df}
G.~Colangelo, J.~Gasser, and H.~Leutwyler.
\newblock {$\pi \pi$ scattering}.
\newblock {\em Nucl. Phys. B}, 603:125--179, 2001.

\bibitem{Xiao:2000kx}
Zhiguang Xiao and H.~Q. Zheng.
\newblock {Left-hand singularities, hadron form-factors and the properties of
  the sigma meson}.
\newblock {\em Nucl. Phys. A}, 695:273--294, 2001.

\bibitem{Descotes-Genon:2006sdr}
S.~Descotes-Genon and B.~Moussallam.
\newblock {The K*0 (800) scalar resonance from Roy-Steiner representations of
  pi K scattering}.
\newblock {\em Eur. Phys. J. C}, 48:553, 2006.

\bibitem{Dai:2014zta}
Ling-Yun Dai and Michael~R. Pennington.
\newblock {Comprehensive amplitude analysis of $\gamma\gamma \rightarrow
  \pi^+\pi^-, \pi^0\pi^0$ and $\overline{K} K$ below 1.5 GeV}.
\newblock {\em Phys. Rev. D}, 90(3):036004, 2014.

\bibitem{Ecker:1988te}
G.~Ecker, J.~Gasser, A.~Pich, and E.~de~Rafael.
\newblock {The Role of Resonances in Chiral Perturbation Theory}.
\newblock {\em Nucl. Phys. B}, 321:311--342, 1989.

\bibitem{Ecker:1989yg}
G.~Ecker, J.~Gasser, H.~Leutwyler, A.~Pich, and E.~de~Rafael.
\newblock Chiral lagrangians for massive spin 1 fields.
\newblock {\em Phys. Lett. B}, 223:425--432, 1989.

\bibitem{Cirigliano:2006hb}
V.~Cirigliano, G.~Ecker, M.~Eidemuller, Roland Kaiser, A.~Pich, and
  J.~Portoles.
\newblock {Towards a consistent estimate of the chiral low-energy constants}.
\newblock {\em Nucl. Phys. B}, 753:139--177, 2006.

\bibitem{Kampf:2006yf}
Karol Kampf, Jiri Novotny, and Jaroslav Trnka.
\newblock {On different lagrangian formalisms for vector resonances within
  chiral perturbation theory}.
\newblock {\em Eur. Phys. J. C}, 50:385--403, 2007.

\bibitem{Portoles:2010yt}
J.~Portoles.
\newblock {Basics of Resonance Chiral Theory}.
\newblock {\em AIP Conf. Proc.}, 1322(1):178--187, 2010.

\bibitem{Kampf:2011ty}
Karol Kampf and Jiri Novotny.
\newblock Resonance saturation in the odd-intrinsic parity sector of low-energy
  {QCD}.
\newblock {\em Phys. Rev. D}, 84:014036, 2011.

\bibitem{Muong-2:2021ojo}
B.~Abi et~al.
\newblock Measurement of the positive muon anomalous magnetic moment to 0.46
  ppm.
\newblock {\em Phys. Rev. Lett.}, 126(14):141801, 2021.

\bibitem{Muong-2:2025xyk}
D.~P. Aguillard et~al.
\newblock {Measurement of the Positive Muon Anomalous Magnetic Moment to
  127~ppb}.
\newblock {\em Phys. Rev. Lett.}, 135(10):101802, 2025.

\bibitem{Muong-2:2006rrc}
G.~W. Bennett et~al.
\newblock {Final Report of the Muon E821 Anomalous Magnetic Moment Measurement
  at BNL}.
\newblock {\em Phys. Rev. D}, 73:072003, 2006.

\bibitem{Hoid:2020xjs}
Bai-Long Hoid, Martin Hoferichter, and Bastian Kubis.
\newblock {Hadronic vacuum polarization and vector-meson resonance parameters
  from $e^+e^-\rightarrow \pi ^0\gamma$}.
\newblock {\em Eur. Phys. J. C}, 80(10):988, 2020.

\bibitem{Benayoun:2021ody}
Maurice Benayoun, Luigi DelBuono, and Friedrich Jegerlehner.
\newblock {BHLS$_2$ upgrade: $\tau $ spectra, muon HVP and the [$\pi ^0,~\eta
  ,~{\eta ^\prime }$] system}.
\newblock {\em Eur. Phys. J. C}, 82(2):184, 2022.

\bibitem{Yi:2021ccc}
Jing-Yu Yi, Zhong-Yu Wang, and C.~W. Xiao.
\newblock {Study of the pion vector form factor and its contribution to the
  muon $g-2$}.
\newblock {\em Phys. Rev. D}, 104(11):116017, 2021.

\bibitem{Hoferichter:2021wyj}
Martin Hoferichter and Thomas Teubner.
\newblock {Mixed Leptonic and Hadronic Corrections to the Anomalous Magnetic
  Moment of the Muon}.
\newblock {\em Phys. Rev. Lett.}, 128(11):112002, 2022.

\bibitem{Colangelo:2022jxc}
G.~Colangelo et~al.
\newblock {Prospects for precise predictions of $a_\mu$ in the Standard Model}.
\newblock 3 2022.

\bibitem{Dai:2013joa}
L.~Y. Dai, J.~Portolés, and O.~Shekhovtsova.
\newblock Three pseudoscalar meson production in $e^{+}e^{-}$ annihilation.
\newblock {\em Physical Review D}, 88(5), September 2013.

\bibitem{Wang:2023njt}
Shi-Jia Wang, Zhen Fang, and Ling-Yun Dai.
\newblock {Two body final states production in electron-positron annihilation
  and their contributions to ${(g-2)_\mu}$}.
\newblock {\em JHEP}, 07:037, 2023.

\bibitem{Qin:2024ulb}
Bing-Hai Qin, Wen Qin, and Ling-Yun Dai.
\newblock {Study of electron-positron annihilation into $K\bar{K}\pi$ within
  resonance chiral theory}.
\newblock {\em Phys. Rev. D}, 111(3):034025, 2025.

\bibitem{Aliberti:2025beg}
R.~Aliberti et~al.
\newblock {The anomalous magnetic moment of the muon in the Standard Model: an
  update}.
\newblock 5 2025.

\bibitem{Roig:2014uja}
P.~Roig, A.~Guevara, and G.~L{\'o}pez~Castro.
\newblock {$VV^\prime P$ form factors in resonance chiral theory and the
  $\pi-\eta-\eta^\prime$ light-by-light contribution to the muon $g-2$}.
\newblock {\em Phys. Rev. D}, 89(7):073016, 2014.

\bibitem{Guevara:2018rhj}
A.~Guevara, P.~Roig, and J.~J. Sanz-Cillero.
\newblock {Pseudoscalar pole light-by-light contributions to the muon $(g-2)$
  in Resonance Chiral Theory}.
\newblock {\em JHEP}, 06:160, 2018.

\bibitem{Estrada:2024cfy}
Emilio~J. Estrada, Sergi Gonz{\`a}lez-Sol{\'\i}s, Adolfo Guevara, and Pablo
  Roig.
\newblock {Improved {\ensuremath{\pi}}$^{0}$, {\ensuremath{\eta}},
  {\ensuremath{\eta}}' transition form factors in resonance chiral theory and
  their $ {a}_{\mu}^{\textrm{HLbL}} $ contribution}.
\newblock {\em JHEP}, 12:203, 2024.

\bibitem{Estrada:2025bty}
Emilio~J. Estrada and Pablo Roig.
\newblock Tensor meson pole contributions to the {HLbL} piece of
  $a_{\mu}^{{\mathrm{HLbL}}}$ within {R$\chi$T}.
\newblock 4 2025.

\bibitem{Colangelo:2014dfa}
Gilberto Colangelo, Martin Hoferichter, Massimiliano Procura, and Peter
  Stoffer.
\newblock Dispersive approach to hadronic light-by-light scattering.
\newblock {\em JHEP}, 09:091, 2014.

\bibitem{Hoferichter:2018kwz}
Martin Hoferichter, Bai-Long Hoid, Bastian Kubis, Stefan Leupold, and
  Sebastian~P. Schneider.
\newblock {Dispersion relation for hadronic light-by-light scattering: pion
  pole}.
\newblock {\em JHEP}, 10:141, 2018.

\bibitem{Holz:2024lom}
Simon Holz, Martin Hoferichter, Bai-Long Hoid, and Bastian Kubis.
\newblock {Precision Evaluation of the {\ensuremath{\eta}}- and
  {\ensuremath{\eta}}'-Pole Contributions to Hadronic Light-by-Light Scattering
  in the Anomalous Magnetic Moment of the Muon}.
\newblock {\em Phys. Rev. Lett.}, 134(17):171902, 2025.

\bibitem{Kadavy:2022scu}
Tomas Kadavy, Karol Kampf, and Jiri Novotny.
\newblock {On the three-point order parameters of chiral symmetry breaking}.
\newblock {\em JHEP}, 03:118, 2023.

\bibitem{Gerardin:2023naa}
Antoine G{\'e}rardin, Willem E.~A. Verplanke, Gen Wang, Zoltan Fodor, Jana~N.
  Guenther, Laurent Lellouch, Kalman~K. Szabo, and Lukas Varnhorst.
\newblock {Lattice calculation of the {\ensuremath{\pi^0}}, {\ensuremath{\eta}}
  and {\ensuremath{\eta}}' transition form factors and the hadronic
  light-by-light contribution to the muon g-2}.
\newblock {\em Phys. Rev. D}, 111(5):054511, 2025.

\bibitem{ExtendedTwistedMass:2022ofm}
Constantia Alexandrou et~al.
\newblock $\eta \to \gamma^*\gamma^*$ transition form factor and the hadronic
  light-by-light $\eta$-pole contribution to the muon $g-2$ from lattice {QCD}.
\newblock {\em Phys. Rev. D}, 108(5):054509, 2023.

\bibitem{ExtendedTwistedMass:2023hin}
C.~Alexandrou et~al.
\newblock Pion transition form factor from twisted-mass lattice {QCD} and the
  hadronic light-by-light $\pi^0$-pole contribution to the muon $g-2$.
\newblock {\em Phys. Rev. D}, 108(9):094514, 2023.

\bibitem{Hayakawa:1997rq}
M.~Hayakawa and T.~Kinoshita.
\newblock {Pseudoscalar pole terms in the hadronic light by light scattering
  contribution to muon $g-2$}.
\newblock {\em Phys. Rev. D}, 57:465--477, 1998.
\newblock [Erratum: Phys.Rev.D 66, 019902 (2002)].

\bibitem{Danilkin:2019mhd}
Igor Danilkin, Christoph~Florian Redmer, and Marc Vanderhaeghen.
\newblock The hadronic light-by-light contribution to the muon's anomalous
  magnetic moment.
\newblock {\em Prog. Part. Nucl. Phys.}, 107:20--68, 2019.

\bibitem{Miranda:2018cpf}
J.~A. Miranda and P.~Roig.
\newblock {Effective-field theory analysis of the $\tau^-\to
  \pi^-\pi^0\nu_\tau$ decays}.
\newblock {\em JHEP}, 11:038, 2018.

\bibitem{Hoferichter:2023sli}
Martin Hoferichter, Gilberto Colangelo, Bai-Long Hoid, Bastian Kubis,
  Jacobo~Ruiz de~Elvira, Dominic Schuh, Dominik Stamen, and Peter Stoffer.
\newblock {Phenomenological Estimate of Isospin Breaking in Hadronic Vacuum
  Polarization}.
\newblock {\em Phys. Rev. Lett.}, 131(16):161905, 2023.

\bibitem{GomezDumm:2013sib}
D.~G{\'o}mez~Dumm and P.~Roig.
\newblock {Dispersive representation of the pion vector form factor in
  $\tau\to\pi\pi\nu_\tau$ decays}.
\newblock {\em Eur. Phys. J. C}, 73(8):2528, 2013.

\bibitem{Keshavarzi:2018mgv}
Alexander Keshavarzi, Daisuke Nomura, and Thomas Teubner.
\newblock {Muon $g-2$ and $\alpha(M_Z^2)$: a new data-based analysis}.
\newblock {\em Phys. Rev. D}, 97(11):114025, 2018.

\bibitem{Colangelo:2019uex}
Gilberto Colangelo, Franziska Hagelstein, Martin Hoferichter, Laetitia Laub,
  and Peter Stoffer.
\newblock {Longitudinal short-distance constraints for the hadronic
  light-by-light contribution to $(g-2)_\mu$ with large-$N_c$ Regge models}.
\newblock {\em JHEP}, 03:101, 2020.

\bibitem{Davier:2019can}
M.~Davier, A.~Hoecker, B.~Malaescu, and Z.~Zhang.
\newblock {A new evaluation of the hadronic vacuum polarisation contributions
  to the muon anomalous magnetic moment and to
  $\mathbf{\boldsymbol\alpha(m_Z^2)}$}.
\newblock {\em Eur. Phys. J. C}, 80(3):241, 2020.
\newblock [Erratum: Eur.Phys.J.C 80, 410 (2020)].

\bibitem{Keshavarzi:2019abf}
Alexander Keshavarzi, Daisuke Nomura, and Thomas Teubner.
\newblock {$g-2$ of charged leptons, $\alpha (M^2_Z)$ , and the hyperfine
  splitting of muonium}.
\newblock {\em Phys. Rev. D}, 101(1):014029, 2020.

\bibitem{Qin:2020udp}
Wen Qin, Ling-Yun Dai, and Jorge Portoles.
\newblock {Two and three pseudoscalar production in $\mathbf{e^+e^-}$
  annihilation and their contributions to $\mathbf{(g-2)_\mu}$}.
\newblock {\em JHEP}, 03:092, 2021.

\bibitem{Fedotovich:2024dpk}
G.~V. Fedotovich.
\newblock {Measurement of the
  $\boldsymbol{e}^{\boldsymbol{+}}\boldsymbol{e}^{\boldsymbol{-}}\boldsymbol{\to}{\pi}^{\boldsymbol{+}}{\pi}^{\mathbf{-}}$
  Cross Section below 1.2 {GeV} with the CMD-3 Detector at VEPP-2000 Collider}.
\newblock {\em Moscow Univ. Phys. Bull.}, 79(Suppl 1):86--92, 2024.

\bibitem{KLOE:2008fmq}
F.~Ambrosino et~al.
\newblock {Measurement of $\sigma(e^+ e^- \to \pi^+ \pi^- \gamma(\gamma))$ and
  the dipion contribution to the muon anomaly with the KLOE detector}.
\newblock {\em Phys. Lett. B}, 670:285--291, 2009.

\bibitem{KLOE:2010qei}
F.~Ambrosino et~al.
\newblock {Measurement of $\sigma(e^+ e^- \to \pi^+ \pi^-)$ from threshold to
  0.85 GeV$^2$ using Initial State Radiation with the KLOE detector}.
\newblock {\em Phys. Lett. B}, 700:102--110, 2011.

\bibitem{KLOE:2012anl}
D.~Babusci et~al.
\newblock {Precision measurement of $\sigma(e^+e^-\rightarrow
  \pi^+\pi^-\gamma)/ \sigma(e^+e^-\rightarrow \mu^+\mu^-\gamma)$ and
  determination of the $\pi^+\pi^-$ contribution to the muon anomaly with the
  KLOE detector}.
\newblock {\em Phys. Lett. B}, 720:336--343, 2013.

\bibitem{KLOE-2:2017fda}
A.~Anastasi et~al.
\newblock {Combination of KLOE
  $\sigma\big(e^+e^-\rightarrow\pi^+\pi^-\gamma(\gamma)\big)$ measurements and
  determination of $a_{\mu}^{\pi^+\pi^-}$ in the energy range $0.10 < s < 0.95$
  GeV$^2$}.
\newblock {\em JHEP}, 03:173, 2018.

\bibitem{SND:2020nwa}
M.~N. Achasov et~al.
\newblock {Measurement of the $e^+e^- \to\pi^+\pi^- $ process cross section
  with the SND detector at the VEPP-2000 collider in the energy region
  $0.525<\sqrt{s}<0.883$ GeV}.
\newblock {\em JHEP}, 01:113, 2021.

\bibitem{BESIII:2015equ}
M.~Ablikim et~al.
\newblock {Measurement of the $e^+ e^- \to \pi^+ \pi^-$ cross section between
  600 and 900 MeV using initial state radiation}.
\newblock {\em Phys. Lett. B}, 753:629--638, 2016.
\newblock [Erratum: Phys.Lett.B 812, 135982 (2021)].

\bibitem{Xiao:2017dqv}
T.~Xiao, S.~Dobbs, A.~Tomaradze, Kamal~K. Seth, and G.~Bonvicini.
\newblock {Precision Measurement of the Hadronic Contribution to the Muon
  Anomalous Magnetic Moment}.
\newblock {\em Phys. Rev. D}, 97(3):032012, 2018.

\bibitem{CMD-2:2005mvb}
V.~M. Aul'chenko et~al.
\newblock {Measurement of the pion form-factor in the range 1.04-GeV to
  1.38-GeV with the CMD-2 detector}.
\newblock {\em JETP Lett.}, 82:743--747, 2005.

\bibitem{Aulchenko:2006dxz}
V.~M. Aul'chenko et~al.
\newblock {Measurement of the $e^+ e^-\to\pi^+ \pi^-$ cross section with the
  CMD-2 detector in the 370-520 MeV c.m. energy range}.
\newblock {\em JETP Lett.}, 84:413--417, 2006.

\bibitem{CMD-2:2006gxt}
R.~R. Akhmetshin et~al.
\newblock {High-statistics measurement of the pion form factor in the rho-meson
  energy range with the CMD-2 detector}.
\newblock {\em Phys. Lett. B}, 648:28--38, 2007.

\bibitem{CMD-3:2023alj}
F.~V. Ignatov et~al.
\newblock {Measurement of the $e^+e^-\to\pi^+\pi^-$ cross section from
  threshold to 1.2~{GeV} with the CMD-3 detector}.
\newblock {\em Phys. Rev. D}, 109(11):112002, 2024.

\bibitem{CMD-3:2023rfe}
F.~V. Ignatov et~al.
\newblock {Measurement of the Pion Form Factor with CMD-3 Detector and its
  Implication to the Hadronic Contribution to Muon (g-2)}.
\newblock {\em Phys. Rev. Lett.}, 132(23):231903, 2024.

\bibitem{BaBar:2013jqz}
J.~P. Lees et~al.
\newblock {Precision measurement of the $e^+e^- \to K^+K^-(\gamma)$ cross
  section with the initial-state radiation method at BABAR}.
\newblock {\em Phys. Rev. D}, 88(3):032013, 2013.

\bibitem{Jegerlehner:2009ry}
Fred Jegerlehner and Andreas Nyffeler.
\newblock The muon g-2.
\newblock {\em Phys. Rept.}, 477:1--110, 2009.

\bibitem{Bijnens:1995cc}
Johan Bijnens, Elisabetta Pallante, and Joaquim Prades.
\newblock {Hadronic Light-by-Light Contribution to the Muon $g-2$}.
\newblock {\em Phys. Rev. Lett.}, 75:1447--1450, 1995.
\newblock [Erratum: Phys.Rev.Lett. 75, 3781 (1995)].

\bibitem{Holz:2024diw}
Simon Holz, Martin Hoferichter, Bai-Long Hoid, and Bastian Kubis.
\newblock {Dispersion relation for hadronic light-by-light scattering:
  {\ensuremath{\eta}} and {\ensuremath{\eta}}$^{\prime}$ poles}.
\newblock {\em JHEP}, 04:147, 2025.

\bibitem{Masjuan:2017tvw}
Pere Masjuan and Pablo Sanchez-Puertas.
\newblock {Pseudoscalar-pole contribution to the $(g_{\mu}-2)$: a rational
  approach}.
\newblock {\em Phys. Rev. D}, 95(5):054026, 2017.

\bibitem{Bijnens:2001cq}
Johan Bijnens, Elisabetta Pallante, and Joaquim Prades.
\newblock {Comment on the pion pole part of the light by light contribution to
  the muon g-2}.
\newblock {\em Nucl. Phys. B}, 626:410--411, 2002.

\bibitem{Bartos:2001pg}
E.~Bartos, A.~Z. Dubnickova, S.~Dubnicka, E.~A. Kuraev, and E.~Zemlyanaya.
\newblock Scalar and pseudoscalar meson pole terms in the hadronic light by
  light contributions to $a^{had}_{\mu}$.
\newblock {\em Nucl. Phys. B}, 632:330--342, 2002.

\bibitem{Kaiser:2000gs}
Roland Kaiser and H.~Leutwyler.
\newblock {Large N(c) in chiral perturbation theory}.
\newblock {\em Eur. Phys. J. C}, 17:623--649, 2000.

\bibitem{Kaiser:2005eu}
Roland Kaiser.
\newblock {Large N(c) in chiral resonance Lagrangians}.
\newblock In {\em {Hadrons and Strings Workshop: A Trento ECT Workshop}}, pages
  144--159, 2 2005.

\bibitem{Ruiz-Femenia:2003jdx}
P.~D. Ruiz-Femenia, A.~Pich, and J.~Portoles.
\newblock {Odd intrinsic parity processes within the resonance effective theory
  of QCD}.
\newblock {\em JHEP}, 07:003, 2003.

\bibitem{Landsberg:1985gaz}
L.~G. Landsberg.
\newblock Electromagnetic decays of light mesons.
\newblock {\em Phys. Rept.}, 128:301--376, 1985.

\bibitem{A2:2016sjm}
P.~Adlarson et~al.
\newblock Measurement of the $\pi^{0}\rightarrow e^{+}e^{-}\gamma$ dalitz decay
  at the mainz microtron.
\newblock {\em Phys. Rev. C}, 95:025202, Feb 2017.

\bibitem{NA62:2016zfg}
C.~Lazzeroni et~al.
\newblock Measurement of the $\pi^0$ electromagnetic transition form factor
  slope.
\newblock {\em Phys. Lett. B}, 768:38--45, 2017.

\bibitem{Adlarson:2016hpp}
P.~Adlarson et~al.
\newblock Measurement of the $\omega \to \pi^0 e^+e^-$ and $\eta \to e^+e^-
  \gamma$ dalitz decays with the {A2} setup at {MAMI}.
\newblock {\em Phys. Rev. C}, 95(3):035208, 2017.

\bibitem{NA60:2016nad}
R.~Arnaldi et~al.
\newblock Precision study of the $\eta\to\mu^+\mu^-\gamma$ and
  $\omega\to\mu^+\mu^-\pi^0$ electromagnetic transition form-factors and of the
  $\rho\to\mu^+\mu^-$ line shape in na60.
\newblock {\em Phys. Lett. B}, 757:437--444, 2016.

\bibitem{BESIII:2024pxo}
M.~Ablikim et~al.
\newblock Improved measurements of the dalitz decays $\eta/\eta'\to\gamma
  e^+e^-$.
\newblock {\em Phys. Rev. D}, 109(7):072001, 2024.

\bibitem{BESIII:2015zpz}
M.~Ablikim et~al.
\newblock Observation of the dalitz decay $\eta' \to \gamma e^+e^-$.
\newblock {\em Phys. Rev. D}, 92(1):012001, 2015.

\bibitem{BESIII:2015jiz}
M.~Ablikim et~al.
\newblock Observation of $\eta^{\prime}\to\omega e^{+} e^{-}$.
\newblock {\em Phys. Rev. D}, 92(5):051101, 2015.

\bibitem{SND:2016drm}
M.~N. Achasov et~al.
\newblock Study of the reaction $e^+e^- \to \pi^0\gamma$ with the {SND}
  detector at the {VEPP-2M} collider.
\newblock {\em Phys. Rev. D}, 93(9):092001, 2016.

\bibitem{Achasov:2000zd}
M.~N. Achasov et~al.
\newblock Experimental study of the processes $e^+ e^- \to \phi \to \eta
  \gamma, \pi^0 \gamma$ at {VEPP-2M}.
\newblock {\em Eur. Phys. J. C}, 12:25--33, 2000.

\bibitem{Achasov:2018ujw}
M.~N. Achasov et~al.
\newblock Measurement of the $e^+e^- \to \pi^0\gamma$ cross section in the
  energy range 1.075-2 {GeV} at {SND}.
\newblock {\em Phys. Rev. D}, 98(11):112001, 2018.

\bibitem{Achasov:2003ed}
M.~N. Achasov et~al.
\newblock Experimental study of the $e^+ e^- \to \pi^0 \gamma$ process in the
  energy region $\sqrt{s} = 0.60 - 0.97$ {GeV}.
\newblock {\em Phys. Lett. B}, 559:171--178, 2003.

\bibitem{Achasov:2006dv}
M.~N. Achasov et~al.
\newblock Study of the $e^+ e^- \to \eta \gamma$ process with spherical neutral
  detector at the {VEPP-2M} $e^+ e^- $ collider.
\newblock {\em Phys. Rev. D}, 74:014016, 2006.

\bibitem{Achasov:2013eli}
M.~N. Achasov et~al.
\newblock Study of the process $e^+e^-\to\eta\gamma$ in the center-of-mass
  energy range $1.07-2.00$ {GeV}.
\newblock {\em Phys. Rev. D}, 90(3):032002, 2014.

\bibitem{CMD-2:2004ahv}
R.~R. Akhmetshin et~al.
\newblock Study of the processes $e^+e^-\to\eta\gamma$, $\pi^0\gamma\to
  3\gamma$ in the c.m. energy range $600-1380$ mev at cmd-2.
\newblock {\em Phys. Lett. B}, 605:26--36, 2005.

\bibitem{CMD-2:2001dnv}
R.~R. Akhmetshin et~al.
\newblock Study of the process $e^+e^-\to\eta\gamma$ in center-of-mass energy
  range $600-1380$ {MeV} at {CMD-2}.
\newblock {\em Phys. Lett. B}, 509:217--226, 2001.

\bibitem{SND:2024qaq}
M.~N. Achasov et~al.
\newblock Search for the process $e^+e^-\to\eta'\gamma$ in the energy range
  $\sqrt{s}=1.075-2$ {GeV}.
\newblock {\em Phys. Rev. D}, 110(7):072004, 2024.

\bibitem{CELLO:1990klc}
H.~J. Behrend et~al.
\newblock A measurement of the $\pi^0$, $\eta$ and $\eta'$ electromagnetic
  form-factors.
\newblock {\em Z. Phys. C}, 49:401--410, 1991.

\bibitem{CLEO:1997fho}
J.~Gronberg et~al.
\newblock Measurements of the meson-photon transition form factors of light
  pseudoscalar mesons at large momentum transfer.
\newblock {\em Physics Review D}, 57(1):33--54, 1997.

\bibitem{BaBar:2009rrj}
Bernard Aubert et~al.
\newblock Measurement of the $\gamma \gamma^* \to \pi^0$ transition form
  factor.
\newblock {\em Phys. Rev. D}, 80:052002, 2009.

\bibitem{Belle:2012wwz}
S.~Uehara et~al.
\newblock Measurement of $\gamma \gamma^* \to \pi^0$ transition form factor at
  {Belle}.
\newblock {\em Phys. Rev. D}, 86:092007, 2012.

\bibitem{Redmer:2019zzr}
Christoph~Florian Redmer.
\newblock {Measurements of Hadronic and Transition Form Factors at BESIII}.
\newblock {\em EPJ Web Conf.}, 212:04004, 2019.

\bibitem{BaBar:2011nrp}
P.~del Amo~Sanchez et~al.
\newblock Measurement of the $\gamma \gamma^*\to \eta$ and $\gamma \gamma*
  \to\eta'$ transition form factors.
\newblock {\em Phys. Rev. D}, 84:052001, 2011.

\bibitem{L3:1997ocz}
M~Acciarri et~al.
\newblock Measurement of $\eta'(958)$ formation in two-photon collisions at
  {LEP1}.
\newblock {\em physics letters b}, 418(3–4):399--410, 1998.

\bibitem{BaBar:2018zpn}
J.~P. Lees et~al.
\newblock Measurement of the $\gamma^{\star}\gamma^{\star} \to \eta'$
  transition form factor.
\newblock {\em Phys. Rev. D}, 98(11):112002, 2018.

\bibitem{Lin:2024khg}
Tian Lin, Mattia Bruno, Xu~Feng, Lu-Chang Jin, Christoph Lehner, Chuan Liu, and
  Qi-Yuan Luo.
\newblock {Lattice QCD calculation of the {\ensuremath{\pi}}$^{0}$-pole
  contribution to the hadronic light-by-light scattering in the anomalous
  magnetic moment of the muon}.
\newblock {\em Rept. Prog. Phys.}, 88(8):080501, 2025.

\bibitem{Gasser:1984gg}
J.~Gasser and H.~Leutwyler.
\newblock Chiral perturbation theory: Expansions in the mass of the strange
  quark.
\newblock {\em Nucl. Phys. B}, 250:465--516, 1985.

\bibitem{Gasser:2020mzy}
J.~Gasser, H.~Leutwyler, and A.~Rusetsky.
\newblock On the mass difference between proton and neutron.
\newblock {\em Phys. Lett. B}, 814:136087, 2021.

\bibitem{Chen:2012vw}
Yun-Hua Chen, Zhi-Hui Guo, and Han-Qing Zheng.
\newblock {Study of $\eta-\eta'$ mixing from radiative decay processes}.
\newblock {\em Phys. Rev. D}, 85:054018, 2012.

\bibitem{Witten:1979vv}
Edward Witten.
\newblock {Current Algebra Theorems for the U(1) Goldstone Boson}.
\newblock {\em Nucl. Phys. B}, 156:269--283, 1979.

\bibitem{ParticleDataGroup:2024cfk}
S.~Navas et~al.
\newblock {Review of particle physics}.
\newblock {\em Phys. Rev. D}, 110(3):030001, 2024.

\bibitem{Feldmann:1998vh}
T.~Feldmann, P.~Kroll, and B.~Stech.
\newblock {Mixing and decay constants of pseudoscalar mesons}.
\newblock {\em Phys. Rev. D}, 58:114006, 1998.

\bibitem{Feldmann:1999uf}
Thorsten Feldmann.
\newblock {Quark structure of pseudoscalar mesons}.
\newblock {\em Int. J. Mod. Phys. A}, 15:159--207, 2000.

\bibitem{Guo:2015xva}
Xu-Kun Guo, Zhi-Hui Guo, Jose~Antonio Oller, and Juan~Jose Sanz-Cillero.
\newblock Scrutinizing the $\eta$-$\eta'$ mixing, masses and pseudoscalar decay
  constants in the framework of {U(3)} chiral effective field theory.
\newblock {\em JHEP}, 06:175, 2015.

\bibitem{Witten:1983tw}
Edward Witten.
\newblock {Global Aspects of Current Algebra}.
\newblock {\em Nucl. Phys. B}, 223:422--432, 1983.

\bibitem{Wess:1971yu}
J.~Wess and B.~Zumino.
\newblock {Consequences of anomalous Ward identities}.
\newblock {\em Phys. Lett. B}, 37:95--97, 1971.

\bibitem{Gasser:1982ap}
J.~Gasser and H.~Leutwyler.
\newblock {Quark Masses}.
\newblock {\em Phys. Rept.}, 87:77--169, 1982.

\bibitem{Husek:2015sma}
Tom\'a\ifmmode \check{s}\else~\v{s}\fi{} Husek, Karol Kampf, and Ji\ifmmode
  \check{r}\else~\v{r}\fi{}\'{\i} Novotn\'y.
\newblock Radiative corrections to the dalitz decay
  ${\ensuremath{\pi}}^{0}\ensuremath{\rightarrow}{e}^{+}{e}^{\ensuremath{-}}\ensuremath{\gamma}$
  revisited.
\newblock {\em Phys. Rev. D}, 92:054027, Sep 2015.

\bibitem{Husek:2017vmo}
Tom\'a\ifmmode \check{s}\else~\v{s}\fi{} Husek, Karol Kampf, Ji\ifmmode
  \check{r}\else~\v{r}\fi{}\'{\i} Novotn\'y, and Stefan Leupold.
\newblock Radiative corrections to the ${\ensuremath{\eta}}^{(\prime)}$ dalitz
  decays.
\newblock {\em Phys. Rev. D}, 97:096013, May 2018.

\bibitem{Afanasev:2023gev}
Andrei Afanasev, Jan~C. Bernauer, Peter Blunden, Johannes Blümlein, Ethan~W.
  Cline, Jan~M. Friedrich, Franziska Hagelstein, Tomá Husek, Michael Kohl, and
  Fred Myhrer.
\newblock Radiative corrections: from medium to high energy experiments.
\newblock {\em The European Physical Journal A}, 60(4), 2024.

\bibitem{Gan:2020aco}
Liping Gan, Bastian Kubis, Emilie Passemar, and Sean Tulin.
\newblock Precision tests of fundamental physics with $\eta$ and $\eta'$
  mesons.
\newblock {\em Phys. Rept.}, 945:1--105, 2022.

\bibitem{Lepage:1979zb}
G.~Peter Lepage and Stanley~J. Brodsky.
\newblock Exclusive processes in quantum chromodynamics: Evolution equations
  for hadronic wave functions and the form-factors of mesons.
\newblock {\em Phys. Lett. B}, 87:359--365, 1979.

\bibitem{Hoferichter:2020lap}
Martin Hoferichter and Peter Stoffer.
\newblock Asymptotic behavior of meson transition form factors.
\newblock {\em Journal of High Energy Physics}, 2020(5):159, 2020.

\bibitem{Braaten:1982yp}
Eric Braaten.
\newblock {QCD CORRECTIONS TO MESON - PHOTON TRANSITION FORM-FACTORS}.
\newblock {\em Phys. Rev. D}, 28:524, 1983.

\bibitem{Chai:2025xuz}
Jian Chai and Shan Cheng.
\newblock Form factors of light pseudoscalar mesons from the perturbative {QCD}
  approach.
\newblock {\em JHEP}, 06:229, 2025.

\bibitem{Roig:2014}
Pablo Roig and Juan~José Sanz~Cillero.
\newblock Consistent high-energy constraints in the anomalous {QCD} sector.
\newblock {\em Physics Letters B}, 733:158–163, June 2014.

\bibitem{Brodsky:2011yv}
Stanley~J. Brodsky, Fu-Guang Cao, and Guy~F. de~Teramond.
\newblock Evolved {QCD} predictions for the meson-photon transition form
  factors.
\newblock {\em Phys. Rev. D}, 84:033001, 2011.

\bibitem{Efron:1979bxm}
B.~Efron.
\newblock Bootstrap methods: Another look at the jackknife.
\newblock {\em Annals Statist.}, 7(1):1--26, 1979.

\bibitem{Escribano:2015nra}
R.~Escribano, P.~Masjuan, and P.~Sanchez-Puertas.
\newblock The $\eta $ transition form factor from space- and time-like
  experimental data.
\newblock {\em Eur. Phys. J. C}, 75(9):414, 2015.

\bibitem{Escribano:2015yup}
Rafel Escribano, Sergi Gonz\`alez-Sol\'\i{}s, Pere Masjuan, and Pablo
  Sanchez-Puertas.
\newblock $\eta$' transition form factor from space- and timelike experimental
  data.
\newblock {\em Phys. Rev. D}, 94(5):054033, 2016.

\bibitem{BESIII:2022cul}
M.~Ablikim et~al.
\newblock Observation of the double dalitz decay $\eta'\to e^+e^-e^+e^-$.
\newblock {\em Phys. Rev. D}, 105(11):112010, 2022.

\bibitem{BESIII:2024ddb}
Medina Ablikim et~al.
\newblock {Search for the double Dalitz decays
  {\ensuremath{\eta}}/{\ensuremath{\eta}}'{\textrightarrow}e+e-{\ensuremath{\mu}}+{\ensuremath{\mu}}-
  and
  {\ensuremath{\eta}}'{\textrightarrow}{\ensuremath{\mu}}+{\ensuremath{\mu}}-{\ensuremath{\mu}}+{\ensuremath{\mu}}-}.
\newblock {\em Phys. Rev. D}, 111(5):052002, 2025.

\bibitem{Escribano:2015vjz}
Rafel Escribano and Sergi Gonz{\`a}lez-Sol{\'\i}s.
\newblock {A data-driven approach to $\pi^{0}, \eta$ and $\eta^{\prime}$ single
  and double Dalitz decays}.
\newblock {\em Chin. Phys. C}, 42(2):023109, 2018.

\bibitem{Zhang:2025}
Yi-Hao Zhang, Shao-Zhou Jiang, and Ling-Yun Dai.
\newblock An anlaysis on $j/\psi\to\pi^0\gamma^*$ within resonance chiral
  theory, 2025.
\newblock arxiv: 2512.11515 [hep-ph].

\bibitem{BESIII:2025xjh}
Medina Ablikim et~al.
\newblock {Study of the electromagnetic Dalitz decay $J/\psi\to e^+e^-\pi^0$}.
\newblock {\em Phys. Rev. D}, 112(1):L011101, 2025.

\bibitem{Roskies:1990ki}
R.~Z. Roskies, Michael~J. Levine, and E.~Remiddi.
\newblock {Analytic evaluation of sixth order contributions to the electron's g
  factor}.
\newblock {\em Adv. Ser. Direct. High Energy Phys.}, 7:162--217, 1990.

\bibitem{GomezDumm:2000fz}
D.~Gomez~Dumm, A.~Pich, and J.~Portoles.
\newblock The hadronic off-shell width of meson resonances.
\newblock {\em Phys. Rev. D}, 62:054014, 2000.

\bibitem{Fodor:2024jyn}
Zoltan Fodor, Antoine Gerardin, Laurent Lellouch, Kalman~K. Szabo, Balint~C.
  Toth, and Christian Zimmermann.
\newblock {Hadronic light-by-light scattering contribution to the anomalous
  magnetic moment of the muon at the physical pion mass}.
\newblock {\em Phys. Rev. D}, 111(11):114509, 2025.

\end{thebibliography}
\end{document}